# Specifying and Implementing Security Policies Using LaSCO, the Language for Security Constraints on Objects

By

JAMES ALLEN HOAGLAND

B.S. (University of California, Davis) 1993
M.S. (University of California, Davis) 1996

DISSERTATION

Submitted in partial satisfaction of the requirements for the degree of

DOCTOR OF PHILOSOPHY

in

COMPUTER SCIENCE

in the

OFFICE OF GRADUATE STUDIES

of the

UNIVERSITY OF CALIFORNIA

DAVIS

Approved:

_______________________________________
Professor Raju Pandey, Chair                          Date

_______________________________________
Professor Karl Levitt                                       Date

_______________________________________
Professor Matthew Bishop                              Date

Committee in charge

2000

# Acknowledgments

Much of the work presented in this dissertation was supported by DARPA and the Intel Research Council, to whom I am grateful. I would like to thank Professors Raju Pandey, Karl Levitt, and Matt Bishop for their guidance on this research pursuit, along with their useful advice and comments on this dissertation. Professor Levitt graciously provided me with a research assistantship position throughout my tenure in the UCD Computer Security Lab, and allowed me to work undistracted in the later stages of this dissertation. Without these factors I could not have completed this work. I would like to thank the members of the Security Lab for their assistance in my research pursuits, including Dr. Christopher Wee for his mentorship, Nicole Carlson for her assistance in applying LaSCO to GrIDS and for reviewing an early draft of this dissertation, and David O'Brien for his attempts to keep the lab's computer systems afloat. Without my family's encouragement and the supportive upbringing they provided me, I could not have gotten this far. Last but not least, I would like to express my sincere appreciation to my numerous friends (especially Claire, Molly, Monica, Rosa, and Virginia; I am sorry that I cannot mention everyone) for their emotional support during my work on this dissertation.

James Allen Hoagland
March 2000
Computer Science

Specifying and Implementing Security Policies Using LaSCO, the Language for Security

Constraints on Objects

***Abstract***

In this dissertation, we present LaSCO, the Language for Security Constraints on Objects, a new approach to expressing security policies using policy graphs and present a method for enforcing policies so expressed. A security policy is a statement about how a system (any executing entity) should behave with respect to a site's particular notion of security. Other approaches for stating security policies fall short of what is desirable with respect to either policy clarity, executability, or the precision with which a policy may be expressed. This results in expressed policies that are ambiguous, are not implementable, or are that are not an accurate reflection of the policy goal, respectively. However, LaSCO is designed to have those three desirable properties of a security policy language as well as: relevance for many different systems, statement of policies at an appropriate level of detail, user friendliness for both casual and expert users, and amenability to formal reasoning. In LaSCO, the constraints of a policy are stated as directed graphs annotated with expressions describing the situation under which the policy applies and what the requirement is. LaSCO may be used for such diverse applications as executing programs, file systems, operating systems, distributed systems, and networks.

Formal operational semantics have been defined for LaSCO. An architecture for implementing LaSCO on any system, consisting of a system-independent policy interpretation engine and a system-specific interface layer, is presented along with an implementation of the engine in Perl. Using this, we have implemented LaSCO for Java programs. Our implementation prevents Java programs from violating policy through instrumented run time checks and includes a GUI to facilitate writing policies. This implementation is analyzed quantitatively and qualitatively. We have studied applying LaSCO to a network as viewed by GrIDS, a distributed intrusion detection system for large networks. A proposed design involves correlating partial policy matches in a hierarchy and sending alerts on violations. We conclude that LaSCO has characteristics that enable its use on different types of systems throughout the process of precisely expressing a policy, understanding the implications of a policy, and implementing it on a system.

# 1 Introduction

This dissertation presents a new approach to security policy specification and enforcement. In particular, we describe a policy language called LaSCO (the Language for Security Constraints on Objects). Using this language, users may write policies for many applications. We demonstrate that the language has features that promote ease of use, promote clear and precise statement of policy, and describe policies at the level of abstraction suitable for the problem. A formal operational semantics for LaSCO is described, which forms the basis for formal reasoning. We present a framework to apply LaSCO policies on their target system[1], and demonstrate its use though an implementation for Java and through a design for implementing it for networks using the GrIDS intrusion detection system. The work presented in this dissertation shows promise as a mechanism through which security policies can evolve, be formally reasoned about, and be implemented.

## 1.1 What is a security policy?

Goguen and Meseguer state in [13]: "A security policy … defines the security requirements for a given system." Thus a security policy is a statement about how a certain system should behave with respect to security. What is desired for security may not always be clear, and will vary by application. The military model of security is to prevent improper disclosure of sensitive information; an end user might want to limit who can access his sensitive files; an e-commerce site may want to prevent modification of transaction records; a financial institution may want to ensure certain procedures are always followed; and a developer of an in-house program may want to ensure that certain combinations of methods are never used for the same object. For some of these needs, formal models have been developed to capture the policy; for others it is up to a policy analyst -- hence the need for LaSCO.

---

1. We use "system" generally in this dissertation, to include any computational entity.

The type of security policy that this dissertation focuses on are those that impose constraints on the events and state of a system. Note that this includes the traditional notion of access control policies, i.e., restricting the access a subject has to an object. Beyond this, constraint policies may include restrictions on the allowed state of an object or system. We have been able to express certain types of non-constraint security policies[1] in LaSCO, but do not have a model of its expressiveness in these cases.

## 1.2 Why state a policy in a security policy language?

The current practice for stating security policies (when they are stated at all) is to use English. Due to the nature of natural language, these are usually vague, which can lead to ambiguity and misunderstanding. The translation to operating procedures or mechanized enforcement components is manual, which can be labor-intensive and subject to error. It is also difficult to reason about policies so stated, or to perform other formal operations on informally stated policies, such as composition.

However, if a security policy language is used, then a policy stated in it can be unambiguous, especially if the language has clearly stated semantics. Furthermore, using a formal language enables automated processing of specifications. Given the appropriate enforcement mechanisms such as is presented in this dissertation, this can enable the policies to be enforced with a minimum of human intervention. Formally stated policies allow the possibility of formally reasoning about cases that are covered and the overall effect of policies on a situation.

We believe that the primary impediment to more specific and mechanizable security policies is the unavailability of a formal language that supports the creation, analysis, understanding, and mechanization of real security policies.

## 1.3 Desirable characteristics of a security policy language

We feel that these are the desirable characteristics of a useful security policy language:

---

1. These include: policies indicating obligation, policies that indicate what should happen in certain situations (including intrusion detection response (IDR) policies), and policies describing trust relationships. (To our knowledge, there has not been any published work towards a formal taxonomy of different types of security policies.)

- **Clear statement of policy.** A policy stated in the language should be clear and be obvious in its meaning. There should be no ambiguity in the meaning of the policy.

- **Executable.** The language should be such that it lends itself to being enforced on a system. A stated policy should be able to be generated into implementation mechanisms. This would allow the expressed security properties to be maintained.

- **Applicable to many systems.** The language should be able to express policies for a variety of systems, both common and custom. This would avoid needing to learn a new language for each system with which one is concerned. In addition, when a new system is created, one would not need to invent a new policy language. Furthermore, one may be able to quickly express a particular policy for one system based on an example from another system if the same language is used.

- **Precise statement of desired policy.** The policy language should be able to express the policy the creator wishes to enforce. There should be a minimum of needs to compromise on intention or to unnecessarily represent it using multiple expressions of policy, as these are failures to express what the policy creator wants. Current models of security should expressible in the language.

- **Descriptive.** The language should be able to express the policy at the level of detail desired for an application. This would allow policies to be stated at multiple levels of abstraction. Thus the policy can convey the security intent of the creator. In addition, this ability would be useful in a framework for translating high level policies into lower level policies and enforcement mechanisms -- the same language can be used to represent requirements throughout.

- **User friendly.** Policies expressed in the language should be easy to write, modify, and understand for both casual and experienced users. This will increase the likelihood that the language will be used and that it will embody the policy the user intended. This is part of what Simon and Zurko [31] call "user centered security."

- **Amenable to formal reasoning.** The language should lend itself to reasoning about the completeness and correctness of policies stated in it and to other applications of formal reasoning. Thus one can be sure of the properties derived for policies stated in the language.

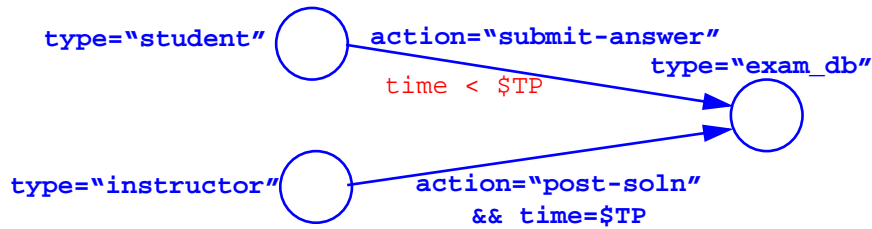

**Figure 1-1.** Policy graph for ordering of exam answer submission and solution posting restriction

## 1.4 LaSCO

LaSCO may be used to state security policies for a wide variety of environments. It may state restrictions on the execution of any system that has been described in terms of its system model. Systems that have been so described include executing programs, file systems, operating systems, distributed systems, and networks of hosts. In our model, a system consists of objects and events. Attributes on a object correspond to variables representing the state of an object. The attributes on an event denote the specifics of the event's execution, e.g., the time of execution and the event type. Events may be concurrent. There are typically several ways to describe a system in terms of the system model. A particular way in which a system is modeled is called a system description.

Policies in LaSCO are stated as policy graphs. These are annotated directed graphs. Consider the policy graph shown in Figure 1-1, depicting a constraint that might be found on a system that handles exam administration. The policy is that "a student submitting exam answers to an exam database should do so before the instructor posts the solution to that database." The annotations on the graph are domain predicates (depicted with **bold text**, for example **action="submit-answer"** above) and requirement predicates (depicted with `standard text`, for example `time < $TP`). Nodes in the policy graph represent any object described by the associated domain predicate. The top left node in Figure 1-1 represents a student, or, more precisely, any system object whose *type* attribute has value "student." Edges represent events in the system. To allow attributes on different objects and events to be interrelated in the policy, policy variables are present in the language. References to policy variables may be found in predicates, denoted prefixed by a dollar sign ("$"). A variable represents the same value throughout the policy.

Collectively, the nodes, edges, and domain predicates form the domain of a policy graph. The domain describes when the policy is in effect, i.e., when it applies. It applies if each of the domain predicates is matched with a different part of the system; this is called a match. Each variable is bound to a singe value in a successful match. In the example, upon a match, the variable TP becomes bound to the time of the posting event. The other part of the policy graph is the requirement, which states the restrictions on the system for whenever the domain matches. The requirement is made up of the requirement predicates found on particular nodes and edges. Each is an expression that must be met on the object or event that matched the node or edge in the domain. The only requirement present in the example policy graph is that the time of the exam solution submission be prior to the time of solution posting, represented in the value of TP from the domain.

The LaSCO system model and language have been formalized, and the language semantics fully defined in logic (Chapter 4). A system-independent policy interpretation engine has been developed (Chapter 5) and a policy enforcement mechanism has been developed for Java programs (Chapter 6). This mechanism has a user interface which facilitates writing policies in the context of a program. A policy compiler inserts the appropriate policy checks into a program; at run time these checks prevent the execution of method invocations that would violate the policies assigned to the program, as decided by the policy engine. A design has been developed in which policy violations on a network are detected in a distributed fashion using that policy engine and the GrIDS intrusion detection system[1] (Chapter 8).

In the next section and Chapter 9, we argue that no other policy statement method has strength in the properties described in the previous section to the degree that LaSCO does.

## 1.5 LaSCO and the desirable properties

LaSCO has characteristics that establish its strength in the desirable properties listed in Section 1.3. We take those individually:

- **Clear statement of policy.** LaSCO's formal semantics say exactly what the policy means, in terms of the system description.

- **Executable.** The system may be modeled as close to the concrete terms of the system as needed due to the generality of the LaSCO system model. This saves needing to translate terms. The executability also is supported by the existence of formal semantics, which allows precise interpretation of the policy with respect to system state and events.

---

1. GrIDS is an intrusion detection system for large networks developed at UC Davis. Novel aspects of GrIDS include its use of graphs as an aggregation method, the use of a hierarchy of departments to address scalability, and the incorporation of external data sources such as host-based IDSs.

- **Applicable to many systems.** The system model employed by LaSCO has a general basis. This allows LaSCO to describe policy for a diverse range of systems, as demonstrated in this dissertation. In fact, we believe any existing system can be described using out model. Furthermore, this also allows the language to express policies in terms customizable for an application, allowing it to be used in unique situations.

- **Precise statement of desired policy.** LaSCO has features to help in expressing policies. These include:

  - A policy's domain describes the type of situation under which it should apply directly rather than enumerating particular objects, avoiding the need to divide a policy along these lines. Along the same lines, the requirement of the policy is an assertion on all the elements of its domain, stated as an expression on the domain's attributes. Thus one can keep the policy at a higher level, rather than subdividing it, e.g., into an enumeration of the particular ways to uphold or violate the policy.

  - Using domain predicates, the objects and events referred to in a particular policy may be described in terms of their (possibly dynamic) attribute values. Included in this is the ability to form expressions composed of comparisons of attributes to other values, including the values of other attributes. Through the use of policy variables, this includes attributes found on other objects or events.

  - Using a multiple edge policy, historical context (both that involving events that have occurred and past state) may be employed in making a policy decision.

  LaSCO can express instances of all of the standard safety models: multi-level security, discretionary access, Clark-Wilson [7], role-based access control, and Chinese Wall [5] as well as custom policies. In addition to these features, that LaSCO expresses policies using policy concepts instead of system primitives means that the policy can be closer to what was intended.

- **Descriptive.** LaSCO can state policy at various levels of detail. The level of detail is determined by the system description. The system description can be customized for a particular application, and the system model employed by LaSCO is flexible enough to allow this. To state a policy at a low level of detail, the system can be modeled close to the system.

- **User friendly.** An intent when developing LaSCO was to make it user friendly and it has some features present to promote this. The graphical basis to the language provides a visual metaphor to help users (especially those with less familiarity) understand policies depicted. In particular, results of the graphical basis are:

  - Represented objects are depicted exactly once in a policy graph, eliminating the need to mentally connect the multiple representations that would otherwise be present.

  - Edges are a metaphor for an event. The edge itself represents a kind of event. The edge also clearly depicts the relationship between the source object and the destination object of the event and their connection to the event.

  - The node depicting an object in the policy serves as a focal point for the representation of the events that surround the object. This means that all the events required for a policy to apply that involve an object are located contiguously and are easy to locate.

  Further supporting the argument that a graphical basis is friendly to users is the fact that graphs are widespread in its use and are familiar formalisms. Some users might find the predicates used to describe the domain and requirement familiar, as it resembles expressions in common programming languages such as Java, C/C++, and Perl and resemble logical expressions. The clear separation in LaSCO between the domain and requirement of a policy conveys the semantics of implication better than if the aspects of the policy were less distinct, which promotes understanding and implementing policies correctly.

- **Amenable to formal reasoning.** Although no overall framework has been devised for formal reasoning about LaSCO policies, we have taken the necessary first steps -- formally describing the language and system model and describing the semantics in full formal detail. Thus it is possible to translate a policy into first order logic. We anticipate that the language is amenable to formal reasoning.

## 1.6 Summary of Contributions

The research discussed in this dissertation provides the following key contributions:

- *A formal language called LaSCO for describing security policies.* LaSCO can be applied to any system that can be modeled as consisting of objects and events, each of which contains labeled attributes. This language is based on graphs and depicts the circumstances (in terms of system state and events, including their history) under which a policy applies and the requirement for when it does. As described in Section 1.4, this language has properties promoting its strength in several desirable characteristics of a policy language.

- *Operational formal semantics for LaSCO.* The formal semantics presented in this dissertation describe the meaning of a LaSCO policy in terms of a system description. The semantics use first order logic.

- *An architecture for applying LaSCO to any system that one can write policies for.* This architecture consists of a generic policy engine, which is not specific to any system, and an interface layer, whose implementation is particular to a system. The generic policy engine maintains a set of policies and responds to reports of changes in the system with any new policy violations that have occurred as a result. The interface layer conveys the objects and events of the underlying system to the policy engine in terms of the system description for the system.

- *An implementation of a generic policy engine for LaSCO.* A generic policy engine is implemented in Perl and can be interfaced with any system that LaSCO can write policies for. For particular policies, it checks either changes in a system history or a whole system history for policy violations.

- *An implementation of a mechanism for applying LaSCO to Java program execution, including means to facilitate writing policies for programs and means to prevent the execution of method invocations that would violate given policies.* A user interface enables a user to write LaSCO policies in the context of a program schema graph -- a view of a program specialized for writing policies. A policy compiler instruments Java source code with calls to a run time system for policy checking.

- *A design for implementing LaSCO for a network using the Graph-based Intrusion Detection System (GrIDS).* Modifications that can be made to GrIDS to monitor a network in a decentralized fashion for violations of LaSCO policies are described in this dissertation. A way of modeling a network for use in LaSCO is also presented.

- *Observations about enforcing policies using IDSs.* Based on experience in applying LaSCO to GrIDS, we conclude that policies and IDSs are mutually beneficial and make observations regarding design considerations for policy languages and IDSs if they are to be used together.

## 1.7 Dissertation Outline

The remainder of this dissertation is organized as follows: Chapter 2 presents how to use LaSCO for different systems and presents a scenario. We introduce LaSCO and its system basis in detail in Chapter 3, including several examples of LaSCO policies. Chapter 4 formalizes LaSCO and its system model and presents its formal semantics. In Chapter 5 we present an architecture for applying LaSCO to systems in general and our implementation of it. Chapter 6 describes our implementation of LaSCO to Java and an analysis of applying LaSCO to Java is presented in Chapter 7. We present a design for implementing LaSCO for networks, using the GrIDS intrusion detection system in Chapter 8. We compare our work to that of others in Chapter 9. Chapter 10 concludes and Chapter 11 is our reference list. The appendix presents details about our implementation that were not presented earlier.

# 2 Using LaSCO

Before going into the details of the LaSCO language in the next two chapters, this chapter takes a small step forward, presenting how LaSCO might be used. The method by which LaSCO is applied to different application domains ("systems") is described in Section 2.1. Section 2.2 presents an extended scenario in which LaSCO is used for Java, using the implementation described in Chapter 6.

## 2.1 How LaSCO can be used for different application domains

The LaSCO language can be used to describe policies for to a variety of systems. However, when a LaSCO policy is applied, it is in the context of a specific system. It is applied to a particular execution of the system, which we refer to as a ***system history***. A LaSCO policy describes patterns in a system history and includes an assertion to evaluate in situations where the pattern is found.

LaSCO may be applied to any system that can be modeled using LaSCO's system model. Thus a system model needs to be constructed for a system. A particular instance of the LaSCO system model is termed a ***system description***.

There can be many ways to model a system within the LaSCO system model. The choice of how to describe a system in accordance with the LaSCO system model effects what policies may be stated in LaSCO and how they are stated, as the elements of the system description constitute the building blocks for the policy specification. It has an effect on the ease with which policies are written. When modeling the system, it is important to keep in mind what is security relevant for the system and to be sure that that is represented in the system description.

One needs to know the correspondence between the terms in the system description and elements of the actual system in order to know the complete implication of a policy on the actual system it is applied to. This would be used in enforcement and reasoning. How best to record this, and in fact how to denote the system description, is likely specific to its intended use and is beyond the scope of this dissertation.

Once an adequate system description has been found for a system, then one can state policies using it. How policies are stated in LaSCO is described in the next chapter.

## 2.2 A session with LaSCO applied to a Java program

Joe Smith, president and founder of Smith Magic Software Solutions, no longer trusts his own software, the lifeblood of his business. While a graduate student, Joe had a burst of inspiration while letting his mind wander in an Advanced Algorithms course. He had figured out how to mechanically generate software packages for a large class of common business applications. Joe recognized the opportunity to pay off his student loans and then some. After a little research to verify the necessity for these applications, Joe decided to go into business for himself to commercially develop his solution (after quickly completing his Ph.D. of course). Though naive to business, he thought he could do okay on his own.

Joe modeled his business as a service bureau, where customers describe their need to Joe and he provides them with the software package that solves their need, typically within a day. He manages this through the use of a proprietary implementation of his "magic" solution in Java. He made a bad business decision though. He decided to outsource much of the implementation to a contract programmer, while he implemented the core functionality himself. However, though written just a couple years before the turn of the millennium, that programmer did not write Y2K compliant code. Joe decided to outsource solving the Y2K problems to a professional firm specializing in year 2000 issues. Unfortunately due to the lateness of his search, the only ones he could find (and afford), were located in the Independent Nation of Kuraq. After the fixes were nearly completed, he heard an FBI warning that foreign countries might conduct industrial espionage through access granted to software to complete Y2K fixes . Now Joe thinks this might be the case for him.

Rather than trying to analyze the complex code worked on by the Kuraqies, Joe decides to employ LaSCO and its implementation for Java (detailed in Section 6) to specify some policies for the part of his code that he does not trust and have them enforced. He decides that he wants to prevent the untrusted code from performing any system level output. He reckons that that will prevent any Trojan horse in his code from leaking out information about the program and its use.

In preparation for using LaSCO on his Java program, he runs the "extract_schema.pl" script on his source code to extract a schema and schema graph (abstract views Java code) for each source file, and runs "merge_schema_graphs.pl" to combine the schema graphs. He also downloads the schemas for the standard Java base classes, thoughtfully provided by the Java development community[1]. Next Joe runs the security policy editor GUI described in Section 6.4. He brings up the combined schema graph representing his program on the left side of the interface for use as reference when writing policies for the program on the right side of the interface.

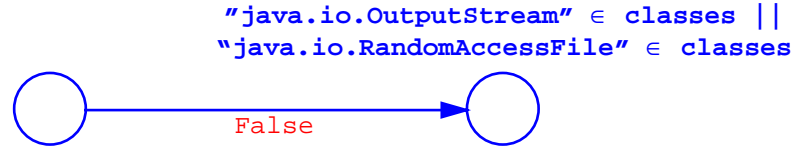

**Figure 2-1.** Policy graph for Joe's first policy

The policy that Joe wishes to state is that no code in the untrusted classes should access (using any method or constructor) the classes provided by Java to perform output. Specifically, these would be a subclass of *java.io.OutputStream* or *java.io.RandomAccessFile* or perhaps a subclass of that. In LaSCO, policies are viewed as having a domain and a requirement. The domain describes the situation under which the policy comes into force and the requirement is an assertion on the code. Joe recognizes that the domain here refers to an attempt is made by any class (or class instance) to use any method on one of the aforementioned classes. Since there is no circumstances under which that is allowed, there is no way to satisfy the requirement. In LaSCO this is denoted by "False". Using the user interface, Joe creates the policy depicted in Figure 2-1. There, the domain is depicted with **bold text** and the requirement with `standard text`. Nodes represent classes and class instances and edges represent method invocations. Domain predicates, when they appear, refine the type of entity that matches with that node or edge. When applied to Java, *classes* denotes the class associated with an class or instance plus its superclasses. He saves the created policy to a file.

Now Joe uses the policy compiler to modify the untrusted code to enforce the policy. To do this, he provides the compiler: the file containing the policy, the source code for the untrusted code, and the schemas for his trusted code and the standard Java classes. This causes the provided source to be instrumented with run time policy checks before key method invocations. This is then recompiled using a standard Java compiler.

---

1. Note that this is not currently available. The most direct way in which it could be derived is running the extract_shema.pl script on its source code. A version of the schemas could also be obtained from the skeletons of those classes present in the Java Language Specification [15] (albeit with certain limitations).

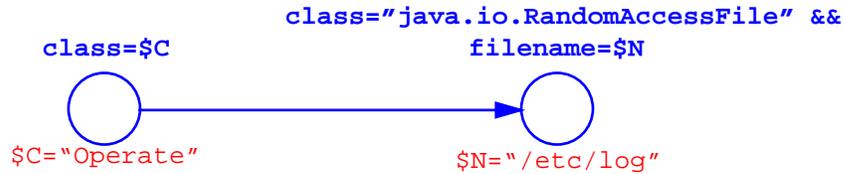

**Figure 2-2.** Policy graph for Joe's RandomAccessFile policy

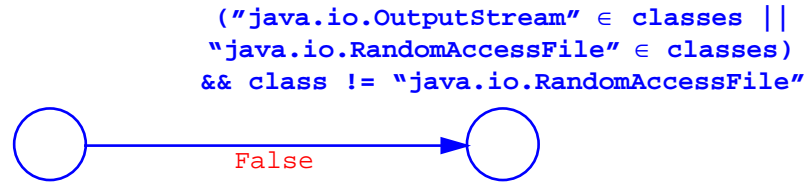

**Figure 2-3.** Policy graph for Joe's revised first policy

When Joe tries using his program after this, he realizes a problem in the policy he devised. A *PolicyViolationException* is triggered when the program writes to the log file. So, in the policy editor, he modifies the policy. Looking at the schema graph displayed, he sees that the class *Operate* makes calls to the class *java.io.RandomAccessFile*. He decides to allow this. So, rather than having a "False" requirement, Joe wishes the source node to be from the class *Operate* and the destination to be an instance of the class *java.io.RandomAccessFile* with *filename* "/etc/log". Since he is going to be making reference to a class member not found in all objects that might match the target in the old policy, he decided to keep things simple and create a separate policy to handle the *java.io.RandomAccessFile* class. The result is shown in Figure 2-2. Variables (the words in the predicate prefixed with a "$") are used here to record values for use in the requirement. He then modifies the domain of the original policy to exclude that class, producing the policy in Figure 2-3.

Saving this, he again runs the policy compiler, but with both the new policy and the revised policy this time. After compiling, he runs the program again and is happy with the policies.

Following this, Joe decides to write a policy for the core part of his program. He is not so much concerned about malicious code as with making sure the program is operating correctly. Thus the policy is very specific to his program. He wants to make sure that when the class *Stage2* and the class *Stage3* both call the *fulfill* method on an instance of the class *Product* and *Stage2*'s call has a *size* parameter at least 100, then the *size* parameter of *Stage3*'s call is larger than *Stage2*'s.

To write the policy, Joe again uses the user interface. To get a quick start at the policy, he decides to use the "create domain constraints" feature of the interface. So, he finds the location in the schema graph of his program displayed where the Stage2 and Stage3 nodes each have an edge labeled "fulfill" to the Product node. He selects these edges and presses the "create domain constraints" button. This creates the policy shown in Figure 2-4. He further refines this by editing the appropriate predicates until he creates the final policy shown in Figure 2-5. He uses the policy compile to add the policy to his code and is happy.

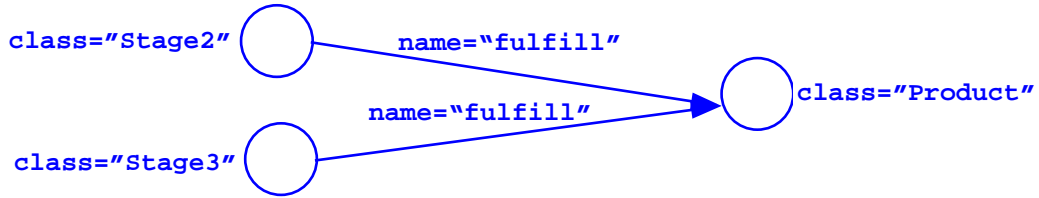

**Figure 2-4.** Initial policy graph created for Joe's "fulfill" policy.

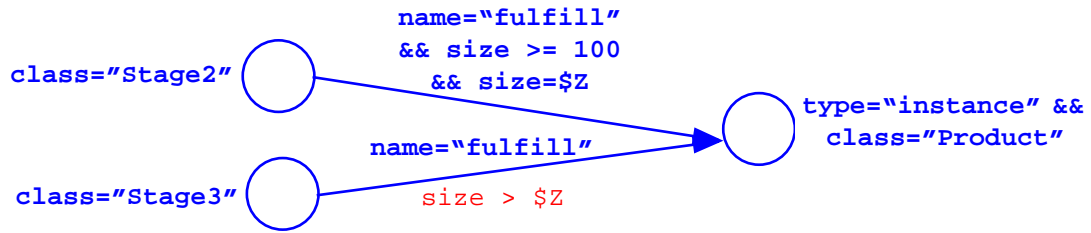

**Figure 2-5.** Final policy graph created for Joe's "fulfill" policy.

# **3** LaSCO Description

This chapter presents the security policy language LaSCO in detail. An overview of the system model is given in Section 3.1 and of LaSCO in Section 3.2. Many examples of LaSCO policies are presented in Section 3.3.

## 3.1 LaSCO system model overview

The LaSCO system model is simple, and closely resembles object-based systems. In the model, the real environment for which a policy is described, which we generically refer to as the ***system***, consists of a set of ***objects*** and a set of ***events***. Both objects and events contain a set of ***attributes***. A particular execution of the system is referred to as a ***system history***.

We now elaborate on the terms just introduced:

- An ***attribute*** is identified by a label. It has a certain value at a given time.

- An ***object*** is a system entity that has state. This state is represented by a set of attributes. An object persists over some duration of time, and the values of its attributes may change over this. All objects contain the attribute *id*, which uniquely identifies a object. This attribute does not change value.

- An ***event*** is momentary activity on a system. It originates at an object and terminates at an object. The details of an event are captured in a set of attributes. An attribute found on all events is *time*, which is the time the event occurs.

- A ***system history*** represents a particular execution of a system over a period of time. Typically this is the duration over which policy is being considered.

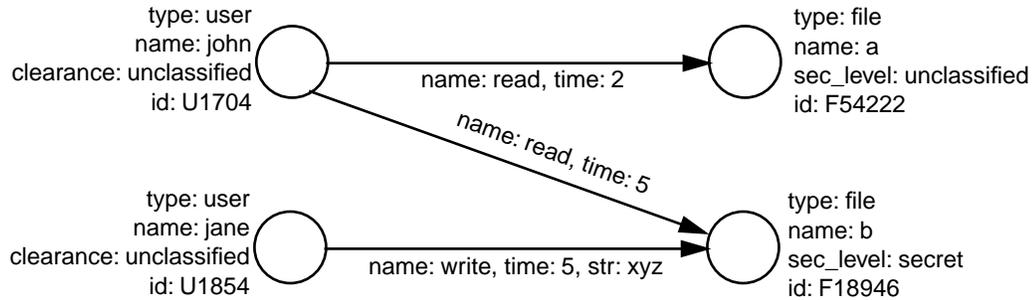

**Figure 3-1.** Example system history depicted as a graph. The nodes represent objects and the edges events.

- From the point of view of our policy language, something existing in the environment to which it is applied but not captured in the system description (Section 2.1) is beyond reference. Therefore, where clear, the term **system** will be used to mean the actual system as viewed from a LaSCO policy, that is, through its system description.

An example system history is presented in Figure 3-1. There are four objects and three events depicted. The user objects have four attributes each. The *type* attribute denotes the type of the object, the value of the *name* attribute is the name of the user, *clearance* represents the clearance level of the user, and *id* is the required unique identifier. Similarly, each event has a set of attribute values that denote the details of its invocation. In this simple example, each event has a attribute *name*, whose value is the name of the event and a attribute *time*, whose value is the time of its execution. The write event also contains an attribute *str*, denoting what is written.

## 3.2 LaSCO Overview

In this dissertation, security policies are described using a formal language based on directed graphs. This language is called LaSCO, the Language for Security Constraints on Objects. LaSCO **policy graphs** describe constraints on a system that must hold for a system execution. In a policy graph, nodes represent system objects and edges represent system events. Each policy graph represents both the situation under which a policy applies (the **domain**) and the constraint that must hold for the policy to be upheld (the **requirement**). Thus a LaSCO policy provides an assertion that indicates that if the system is in a specific state, the events and objects of the system must satisfy a set of properties.

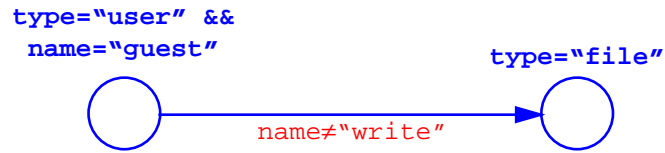

**Figure 3-2.** Policy graph for guest write access restriction.

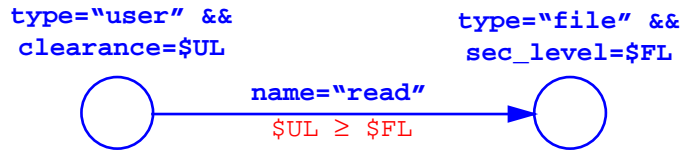

**Figure 3-3.** Policy graph for the simple security property

Figure 3-2 is an example policy graph. The policy depicted is that guests cannot write to any file. The left node represents the user named "guest" and the right node represents a file. The edge here represents an access by the guest to a file and the requirement is that the type of access not be a write.

Another example policy is shown in Figure 3-3. This policy graph depicts the simple security property of Bell-LaPadula [3]: if a user is reading a file, the security level of the user's clearance must be at least as great as that of the file's classification.

### 3.2.1 Predicates

Policy graphs are annotated directed graphs. The annotations on a policy graph are called ***predicates***. There is a ***domain predicate*** (depicted with **bold text** in this dissertation) and a ***requirement predicate*** (denoted with `standard text`) for each node and edge in the policy graph. For instance, there are three domain predicates (`type="user" && clearance=$UL`, `method="read"`, and `type="file" && sec_level=$FL`) and one explicit requirement predicate (`$UL ≥ $FL`) in Figure 3-3. (Nodes and edges without an explicit domain or requirement predicate have an implicit "True" predicate.) Domain predicates determine the objects and events that are relevant to a part of a policy. Requirement predicates describe what must hold once the domain is satisfied. Although they serve different roles, both types of predicates are evaluated in the same way. In either case a predicate is a pattern on the attributes of an object or event.

As a simplification, for the moment let us consider only predicates without variables, which we will term ***simple predicates***. Policy variables will be introduced in Section 3.2.3. A simple predicate is a boolean expression formed from attribute names and literals, combined with operators

**Logical Operators**
- **&&:** logical and
- **||:** logical (non-exclusive) or
- **!:** logical not (unary)

**Comparison Operators**
- **=:** numeric and string equality
- **!=:** numeric and string inequality
- **>:** numeric greater than
- **>=:** numeric greater than or equal to
- **<:** numeric less than
- **<=:** numeric less than or equal to
- **⊂:** proper subset
- **⊆:** subset
- **∈:** set membership

**Arithmetic Operations**
- **+:** addition
- **-:** subtraction
- **\*:** multiplication
- **/:** division
- **%:** modulo

**Set Operations**
- **∩:** intersection
- **∪:** intersection

**Nesting Operation**
- **():** nesting of contained expression

**Notes:**
- all operators are infix binary except as noted
- all binary operators are left-associative or non-associated
- "!" is right-associative
- the order of operator precedence is, from loosest to tightest (most immediate):
  - "&&", "||"
  - "=", "!="
  - "<", ">", "<=", ">="
  - "union", "intersect"
  - "pcont", "cont"
  - "in"
  - "+", "-"
  - "\*", "/", "%"
  - "!"

**Figure 3-4.** LaSCO predicate operators.

shown in Figure 3-4. Predicates are evaluated in the context of a set of attributes; applying the predicate consists of substituting in the corresponding value for each name and resolving the resulting boolean expression to be true or false. If it is true, then attributes satisfy the predicate.

### 3.2.2 Domain matching

The domain of a policy describes the conditions under which the policy applies. It is a set of domain predicates, nodes, and edges. The way in which a LaSCO policy is evaluated is identifying the locations in the system history where the domain matches, and checking the policy requirement for each match. Let us now consider the process of matching the domain of a policy to a particular elements of the system history for simple domain predicates.

The domain pattern is satisfied when each node and edge in the policy graph is satisfied by corresponding elements of a system history. This correspondence is called a ***policy to system (ps) map***. It consists of an object for each node and an event for each edge, in a one-to-one association. In the simple predicate case, a ps map is sufficient to constitute what is termed a ***policy to system match*** (***match*** for short). Section 3.2.4 will describe what else is needed if variables are present. A match represents a particular application of a domain to a system history. As the domain may apply in several ways in the system history or not at all, applying the domain to the system history produces a set of these matches, each of which must satisfy the requirement.

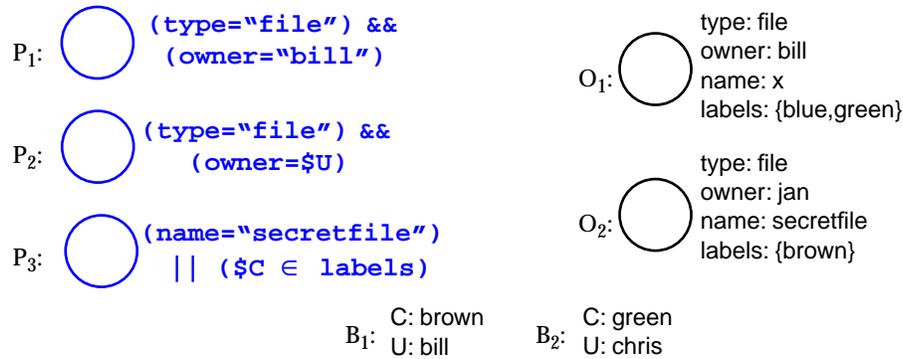

**Figure 3-5.** Example policy nodes, system objects, and variable bindings. $P_1$, $P_2$, and $P_3$ are policy nodes, $O_1$ and $O_2$ are system objects, and $B_1$ and $B_2$ are variable bindings.object.

Consider a policy edge $E$ being considered as matching a system event $e$ that occurs at time $t$. For $E$ to match $e$, these criteria must be met:

1. the domain of $E$ matches $e$
2. the domain of the source of $E$ matches the source of $e$ at time $t$
3. the domain of the destination of $E$ matches the destination of $e$ at time $t$

Note that due to the one-to-one relationship of a ps map, for each event incident with a particular object, the same policy node must be matched. Isolated policy nodes, those with no incident edges, are easier to match. They can match a system object at any time.

### 3.2.3 Variables

LaSCO policies may contain a set of ***policy variables***. A policy variable represents a value of an attribute and relates attribute values associated with different objects and events. Variables may appear as operands in domain and requirement predicates. They are denoted by a "$" prefix. The scope of a variable is a single LaSCO policy graph. Within this scope, each variable has a particular value. Variables are present in the language in order to permit the interrelation of attributes of different objects and events.

***Variable bindings*** represent a set of policy variables that have a bound value. Predicates are evaluated in the context of a set of variable bindings. We demonstrate predicate evaluation through a simple example. Figure 3-5 presents several example policy nodes, system objects, and variable bindings. The table in Figure 3-6 uses these to show the result of evaluating predicates of a domain pattern node in the context of a system objects and variable bindings. For example, P$_2$ evaluated in the context of O$_1$ and B$_1$ yields (`"file"="file"`) `&&` (`"bill"="bill"`), which is true, so the predicate is satisfied. This means that P$_2$ would match O$_1$ given B$_1$. However P$_3$ evaluated in the context of O$_1$ and B$_1$ yields (`"x"="secretfile"`) `||` (`"brown"` $\in$ {`"blue"`,`"green"`}), which is false, so P$_3$ is not satisfied. This means that P$_3$ would not match O$_1$ given B$_1$.

|                | O$_1$                              | O$_2$                              |
|----------------|------------------------------------|------------------------------------|
| P$_1$          | satisfied with any variable binding | not satisfiable by any variable binding |
| P$_2$          | satisfied by B$_1$ but not B$_2$   | not satisfied by B$_1$ nor B$_2$   |
| P$_3$          | satisfied by B$_2$ but not B$_1$   | satisfied by either B$_1$ or B$_2$ |

**Figure 3-6.** Predicate evaluation example. The table depicts which variable bindings satisfy each policy node's predicate when evaluated in the context of the object.

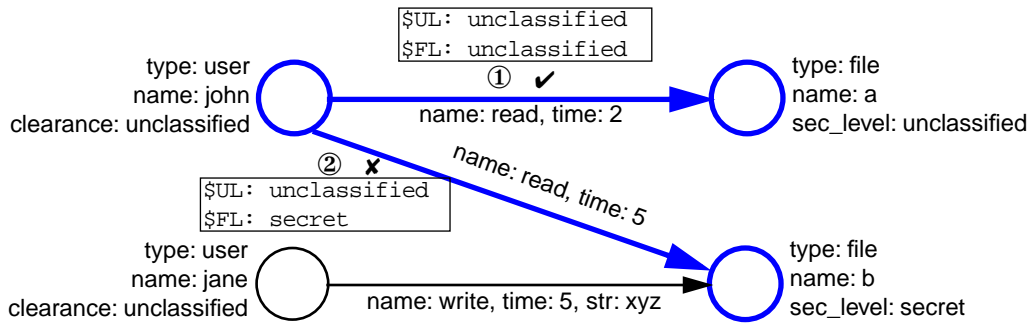

**Figure 3-7.** Depiction of simple security property applied to the example system. The two places where the domain applies is noted along with the necessary variable bindings.

### 3.2.4 Domain matching with variables

A policy domain is satisfied when all of its nodes and edges can simultaneously be satisfied by a set of variable bindings. That is, all of the domain predicates evaluate to true given a certain set of variable bindings and a ps map. Recall that a policy to system match as described for simple predicates contained only a ps map. Now that variables have been introduced, it must also contain the set of variable bindings that enable the map.

Let us now consider domain satisfaction for the policy of Figure 3-3 in the context of the example system history depicted in Figure 3-1. Figure 3-7 is a diagram that depicts the application of the policy domain to the system history by overlaying the policy graph on the system history graph.

We also show the necessary variable binding. When the domain of the policy is applied to the system history, the policy graph matches in two locations, indicated by the ① and the ② and the thick lines. Note that the domain did not match for the third event, which means that the event is not covered by the policy. This is because the given policy only covers read accesses.

Generally, predicates can contain any valid logical expression formed as described in Section 3.2.1. However, there are two restrictions. First, each variable present in a policy must appear in at least one domain predicate subexpression of the form `<variable>=<value>` or `<value>=<variable>` where `<value>` is a single value This restriction ensures that all variables have a single value for a domain.. To qualify, this subexpression with the "=" cannot be part of a disjunction or a negation. (This additional restriction is in place since otherwise that subexpression would be allowed to be false, which would deny us the single value binding.) Note that while could conceptually allow variables to become bound in the requirement, this would add complexity to the semantics and an implementation without additional expressability. It is simpler and cleaner to have all variables bound in a match. There is no additional ability to express policies since the domain predicates offer sufficient opportunity for any attribute value to be captured.

Second, node requirement predicates may not contain attributes as operands. The reason for this is that the attributes of an object might vary at the times of the events that match incident events, e.g., an object might have attribute $t$ with value 1 when one incident event matches a policy edge, but value 2 when a second incident event matches a policy edge. This would lead to ambiguity as to the correct value if the attribute is mentioned in a requirement predicate. Thus, this restriction is imposed to preclude this ambiguity. Variables may be bound to an attribute value in the domain and referred to in the requirement, which ensures that they hold a single value.

### 3.2.5 Requirement checking

The *requirement* of a policy is the set of requirement predicates in a policy graph. A requirement predicate is evaluated against a policy to system match. If each of the requirement predicates evaluates to true for a match, the policy has been upheld. Otherwise the policy has been violated by the match. Edge requirement predicates are evaluated in the context of the parameters of the event that matched the edge in the match and the variable bindings in a match. A node requirement predicate is evaluated in the context of the variable bindings from a match. The policy does not dictate a particular response to a policy violation.

Consider the matches in Figure 3-7 noted earlier. The policy requirement (`$UL` ≥ `$FL`) is satisfied by match ① but not ②. ① succeeds because `"unclassified"` ≥ `"unclassified"` is true but ② does not because `"unclassified"` ≥ `"secret"` is false. Thus, ① is allowed by the policy. However, ② is not.

A policy is said to be ***upheld*** by a system history if and only if all matches of the domain to the system history have their requirement satisfied. A policy is ***violated*** if there is a match where the requirement is not satisfied. This is the logical negation of a policy being upheld on the system.

A set of policies are often in effect on a system. The policies are composed to form an overall policy for the system. Under typical semantics, the overall policy is violated if and only if any of the individual policies are violated. The composed policy is upheld if each of the individual policies are upheld.

## 3.3 Examples

This section presents examples of LaSCO policy graphs. These policies are based on the traditional models as well as models for a particular application. To know the precise implication of the policies, a system description needs to be constructed to describe the correspondence between the LaSCO system model and the system upon which it is applied. This is generally left implicit here; it is assumed that the reader can infer the meaning of an attribute from its name. Section 3.3.1 through Section 3.3.3 present several examples roughly organized by type. Representing an Adage policy in LaSCO is discussed in Section 3.3.4.

### 3.3.1 Single-event restriction policies

The policies in this section involve a single event and its associated objects. This corresponds to a LaSCO policy graph containing a single edge and adjacent nodes. Policies like this do not make use of historical context (except inasmuch as the object state records it) and have particularly low complexity for checking whether they are violated or not.

### 3.3.1.1 State restriction for an event policies

The kind of policy discussed here restricts the state of objects on a system or requires a certain relationship between the source and destination of an event, for a certain type of event. The type of access that is restricted is represented in LaSCO in the domain predicates. A node's requirement predicate is the restriction on the state of objects it denotes. Variables in these predicates may be used to represent a condition between objects.

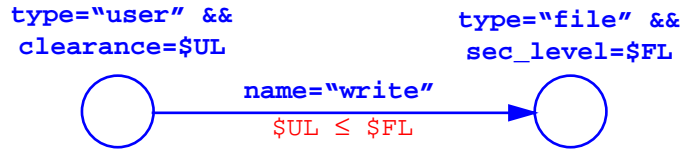

**Figure 3-8.** Policy graph for the star property of Bell-LaPadula.

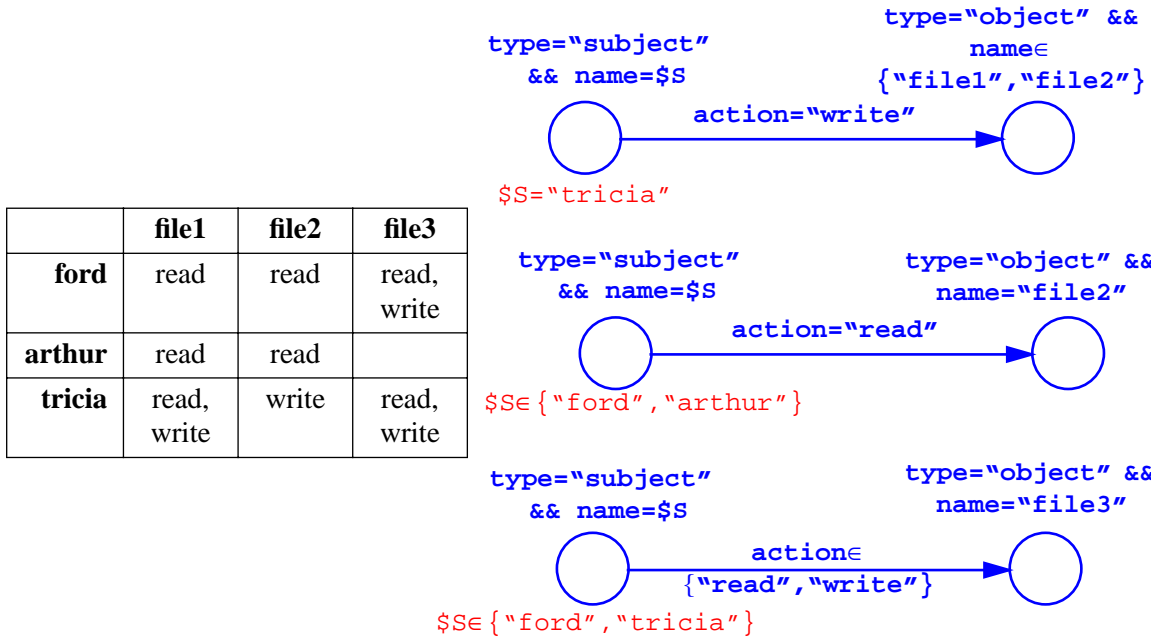

| | **file1** | **file2** | **file3** |
|---|---|---|---|
| **ford** | read | read | read, write |
| **arthur** | read | read | |
| **tricia** | read, write | write | read, write |

**Figure 3-9.** Access control matrix example indicating subject access rights over objects and a set of equivalent policy graphs.

The access restrictions of Bell-LaPadula [3] and Biba [4] fall into this category. Recall that the simple security property was depicted in Figure 3-3 on page 17, representing that if a user is reading a file, the security level of the user's clearance must be at least as great as that of the file's classification. The star property is depicted in Figure 3-8, denoting that a user's clearance must be at least as low as a files security classification for a write to be allowed.

### 3.3.1.2 ACM entry type policies

Policies of the type described in this section are restrictions for subjects regarding their access to objects. This is the type of restriction found in access control matrices (ACM) [23] and access control lists (see [11]).

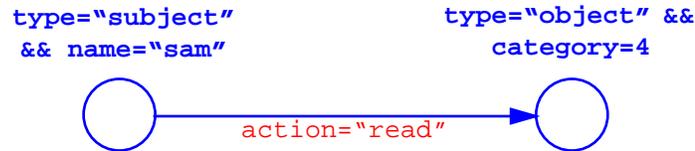

**Figure 3-10.** Policy graph for attribute ACL example

Consider the access control matrix in Figure 3-9. A set of policy graphs in that figure demonstrate one way to describe the restrictions in place using LaSCO policy graphs. Note that there are several such ways; this way focused on restricting the subject given a action and object. Translation of an access control matrix (or access control list) to policy graphs could easily be automated.

Policies generated from access control matrices do not take full advantage of the ability to select subjects and objects. Those are limited to listing the subjects or objects wanted. LaSCO can select objects based on the values of arbitrary attributes. A domain predicate that refers to local attributes limits the policy's applicability to objects or events that meet that restriction. For example, Figure 3-10 depicts the restriction that if subject "sam" is accessing an object in *category* 4, then it must be a read. No other types of access is allowed in this case. The case is similar for the actions. For access control list entries with wildcards that makes a policy applicable to all subjects, objects, or accesses is permitted in LaSCO by having that policy entity have no explicit domain predicate. Other types of generalized ACMs such as [25], additional predicate expressions can be added to restrict the domain or requirement appropriately.

### 3.3.1.3 Role-based access control policies

Role-based access control (RBAC) (see a review in Sandhu, *et.al.* [27]) is similar to access-matrix type restrictions discussed in Section 3.3.1.2. The major difference is in the subject. Whereas the subject in an ACM is a user, program, or process, the subject in RBAC is one of a defined set of roles. Every user with a certain role is treated the same with respect to access control. One might also select objects by attributes.

RBAC is used on a system with defined roles and roles that are (possibly dynamically) assigned to users. We will assume the system description for an example system denotes the roles a user currently has active by a *roles* set attribute on objects that are of the *type* subject. This is part of the way the system is modeled. Figure 3-11 denotes the RBAC policy that only subjects that have the "paymaster" role can issue an object of document type paycheck.

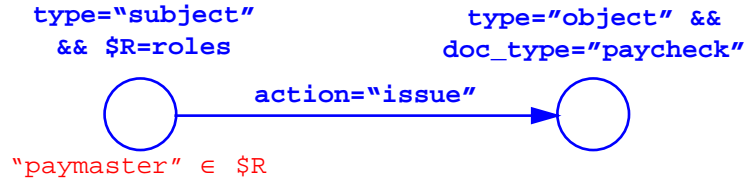

**Figure 3-11.** Policy graph for payroll RBAC example

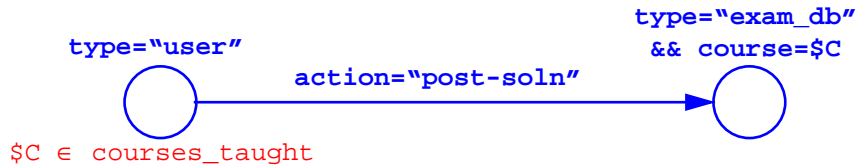

**Figure 3-12.** Policy graph for restriction on solution posting.

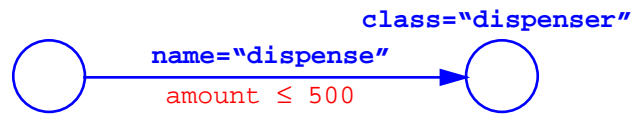

**Figure 3-13.** Policy graph for the ATM example property.

Roles may not explicitly denoted in some systems, but they may still have a similar effect. For example, consider Figure 3-12 which is a policy that might be in effect on a system used to administer exams. The restriction is that users can only post solutions for exams they teach.

### 3.3.1.4 Restricted event parameter access policies

Restricted event parameter access policies limit the values of the parameters to an event. This is denoted in LaSCO by an edge requirement predicate that states the restriction on event attributes. An edge domain predicate describes the kind of event being restricted. An additional limitation can be placed on the policy, to apply it to just a certain type of source or destination object. This is achieved through domain predicates on the edge and nodes.

As an example of this type of policy, consider a bank that operates an automated teller machine (ATM). One of their security interests is making sure that the machine does not give out too much money, either through misuse or a flaw in the control system. The ATM control system is written in Java. It includes a class "dispenser" which interfaces the cash dispenser hardware to the rest of the program. The dispenser class contains a "dispense" method, which tells the dispenser to release a certain amount of money. One policy the bank may wish to impose on the system is that no part of the control system should call the dispense method with an amount parameter greater than 500. This policy is shown in Figure 3-13.

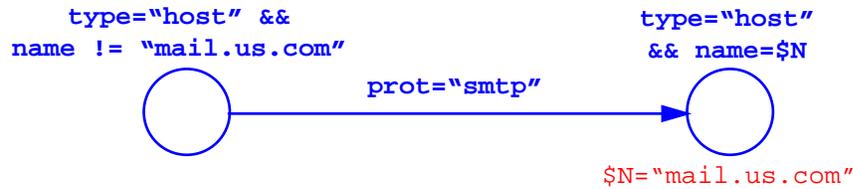

**Figure 3-14.** Policy graph for mail gateway policy.

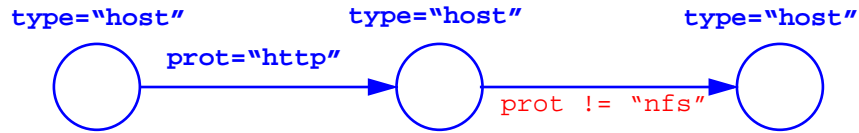

**Figure 3-15.** Policy graph for prohibition of HTTP-NFS chain.

### 3.3.1.5 Moderation policies

One may wish to restrict the source or destination of certain events to a particular object or set of objects. Consider, for example, the policy for a network shown in Figure 3-14. Here the destination is being restricted. The policy comes into effect when a host other than mail.us.com is sending mail. The restriction is that the destination must be mail.us.com. That is, all mail must be sent through mail.us.com.

### 3.3.2 Pattern of events policies

The policies in the previous section all had one edge and therefore matched a single event. Using multiple edges, one can describe policies that come into effect when multiple events occur. These make use of a part of the history of the system. This can be viewed as a pattern of events, since the policies indicate which the objects adjacent the different events are the same and which are not. A few subclasses of this are presented in the following subsections.

### 3.3.2.1 Event type restriction

One of the simpler type of pattern of event policies is when a restriction is placed on what one of the events can be. One example is in Figure 3-15, which is a policy that might be applied to a network. It indicates that when a host is accessed through HTTP, it should not access another computer through NFS. (The idea behind this might be to limit the possible access by someone accessing the web server.) The restriction imposed by this policy is that the protocol of any connection from the host that was accessed by HTTP to any other host cannot use NFS.

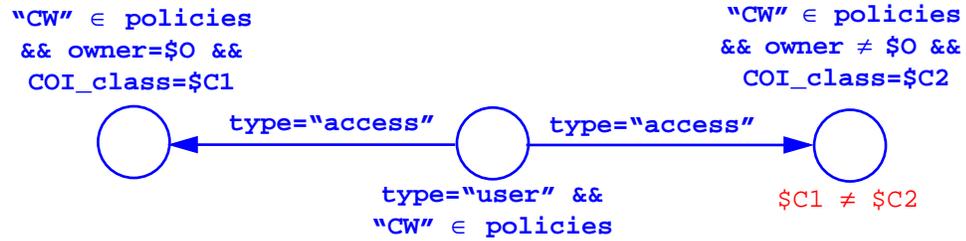

**Figure 3-16.** Policy graph for the Chinese Wall policy

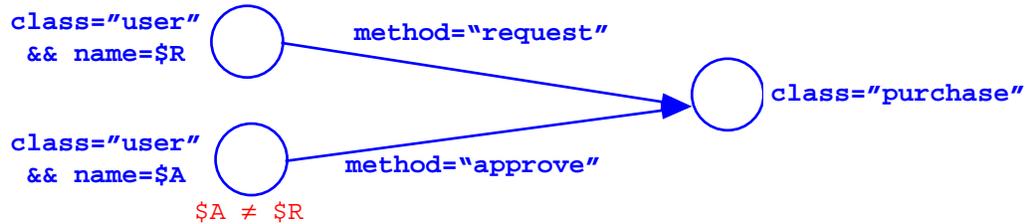

**Figure 3-17.** Policy graph for purchase request and approval separation of duty

### 3.3.2.2 Object comparison policies

A type of pattern of event policy is one in which two or more of the objects related by events are required to have a certain relationship. As an example in which there are two edges with the same origin, consider the Chinese Wall policy [4]. The idea behind the Chinese Wall policy is to prevent conflict of interest situations by consultants that may be employed by several parties with competing interests. The policy specifies separation of interests by forbidding any consultant from accessing data from different parties where the parties are in the same conflict of interest class. The policy is depicted in LaSCO by a node with two edges originating from it as shown in Figure 3-16. The middle node represents a "consultant" whose accesses are limited by the Chinese wall policy. The edges from the consultant node represent accesses to sensitive objects with different owners that are subject to the Chinese Wall policy. The constraint is that the owners of these objects cannot be in the same conflict of interest class, stored in the attribute "COI_class".

Separation of duty policies require different peoples' involvement in some transaction. In LaSCO this is depicted by two edges with different sources leading to either a single node or to nodes that are somehow linked through their domain predicates. An example is shown in Figure 3-17. This depicts a policy for a system where there is a separate function for requesting policies and having them approved. The policy states the restriction that the name of the "request" user must be different than the name of the "approve" user.

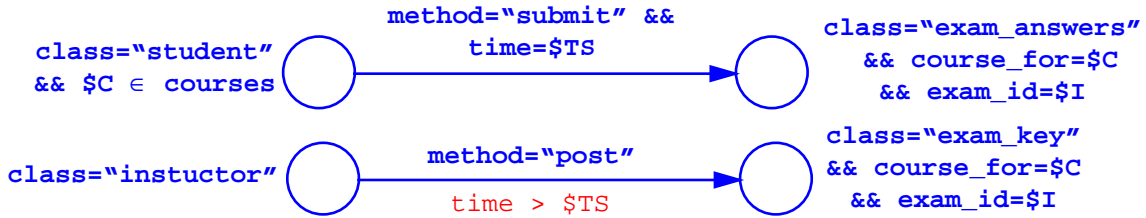

**Figure 3-18.** Policy graph for ordering of exam submission and key posting restriction

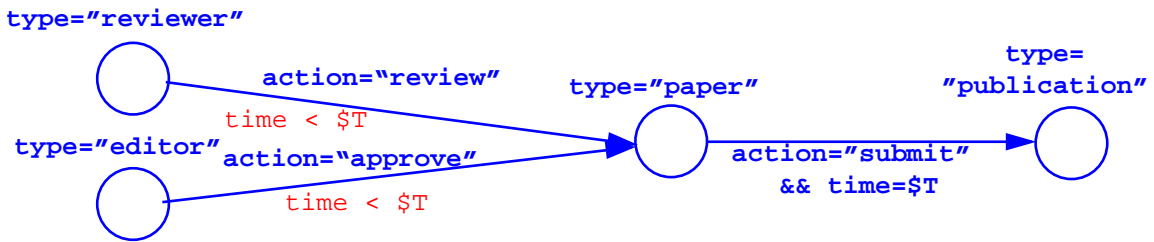

**Figure 3-19.** Policy graph for review and publication policy

### 3.3.2.3 Time ordered access policies

In LaSCO policies, time is always available through references to attributes on edges. Pattern of event policies may impose a certain restriction is placed on the order of events. This is achieved by setting a variable in the domain of one of the edges and making use of that in the requirement of another edge. Multiple variable and multiple restrictions may be used in policies with more than two edges. Consider two examples.

Figure 3-18 shows a policy that might be in effect over a system for electronically processing exams. The restriction is that when a student in a course submits his or her solutions for an exam in the course, then that should take place before the time that an instructor posts the exam key. Here variable C is used to represent the course and variable I to represent the exam identifier. Variable TS records the time of the submit for use in comparing it to the time of the post.

The three edge policy in Figure 3-19 specifies a policy that when a reviewer reviews a paper, an editor approves the paper, and the paper is submitted to a publication, then the time of the submit should be after both the review and the approval. The variable T is used to record the time of submission. Note the use of two requirement predicates. As per the LaSCO semantics (Section 4.4), these both must be satisfied or the policy is violated.

### 3.3.2.4 Forbidden pattern policies

Sometimes what a policy wishes to indicate is that a certain pattern of events should never occur. For example, a certain signature of events on a network or on a host may indicate that an attack is underway. The way that this type of policy may be represented in LaSCO is that the pattern that is forbidden is described in the domain and the (arbitrarily placed) requirement predicate is `False`. Since `False` never evaluates to true, whenever the signature is matched, the policy is violated.

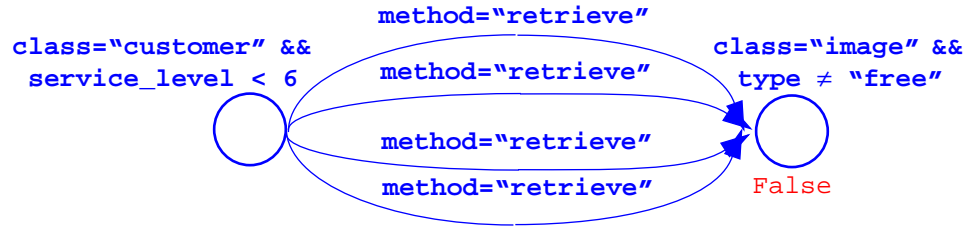

**Figure 3-20.** Policy graph for image retrieval quantity restriction

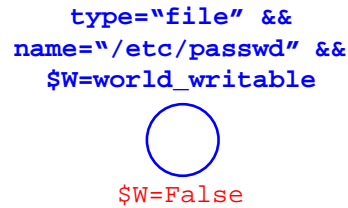

**Figure 3-21.** Policy graph for password file

As an example, consider the LaSCO policy depicted in Figure 3-21. This is a policy that might be desired on a program that serves images from a database. The restriction is that, for a customer with assigned service level less than 6 and for images that are not free, the image cannot be retrieved by the customer more than three times. By the semantics, if the *service_level* attribute of the source changes while the policy is being considered, the policy will only count retrieves that occur while the object has *service_level* less than 6. (An extension to LaSCO that might be useful as syntactic sugar for policies such as this is an iteration count for edges.)

### 3.3.3 Object state policies

The policies stated thus far do not contain any isolated nodes. Isolated nodes may be used to state restrictions on objects that apply to all times or may be used to require certain historical context for a policy to apply. An example of each is presented here. An example of the former is in Figure 3-21. The policy in this figure states that the "/etc/passwd" file should never be world writable.

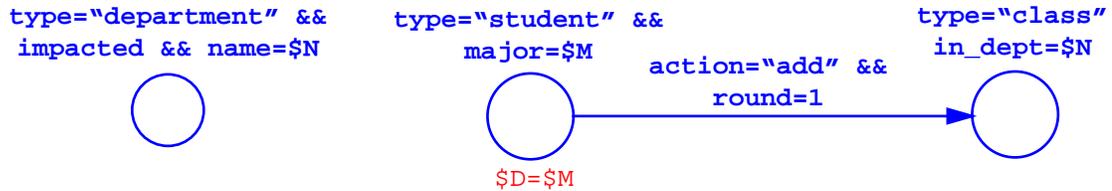

**Figure 3-22.** Policy graph for course adding policy.

Figure 3-22 presents a policy that might be in effect at a university. Here, if a department has been declared to impacted and student is adding a class in the department during round 1 of enrollment, then the student must have that department as their major.

### 3.3.4 Implementing Adage Policies in LaSCO

Adage [31,2] is a system for supporting the definition and distribution of policies in a distributed environment. Authorization rules (Adage's notion of policies) can be translated into a one edge LaSCO policy. One way to describe the system in LaSCO terms is to have the individual actors and targets be the nodes and to have the actions be the events, which occur between an actor and a target. All objects can have the *labels*, *secrecyLevel*, and *integrityLevel* attributes as represented in the Adage Authorization Language (AL). (Levels can be converted to numbers.) Also as in the AL representation, actor objects can have a *teams* attribute and target objects can have a *collections* attribute. (Other parts of a actor or target in the AL that would seem to be candidate for attributes, such as fullName, are never a basis for policy decisions in Adage, so are not included here.) Events would have few attributes since actions in Adage are not parametrized. Their label, the roles that they are a part of, and the time of the action would be encoded as attributes. In addition, an attribute for each role it is part of is present to indicate the number of principals acting in the role at the time of the action.

The history mechanism in Adage is both different and more limited than that of LaSCO. Whereas in LaSCO event history is used as part of domain, in Adage, action history can be referred to as part of the requirement (but not the domain). The restrictions that Adage can employ regarding history are requirements that a certain type of action has had to have occurred or not occurred. The type of action can only be selected by the principal, action, and target involved. As had been shown, LaSCO is far more general in its ability to refer to history. Regardless, LaSCO (at least at present) cannot directly require the existence or non-existence of an event. However the capability found in Adage can be simulated by making history available to the policy through the principal object as an attribute, *history*. (This seems reasonable, as the Adage system would need to record this information anyway.) The history attribute is a set reflecting what has occurred involving the principal represented. Three types of string may be found in that set. The name of the action is found if the principal has ever performed that action. The name of a target is found if the target has ever been used. A string that is the is the concatenation of an action, a colon, and a target indicates that the principal has done the action on the target. (If new data types or new operators were introduced to LaSCO, this representation could be made more efficient.)

Authorization rules consist of a principal scope, an action scope, a target scope, and a constraint. The action and target scopes correspond to the domain of a LaSCO policy and the principal score and the constraint corresponds to the requirement. The action scope is either the label of an action or a role. If it is a label, the edge domain predicate is of the form `label=<action-label>`. If it is a role it is in the form `role∈<role-name>`. The target scope is a collection of targets and would be represented by the destination node domain predicate. The predicate would be an expression involving a reference to one of the required teams being a member of the *collections* attribute or, if the target is an label, possibly in a more compact expression. The principal scope is a set of team names. This would be represented in on the source node requirement predicate as a expression stating that the the current team name is one of those in the principal scope.

The constraint part of authorization rules can be represented as an additional part of the requirement predicate on the source node, conjuncted with the requirement from the principal scope. Though the references in the constraint are not localized to that node, attributes on the edge and destination node can be referred to through variables that are introduced as needed. The constraint contains three types of comparisons which can be connected by conjunction, disjunction, and negation. To represent these in the predicate in a direct translation, join the representations of these types as described here with "&&", "||", and "!". The first of these types are several relationship operators to compare secrecy level, integrity level, and category sets. These can be straightforwardly translated into expressions with "=", "≤", "≥", "⊂", and "⊆", and make reference to the *labels*, *secrecyLevel*, and *integrityLevel* attributes on objects. The second type is activation rules. There, "notwith" indicates that two specific teams cannot both be in the principal's labels, a straightforward expression in LaSCO. "Atmost" uses the edge attribute holding the number of principals in a role to test that it is not greater than an certain integer. The remaining type is history expressions. The design of the *history* attribute makes it easy to test the history as required in the constraint. the form of these history comparison relations in Adage are: <principal> "hasdone" <action>, <prin-

| Description of Policy in Adage AL | |
|---|---|
| **principal scope** | Proj2 **OR** Web-Admin |
| **action scope** | Write |
| **target scope** | [Internal,(Proj2,Web-Page)] |
| **constraint** | **PRINCIPAL HASDONE**<br>Read **TO** web-content-policy |

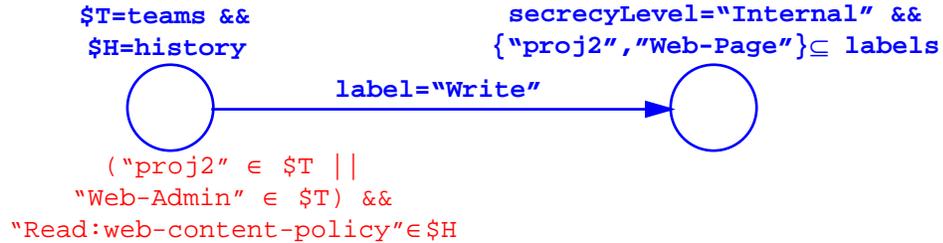

**Figure 3-23.** Adage policy and it converted to a LaSCO policy graph.

cipal> "hasdone" <action> "to" <target>, <principal> "neverdid" <action>, <principal> "neverdid" <action> "to" <target>, and <principal> "neverused" <target>.

Figure 3-23 shows a policy in Adage and its translation into a LaSCO policy. The policy is based on one in [31], but made more complicated to demonstrate more features. The policy requires writes to the internal project 2 web pages to be made by a member of project 2 or a member of the Web-Admin team and that the writer have read the web-content-policy. Translation of an Adage authorization to an LaSCO policy graph could be automated.

# 4 LaSCO Formal Description

The LaSCO system model, language, and semantics is formalized in this chapter. Section 4.1 formalizes the system model, describing how system histories are denoted by nested tuples. LaSCO is formally described in Section 4.2. Some supporting concepts for the language semantics are introduced in Section 4.3. Section 4.4 presents the LaSCO formal semantics, providing functions in first order logic to evaluate tuples describing system history against the representation of a policy. Finally, a formal concept useful in applying and implementing LaSCO is introduced in Section 4.5.

## 4.1 Formal system model

The LaSCO system model, informally described in Section 3.1, is formalized in this section.

### 4.1.1 LaSCO system model

A system history consists of a sequence of **system instances**, each representing the system at a particular time. (The only assumptions about the time representation is that it is monotonically increasing and the different event sources are synchronized.) A system instance is a snapshot of the system at a moment in time. It contains the set of objects present in the system and the set of events that are pending. There can be infinitely many of these in a system history.

Some notation for attributes, objects, and events is defined here.

**Definition 1.** Attribute binding.
   An attribute binding is a tuple of the form ⟨$\alpha,\mu$⟩ where:
      $\alpha$ is the attribute name, and
      $\mu$ is the value of $\alpha$
**Definition 2.** System object.
   Let o denote an object in a system instance:
      o is the set of attribute bindings associated with the object, each as in Definition 1.
      For system object o:

id(o)=i where ⟨'id',i⟩ ∈ o

**Definition 3.** System event.

Let ⟨s,d,a⟩ denote an event in system instance $I_t$, where:

s is the id of the source object of the event,

d is the id of the destination object of the event, and a is the set of attribute bindings associated with the event, including ⟨'time',t⟩, each as in Definition 1.

Let e=⟨s,d,a⟩:

src(e)=s

dest(e)=d

attrs(e)=a

time(e)=t where ⟨'time',t⟩ ∈ a

System history and a system instance are formally defined here:

**Definition 4.** System instance.

Let $I_t$=⟨$M_t$,$O_t$⟩ denote the system instance at time t, where:

$M_t$ is a set of pending events and

$O_t$ is a set of objects present in the system at time t.

**Definition 5.** System history.

Let H=⟨I,<⟩ denote a system history, where:

I={⟨$M_t$,$O_t$⟩| t∈$N$} is the set of all system instances in the history (where $N$ is the set of natural numbers,

< is a relation that totally orders the (discrete) system instances in I by the time they occur (⟨$M_i$,$O_i$⟩ < ⟨$M_j$,$O_j$⟩ iff i < j), and

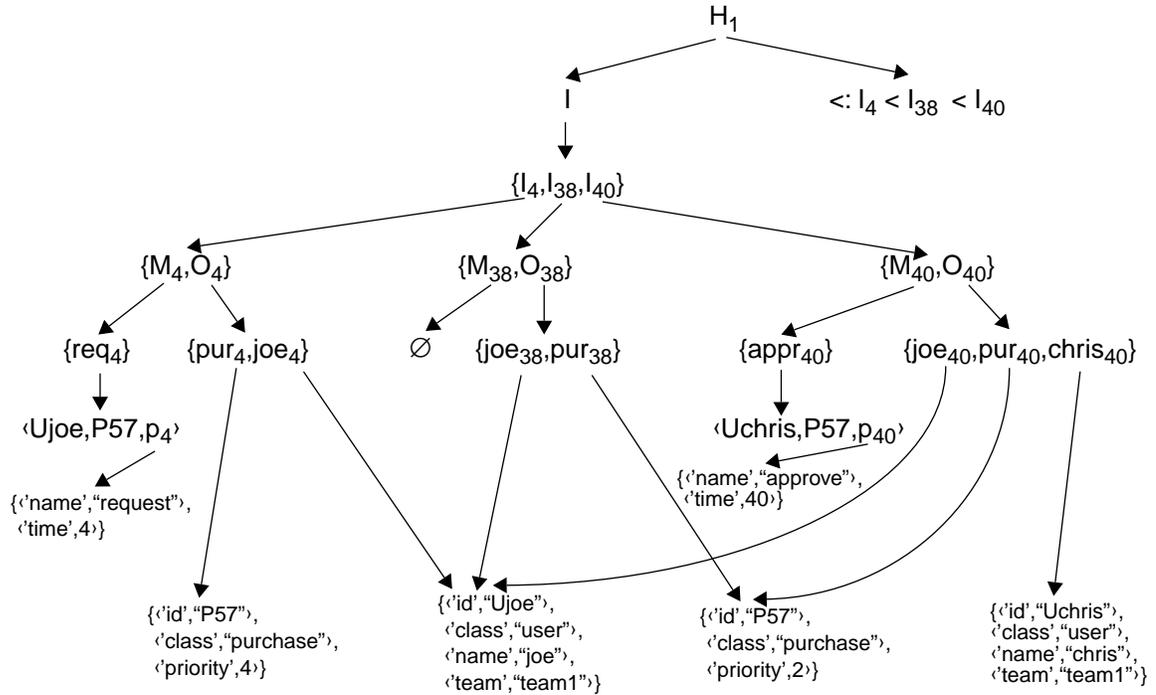

**Figure 4-1.** System history $H_1$. Arrows run from a term to its definition.

A simple example system history is depicted in Figure 4-1. (This history is depicted as a graph in Figure 4-2.)

### 4.1.2 System Graph

In a ***system graph***, the nodes and edges are the objects and events in a system history. Figure 3-1 presented an example system graph. The event attributes are annotations on edges. System graph nodes contain a set of attribute sets. One is present for each event it is involved with and for each instance in which the attributes change. We define some system graph notation in Definition 6.

**Definition 6.** System graph for the times in T.
 Let ⟨O,M⟩ denote a system graph for the times in T, where:
  O={o | t∈T ∧ o ∈ $O_t$} is the set of nodes of the graph, and
  M={e | t∈T ∧ e ∈ $M_t$} is the set of edges of the graph.
 For each system graph edge e∈M:
  src_attr(e)=o where o ∈ O ∧ time(o)=time(e) ∧ src(e)=id(o) is the set of attribute values on src(e) that is from the same system instance as e
  dest_attr(e)=o where o ∈ O ∧ time(o)=time(e) ∧ dest(e)=id(o) is the set of attribute values on dest(e) that is from the same system instance as e

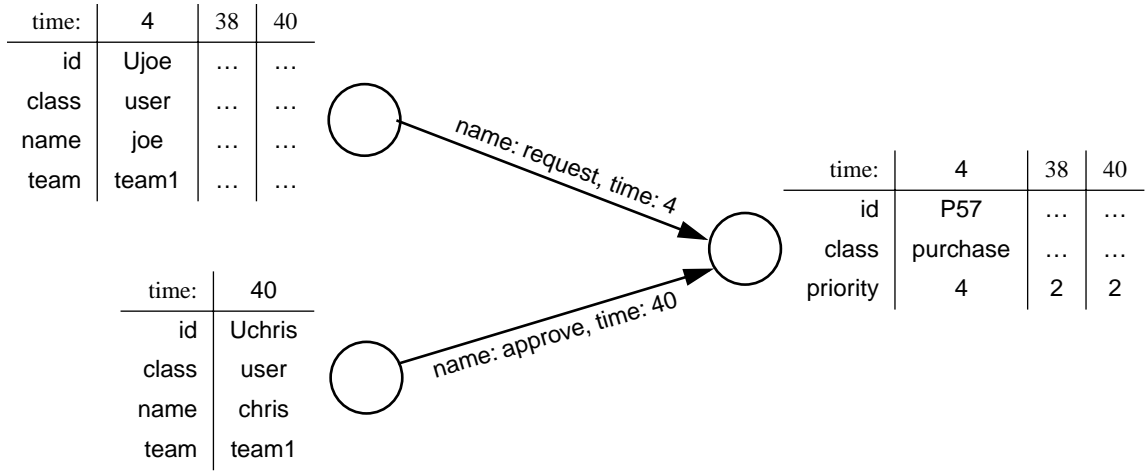

**Figure 4-2.** System graph S$_1$, depicting system history H$_1$ as a system graph. The object attributes at different times are depicted. The "…" denotes that the attribute value is the same as in the previous column.

Figure 4-2 is a depiction of H$_1$ (Figure 4-1) as a system graph, S$_1$. This is formally denoted in Example 1:

> **Example 1.** System graph S$_1$, a representation of H$_1$.
> S$_1$=⟨O$_4$ ∪ O$_{38}$ ∪ O$_{40}$,M$_4$ ∪ M$_{38}$ ∪ M$_{40}$⟩, where
> > O$_4$, O$_{38}$, O$_{40}$, M$_4$, M$_{38}$, M$_{40}$ are as defined in Figure 4-1.

## 4.2 Formal language description

This section presents the formal notation that we use to represent LaSCO policies.

### 4.2.1 Graphs

In this dissertation several kinds of graphs are used for LaSCO and for defining its semantics. System graphs were introduced in Section 4.1.2. To keep the terminology clear, the term ***basic graph*** will refer to a conventional directed graph, which consists of ***basic nodes*** and ***basic edges***. Definition 7 gives the notation associated with basic edges and Definition 8 that with basic graphs.

> **Definition 7.** Basic edge.
> Let the tuple e=⟨s,d⟩ denote a basic edge, where:
> > s is the source node and
> > d is the destination node
> For e=⟨s,d⟩:
> > src(e)=s
> > dest(e)=d

> **Definition 8.** Basic graph.
> Let the tuple ⟨N,E⟩ denote a basic directed graph, where:
> > N is the set of basic nodes in the graph, and

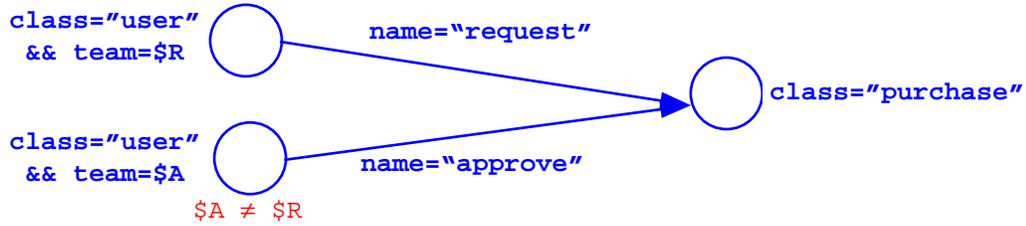

**Figure 4-3.** Policy graph for purchase request and approval separation of duty

> E is the set of basic edges in the graph.
> For a node n, isolated(n)=⟨∀e : e ∈ E : src(e) ≠ n ∧ dest(e) ≠ n⟩ is true if n is a isolated
> node

An ***isolated node*** is a node that has no incident edges. An example basic graph is given in Example 2:

> **Example 2.** Basic graph $G_1$.
> $G_1$=⟨{$n_1$,$n_2$,$n_3$},{$e_1$,$e_2$}⟩, where:
>    $e_1$=⟨$n_1$,$n_2$⟩, and
>    $e_2$=⟨$n_3$,$n_2$⟩.

### 4.2.2 Language description

Definition 9 gives the notation for a policy graph, the basic unit of policy representation:

> **Definition 9.** Policy graph.
> Let a policy graph be represented by the tuple ⟨G,γ,Λ,V⟩, where:
>    G is the basic graph in the policy graph,
>    γ is a function mapping nodes and edges in G to their domain predicate,
>    Λ is a function mapping nodes and edges in G to their requirement predicate, and
>    V is the set of variables for the policy graph.

Figure 4-3 gives an example policy graph. Example 3 shows how this is represented formally.

> **Example 3.** Policy graph $P_1$ describing the graph in Figure 4-3.
> $P_1$=⟨$G_1$,$γ_1$,$Λ_1$,{A,R}⟩, where:
>    $G_1$ is described in Example 2,
>    $γ_1$={$n_1$⇒$nd_1$, $n_2$⇒$nd_2$, $n_3$⇒$nd_3$, $e_1$⇒$ed_1$, $e_2$⇒$ed_2$},
>    $Λ_1$={$n_1$⇒true, $n_2$⇒true, $n_3$⇒$nr_3$, $e_1$⇒true, $e_2$⇒true},
>    $nd_1$ is a representation of the predicate `class="user" && team=$R`,
>    $nd_2$ is a representation of the predicate `class="purchase"`,
>    $nd_3$ is a representation of the predicate `class="user" && team=$A`,
>    $ed_1$ is a representation of the predicate `name="request"`,
>    $ed_2$ is a representation of the predicate `name="approve"`,
>    true is a representation of the predicate `True`, and
>    $nr_3$ is a representation of the predicate `$A ≠ $R`.

The domain is a pattern for the system which describes when policy is in effect. The requirement, on the other hand, is a pattern which indicates the restrictions imposed on the system by the policy. Both the domain and requirement consist of nodes and edges annotated with predicates.

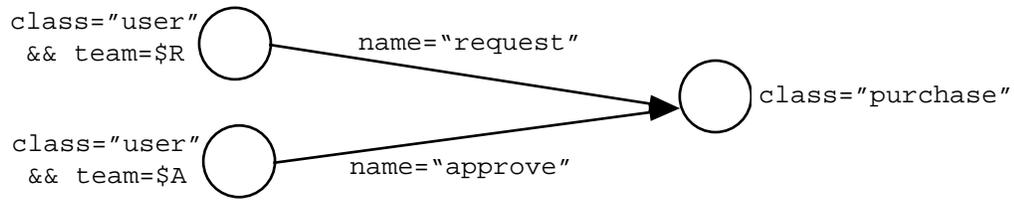

**Figure 4-4.** Domain pattern graph for purchase request and approval separation of duty

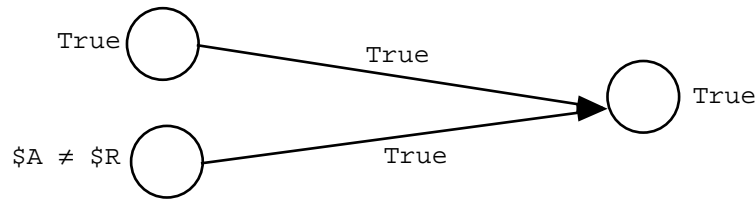

**Figure 4-5.** Requirement pattern graph for purchase request and approval separation of duty

From a formal point of view, it convenient to view the domain and requirement each as a ***pattern graph***, a pattern for part of a system. Definition 10 gives the notation associated with pattern graphs.

**Definition 10.** Pattern graph.
Let a pattern graph be represented by the tuple ‹G,χ,V›, where:
G is a basic graph,
χ is a function mapping each node and edge in G to its predicate, and
V is a set of policy variables associated with the pattern graph.

The pattern graph for the domain of the policy graph in Figure 4-3 is given in Figure 4-4 and for the requirement in Figure 4-5. As examples, the domain pattern graph is formalized in Example 4 and the requirement pattern graph in :

**Example 4.** Pattern graph $D_1$, describing the pattern graph in Figure 4-4.
$D_1$=‹$G_1$,$γ_1$,{A,R}›, where:
$G_1$ is described in Example 2, and
$γ_1$ is described in Example 3.

**Example 5.** Pattern graph $R_1$, describing the pattern graph in Figure 4-5.
$R_1$=‹$G_1$,$Λ_1$,{A,R}›, where:
$G_1$ is described in Example 2, and
$Λ_1$ is described in Example 3.

A node n in G along with its predicate χ(n) is termed a ***pattern node***. A pattern node represents an entity in a LaSCO policy that matches a set of objects in a system. For example, the right pattern node in Figure 4-4 represents all users and would match any object with *class* attribute equal to "purchase". A ***pattern edge*** is an edge in G with its predicate. A pattern edge matches a set of events. It has a pattern node as its source that represents the originator of the event that the edge represents and a pattern node as its destination that represents the target object of the event. There are two pattern edges in Figure 4-4.

The domain pattern graph consists of a basic graph, the domain predicates, and the set of variables mentioned in the policy. Similarly, the requirement pattern graph consists of a basic graph, the requirement predicates, and the set of variables mentioned in the policy. The domain and requirement pattern graphs are related in that they have the same basic graph and the same set of variables are used in both graphs. (Since they have the same basic graph, we show them composed.) The domain for a policy is ‹G,γ,V› and the requirement is ‹G,Λ,V›. Note that together these two pattern graphs fully describe a policy graph ‹G,γ,Λ,V› as defined in Definition 9.

Variable bindings represent a set of policy variables that are bound to specific values.

> **Definition 11.** Variable bindings B.
> B={‹υ,μ›| υ is the variable name
> μ is the value bound to υ.
> vars(B)= { υ | ‹υ,μ› ∈ B} is the set of variable names in B
> val(B,υ)= μ where ‹υ,μ› ∈ B is the value in B of υ

A set of variable bindings is ***complete*** if every policy variable has a value bound to it.

Logically, a LaSCO predicate is a pattern on actual objects or events. The pattern specifies which objects and events a pattern node or edge represents. Syntactically, a predicate is a boolean expression containing the operators described in Figure 3-4. Formally, a predicate is a predicate expression, as defined in Definition 12:

> **Definition 12.** Predicate expression.
> Let a predicate expression be represented by the tuple ‹l,$p_1$,$p_2$›, where:
> > β= {'&&', '||', '=', '!=', '<', '>', '<=', '>=', '∈', '⊂', '⊆', '∩', '∪', '+', '-', '*', '/', '%'}, the set of binary operators
> > l is a label indicating the form of the predicate expression, l ∈ ({'literal', 'attrname', 'varname', '()', '!'} ∪ β),
> > $p_1$ is dependent on the value of l:
> > > l='literal': $p_1$ ∈ A is the literal value
> > > l='attrname': $p_1$ is the attribute name
> > > l='varname': $p_1$ is the variable name
> > > l='()': $p_1$ is the parenthesized predicate expression
> > > l='!': $p_1$ is the predicate expression operand
> > > l∈β: $p_1$ is the first operand of the operator, a predicate expression
> > $p_2$ is dependent on the value of l:
> > > l∈β: $p_2$ is the second operand of the operator, a predicate expression
> > > l∉β: $p_2$ is vacant (denoted by λ)
> > true=‹'literal',true› is a literal representing the expression `true`
> > false=‹'literal',false› is a literal representing the expression `false`
> > p ∧ q=‹'&&',p,q›, where p and q are predicate expressions

For example, consider nd$_1$:

> **Example 6.** Predicate expression, nd$_1$ (representing `class="user" && team=$R`)
> nd$_1$=‹'&&',left,right›, where:
> > left=‹'=', ‹'attrname', "class", λ›, ‹'literal', "user", λ››, and
> > right=‹'=', ‹'attrname', "team", λ›, ‹'varname', "R", λ››.

## 4.3 Formal semantics supporting concepts

This section presents some additional concepts and notation used to define LaSCO semantics.

### 4.3.1 Notation for bindings, maps, and function

We define the following notation for use in formal LaSCO semantics:

> **Definition 13.** Map notation.
> Let domain(m) denote the (mathematical) domain of a binding, map, or function m.
> Let m(e) the value of the binding, map, or function m for e, for e∈ domain(m).

This *domain* should not be confused with the domain of a policy.

### 4.3.2 Map and function combination

A pair of mathematical maps (therefore functions or bindings) is consistent if they do not disagree on the value associated with any element of their overlapping domain.

> **Definition 14.** consistent.
> For maps m$_1$ and m$_2$:
> > consistent(m$_1$,m$_2$)=‹∀e ∈ (domain(m$_1$) ∩ domain(m$_2$)) : m$_1$(e) = m$_2$(e)›

map_combine() denotes the combination of two maps. The domains are merged and the first operand takes precedence over the second operand if the value of any range element disagrees (which they do not if the are consistent).

> **Definition 15.** map_combine.
> For maps m$_1$ and m$_2$:
> > map_combine(m$_1$,m$_2$)={‹e,m$_1$(e)› | e∈ domain(m$_1$)} ∪ {‹e,m$_2$(e)› | e∉ domain(m$_1$) ∧
> > e∈ domain(m$_2$)}

### 4.3.3 Policy to system match

A policy to system match consists of a ps map and a set of variable bindings. The variable bindings are the ones needed to satisfy the domain predicates when evaluated with their corresponding system node or edge the ps map. As formally denoted, a ps map is two functions: ϖ, which maps basic edges to system edges, and ω, which maps isolated basic node to system nodes. Definition 16 formalizes the definition of a policy to system match:

**Definition 16.** Policy to system match.
Let $\Theta_P = \langle \varpi, \omega, B \rangle$ denote a complete policy to system match, where:
P=$\langle G,\gamma,\Lambda,V \rangle$ is a policy graph as defined in Definition 9,
$\varpi$ is a one-to-one function mapping each edge in G to a system edge (E→M),
$\omega$ is a one-to-one function mapping each isolated node in G to a system node (N→O), and
B is a complete set of variable bindings.
$\Theta_P$ is shortened to $\Theta$ when the policy referred to is clear.

Since the edge binding function $\varpi$ is one-to-one, that each policy edge matches a distinct event. An example match of $P_1$ (Example 3) to $S_1$ (Example 1), $\Theta_1$ is given in Example 7:

**Example 7.** Policy to system match $\Theta_1$.
$\Theta_1 = \langle \varpi_1, \omega_1, \{\langle R, \text{"team1"} \rangle, \langle A, \text{"team1"} \rangle \} \rangle$, where
$\varpi_1 = \{e_1 \Rightarrow \text{req}_4, e_2 \Rightarrow \text{appr}_{40}\}$,
$\omega_1 = \{n_1 \Rightarrow \text{Ujoe}, n_2 \Rightarrow \text{P57}, n_3 \Rightarrow \text{Uchris}\}$, and
other terms are as in Example 1 and Example 3.

### 4.3.4 Variable conditions

For the purposes of formalizing LaSCO's semantics, we introduce the concept of ***variable conditions***. This is the logical conditions for policy variables under which a predicate is satisfied.

### 4.3.4.1 Condition expressions

A ***condition expression***, a part of variable conditions, is a special case of a predicate expression. It is a predicate expression without any attribute references, i.e., the only operands are variables and literals. Notationally it is identical to the notation for predicate expression in Definition 12 (page 39), except that no label in the expression may have the label "attrname".

An example condition expression is `$x > 17 && $y ∈ {"a","b"}`, which indicates that x must have a value greater than 17 and that y must be either "a" or "b". Another is `$x > $y`, indicating that x must have a value greater than y.

Two test functions on condition expressions are defined here. has_vars checks whether a condition expression contains any variables. If this is not the case, then the condition expression can be completely evaluated without additional context.

**Definition 17.** has_vars.
Where c=$\langle l, c_1, c_2 \rangle$ is a condition expression:

| has_vars(c)= | {true | if l='varname' |
|---|---|---|
| | {false | if l='literal' |
| | {has_vars($c_1$) | if l='!' |
| | {has_vars($c_1$) ∧ has_vars($c_2$) | if l∈ β |

may_be_sat reports if it is possible for a certain condition expression to be true with some bindings for variables. This is considered the case if it contains variables or has been evaluated to true.

> **Definition 18.** may_be_sat.
> may_be_sat(c) denotes whether the condition expression c might be satisfiable, where:
> may_be_sat(c)=has_vars(c) $\vee$ p=true

### 4.3.4.2 Variable conditions

Variable conditions consist of variable bindings and a condition expression.

> **Definition 19.** Variable conditions.
> Let variable conditions c be represented by the tuple ‹$B_c$,$C_c$›, where:
> $B_c$ is the (possibly not complete) set of variable bindings for c as defined in Definition 11, and
> $C_c$ is the condition expression for c.
> true_expr(‹$B_c$,$C_c$›)= ($C_c$=true) is checks if variable conditions have a true (Definition 12) condition expression

A variable is bound and is present in the variable bindings of a variable conditions only if it can be satisfied by a single value. The condition expression in a variable conditions indicates the possible values for those variables that are not bound to a single value. This expression, when given particular values to use for the currently unbound variables, will evaluate to true if the new bindings are acceptable and false otherwise. Two examples of formally represented variable conditions are given in Example 8:

> **Example 8.** Variable conditions $c_1$ and $c_2$.
> $c_1$=‹{‹A,4›}, ‹'!=', ‹'varname', "B", $\lambda$›, ‹'literal', "user", $\lambda$›››, representing that \$A is bound to 4 and \$B != "user".
> $c_2$=‹$\varnothing$,‹'>', ‹'varname', "A", $\lambda$›, ‹'literal', 1, $\lambda$›››, representing that there are no known bindings and \$A > 1.

### 4.3.4.3 Merging variable conditions

As will be seen in Section 4.4.2, we sometimes need to merge two variable conditions using logical "and" semantics. This results in a new variable conditions representing the combined restriction on variables of the originals. If there is no way to satisfy both conditions simultaneously (i.e., a variable found in both variable conditions has different bound values), the result is ‹$\varnothing$,false›. Otherwise, the resulting condition is reduced. That is, all bound variables found in the condition expression are moved to the variable bindings. The function merge_conds merges several variable conditions. To support the definition of merge_conds, we first introduce two functions and a operation. Only variable substitution, simultaneously replacing variable names in a predicate with their values, is used elsewhere in this dissertation.

**Definition 20.**  Variable substitution.

Substitution of a set of variable names for their values is denoted by p•B, where:

B is a set of variable bindings (as defined in Definition 11), and

p is a predicate or condition expression (as defined in Definition 12)

Let ⟨l,$p_1$,$p_2$⟩=p

⟨l,$p_1$,$p_2$⟩•B={⟨'literal', val(B,$p_1$), $\lambda$⟩   if l='varname'

{⟨l, $p_1$•B, $p_2$•B⟩        otherwise

**Definition 21.**  extract_bound.

For predicate expression p (as defined in Definition 12):

Let:

⟨l,$p_1$,$p_2$⟩=p

⟨$l_1$,$p_{11}$,$p_{12}$⟩=$p_1$

⟨$B_1$,$p_1$´⟩=extract_bound($p_1$)

When $p_2$≠$\lambda$:

⟨$l_2$,$p_{21}$,$p_{22}$⟩=$p_2$

⟨$B_2$,$p_2$´⟩=extract_bound($p_2$)

extract_bound(p)= {⟨∅,p⟩                    if l∈{'varname','attrname','literal'}

{⟨{⟨$p_{11}$,$p_{21}$⟩},true⟩        if l='='∧ $l_1$='varname'∧ $l_2$='literal'

{⟨{⟨$p_{21}$,$p_{11}$⟩},true⟩        if l='='∧ $l_2$='varname'∧ $l_1$='literal'

{⟨$B_1$,⟨'!',$p_1$´,$\lambda$⟩⟩         if l='!'

{⟨$B_1$ ∪ $B_2$,⟨l,$p_1$´,$p_2$´⟩⟩        otherwise

**Definition 22.**  reduce_cond.

For variable conditions ⟨$B_c$,$C_c$⟩:

Let ⟨B,C⟩=extract_bound(⟨$B_c$,$C_c$•$B_c$⟩)

reduce_cond(⟨$B_c$,$C_c$⟩)= ⟨$B_c$ ∪ B, C⟩

**Definition 23.**  merge_conds.

merge_conds(c1,c2, …, cn)= merge_conds(merge_conds(c1,c2), …, cn)

merge_conds(c1,c2)={⟨∅,false⟩                         if ¬consistent($B_{c1}$,$B_{c2}$)

{reduce_cond(⟨$B_{c1}$ ∪ $B_{c2}$,$C_{c1}$ ∧ $C_{c2}$⟩)        otherwise

For example, using $c_1$ and $c_2$ from Example 8, merge_conds($c_1$,$c_2$)= reduce_cond(⟨{⟨A,4⟩} ∪ ∅,

⟨'!=', ⟨'varname', "B", $\lambda$⟩, ⟨'literal', "user", $\lambda$⟩⟩ ∧ ⟨'varname', "A", $\lambda$⟩, ⟨'literal', 1, $\lambda$⟩⟩⟩)=

⟨{⟨A,4⟩},⟨'!=', ⟨'varname', "B", $\lambda$⟩, ⟨'literal', "user", $\lambda$⟩⟩⟩.

## 4.4 LaSCO semantics

The meaning of LaSCO policies when applied to a system history is formally expressed in this section. This have a bottom-up up construction, starting with evaluating predicates and ending with the semantics of a policy graph and a set of policy graphs.

### 4.4.1 Predicate evaluation

Predicates are evaluated with respect to variable bindings and either a system object or event. After instantiating a predicate with variable bindings as described in Definition 20, the resulting ex-pression is evaluated with respect to attribute bindings from the system object or event, resulting in a boolean value. The function eval_pred is formally defined in Definition 24. The meaning is that, given a predicate p, a set of attribute values values $\sigma$, and a set of variable bindings B, eval_pred(p,$\sigma$,B) is the variable conditions under which p evaluates to true given $\sigma$ and B:

**Definition 24.** eval_pred.

$$\text{eval\_pred}(p,\sigma,B) = \begin{cases} \langle B,\text{false} \rangle & \text{if } sp(p \bullet B, \sigma) = \lambda \parallel \neg sp(p \bullet B, \sigma) \\ \langle B,\text{true} \rangle & \text{otherwise} \end{cases}$$

where:

$$sp(\langle l,p_1,p_2 \rangle, \sigma) = \begin{cases} val(\sigma,p_1) & \text{if } l = \text{'attrname'} \wedge p_1 \in \text{domain}(\sigma) \\ \lambda & \text{if } l = \text{'attrname'} \wedge p_1 \notin \text{domain}(\sigma) \\ p_1 & \text{if } l = \text{'literal'} \\ sp(p_2,\sigma) & \text{if } l = \text{'||'} \wedge sp(p_1,\sigma) = \lambda \\ sp(p_1,\sigma) & \text{if } l = \text{'||'} \wedge sp(p_2,\sigma) = \lambda \\ \lambda & \text{if } sp(p_1,\sigma) = \lambda \\ sp(p_1,\sigma) & \text{if } l = \text{'()'} \\ \neg sp(p_1,\sigma) & \text{if } l = \text{'!'} \\ \lambda & \text{if } sp(p_2,\sigma) = \lambda \\ sp(p_1,\sigma) \vee sp(p_2,\sigma) & \text{if } l = \text{'||'} \\ sp(p_1,\sigma) \wedge sp(p_2,\sigma) & \text{if } l = \text{'\&\&'} \\ sp(p_1,\sigma) = sp(p_2,\sigma) & \text{if } l = \text{'='} \\ sp(p_1,\sigma) \neq sp(p_2,\sigma) & \text{if } l = \text{'!='} \\ sp(p_1,\sigma) < sp(p_2,\sigma) & \text{if } l = \text{'<'} \\ sp(p_1,\sigma) > sp(p_2,\sigma) & \text{if } l = \text{'>'} \\ sp(p_1,\sigma) \leq sp(p_2,\sigma) & \text{if } l = \text{'<='} \\ sp(p_1,\sigma) \geq sp(p_2,\sigma) & \text{if } l = \text{'>='} \\ sp(p_1,\sigma) \in sp(p_2,\sigma) & \text{if } l = \text{'}\in\text{'} \\ sp(p_1,\sigma) \subset sp(p_2,\sigma) & \text{if } l = \text{'}\subset\text{'} \\ sp(p_1,\sigma) \cap sp(p_2,\sigma) & \text{if } l = \text{'}\cap\text{'} \\ sp(p_1,\sigma) \cup sp(p_2,\sigma) & \text{if } l = \text{'}\cup\text{'} \\ sp(p_1,\sigma) + sp(p_2,\sigma) & \text{if } l = \text{'+'} \\ sp(p_1,\sigma) - sp(p_2,\sigma) & \text{if } l = \text{'-'} \\ sp(p_1,\sigma) * sp(p_2,\sigma) & \text{if } l = \text{'*'} \\ sp(p_1,\sigma) / sp(p_2,\sigma) & \text{if } l = \text{'/'} \\ sp(p_1,\sigma) \bmod sp(p_2,\sigma) & \text{if } l = \text{'\%'} \end{cases}$$

Generally, after the variable values are substituted into the predicate, the attribute values are also substituted and the operators in the expression may be evaluated since there are only literals present. The case in which an attribute name is mentioned in a predicate but is not found in the attribute bindings requires special attention. As fully described above, the most immediate boolean expression in which the attribute name without a value appears evaluates to false, regardless of any other part of that expression. This implies that, unless that boolean expression is within an expression with a disjunction, the predicate is not satisfied by the object or event. For example, using $nd_1$ from Example 6, eval_pred(nd$_1$,{‹R,"team1"›},{‹class,"user"›}) = ‹∅,false›, since the *team* attribute is not defined.

### 4.4.2 Pattern node and pattern edge matching

The ***semantic pieces*** of a pattern graph are defined to be each pattern edge (and its incident nodes) and each isolated pattern node. It is from matches to these pieces that a pattern graph match is derived. Each match of a semantic piece in a policy domain is concerned occurs within one particular time instance. Separate semantic pieces may occur at separate time instances though. This design allows an implementation of LaSCO to look for matches to the system at the semantic entity level, which may be later composed into a full match of the domain.

Match_node(p,σ,B) denotes the variable conditions under which a pattern node with the predicate p matches the system node attribute bindings σ with variable bindings B.

> **Definition 25.** match_node.
>   match_node(p,σ,B)= eval_pred(p,σ,B)

As an example, consider nd$_1$ from Example 6, joe$_4$ from Figure 4-1 and the variable R bound to "team1". match_node(nd$_1$, joe$_4$, {‹R,"team1"›})= ‹{‹R,"team1"›},true›.

The variable conditions under which a pattern edge with predicate p matches the system edge attributes σ with variable bindings B are given by match_edge(p,σ,B).

> **Definition 26.** match_edge.
>   match_edge(p,σ,B)= eval_pred(p,σ,B)

As an example, consider ed$_1$, representing `name="request"` as per Example 3, req$_4$ from Figure 4-1 and no variable bindings. match_node(nd$_1$, attrs(req$_4$), ∅)= ‹∅,true›.

A pattern edge together with its source and destination nodes match an event using variable bindings B if:

1. using B, the predicate of the edge evaluates to true for the event,
2. the source pattern node matches the source object of the event using B and the attribute values that were on the source object at the time of the event, and
3. the destination pattern node matches the destination object of the event using B and the attribute values that were on the destination object at the time of the event.

This is detailed in Definition 27. That definition presents match_edge_area, the variable conditions that result from attempting to match a pattern edge and its adjacent nodes to a corresponding part of the system, given certain variable bindings.

**Definition 27.** match_edge_area.
    match_edge_area(e,m,$\chi$,B)= merge_conds(
        match_edge($\chi$(e),attrs(m),B),
        match_node($\chi$(src(e)),src_attr(m),B),
        match_node($\chi$(dest(e)),dest_attr(m),B) )
    where:
        e is a pattern edge,
        m is a system event,
        $\chi$ is a function mapping pattern graph nodes and edges to their predicates, and
        B is a complete set of variable bindings

For this example, consider matching the request policy edge in $P_1$ to the $req_4$ edge in $S_1$. There, corresponding incident nodes must also match.

**Example 9.** Applying match_edge_area to $e_1$ in $P_1$ to the $req_4$ in $S_1$.
    Given:
        $\gamma_1$ is as defined in Example 3,
    match_edge_area($e_1$,$req_4$,$\gamma_1$,$\varnothing$)=
        merge_conds(
            match_edge($ed_1$,attrs($req_4$),$\varnothing$),
            match_node($nd_1$,$joe_4$,$\varnothing$),
            match_node($nd_2$,$pur_4$,$\varnothing$) )=
        merge_conds(‹$\varnothing$,true›, ‹{‹R,"team1"›},true›, $\varnothing$,true›)=
        ‹{‹R,"team1"›},true›.

### 4.4.3 Pattern and policy graph matching

Given a pattern graph ‹G,$\chi$,V›, a map of edges in G to edges in a system graph $\varpi$, a map of isolated nodes in G to nodes in a system graph $\omega$, and a complete set of variable bindings B, match_graph(G,$\varpi$,$\omega$,$\chi$,B) returns true if and only if the pattern graph matches given the pattern graph to system graph maps and the variable bindings. The edges and isolated nodes match if variable conditions that result from the match satisfy true_expr (Definition 19).

**Definition 28.** match_graph.
    match_graph(G,$\varpi$,$\omega$,$\chi$,B)= ‹$\forall$e : e $\in$ E : true_expr(match_edge_area(e,$\varpi$(e),$\chi$,B))›
      $\wedge$ ‹$\forall$n : n $\in$ N $\wedge$ isolated(n): true_expr(match_node($\chi$(n),$\omega$(n),B))›

To demonstrate this, consider the pattern graph $D_1$ from Example 4 as applied to $S_1$.

**Example 10.** match_graph applied to $D_1$ and $S_1$.
    Given:
        $D_1$=‹$G_1$,$\gamma_1$,{A,R}› as per Example 4,
        $\varpi_1$={$e_1 \Rightarrow req_4$, $e_2 \Rightarrow appr_{40}$},
        $\omega_1$={$n_1 \Rightarrow Ujoe$, $n_2 \Rightarrow P57$, $n_3 \Rightarrow Uchris$}, and
        B={‹R,"team1"›,‹A,"team1"›},
    match_graph($G_1$,$\varpi_1$,$\omega_1$,$\gamma_1$,B)=
        ‹$\forall$e : e $\in$ {$e_1$,$e_2$} : $C_c$=true where c=match_edge_area(e,$\varpi_1$(e),$\gamma_1$,B)› $\wedge$
            ‹$\forall$n : n $\in$ $\varnothing$: $C_c$=true where c=match_node($\gamma_1$(n),$\omega_1$(n),B)›=

$C_{c1}$=true $\wedge$ $C_{c2}$=true (where:
  c1=match_edge_area($e_1$,$req_4$,$\gamma_1$,B)=‹B,true› and
  c2=match_edge_area($e_2$,$appr_{40}$,$\gamma_1$,B)=‹B,true›) =
true

Note that node and edge predicates can be evaluated in any order.

### 4.4.4 Domain matching

The application of the domain with respect to a part of a system history, when it succeeds, produces a policy to system match (see Section 4.3.3). A match represents the location and the way in which the domain is satisfied. As the domain may apply in several ways in a system history or not at all, applying the domain to a history produces a set of matches.

A policy to system match completely describes a match of a policy domain to the system. Note that at most one set of variable bindings will cause the domain to match a particular part of the system due to the requirement mentioned in Section 3.2.4 that each variable be on one side of an "=" operator in some domain predicate. This has benefits toward implementation efficiency since one does not need to search for variable bindings, one can just accumulate variable conditions as a match is formed. If the variable conditions become unsatisfiable, then the considered match was not meant to be.

The domain of a policy is satisfied if the domain pattern graph is satisfied by a match, as defined in Section 4.3.3. Satisfaction of the domain of policy graph P is denoted by $D_P$(‹$\varpi$,$\omega$,B›). Matches(P,S) is all the ways to satisfy $D_P$ with the system S.

**Definition 29.** Domain satisfaction.
  $D_P$(‹$\varpi$,$\omega$,B›)= match_graph(G,$\varpi$,$\omega$,$\gamma$,B)
  matches(P,S)= {‹$\varpi$,$\omega$,B› | $\varpi$ $\in$ $2^{E \rightarrow M}$ $\wedge$ $\omega$ $\in$ $2^{N \rightarrow O}$ $\wedge$ B $\in$ $2^{V \rightarrow A}$ $\wedge$ $D_P$(‹$\varpi$,$\omega$,B›)}

The effect of the semantics presented here is that nodes must match for the attribute values for the system node for each system instance in which an incident edge matches. Since isolated nodes have no incident edges, they can match the system node in any system instance. An interesting question is: what happens if the value of a object attribute changes over time such that it no longer matches a node as it did for when an adjacent edge matched a previous event. The semantics indicate it does not matter if the attribute values change. The domain will match (for the part of the policy involving that node) when each of the edges find a match given the value of the object attributes at the time of the event that matches the edge.

### 4.4.5 Requirement satisfaction

Formally, in terms of pattern graphs, whenever the domain pattern graph matches, one checks to see that the requirement pattern graph also matches. The variable bindings and map between the basic graph and system graph has been determined by this point and are represented in a match. Formally noted, $R_P$ is whether the requirement of policy P (as defined in Definition 9) is satisfied by a match $\langle\varpi,\omega,B\rangle$.

> **Definition 30.** Requirement satisfaction.
> For a policy P=$\langle G,\gamma,\Lambda,V\rangle$, as defined in Definition 9:
> $R_P(\langle\varpi,\omega,B\rangle)$= match_graph(G,$\varpi,\omega,\Lambda,$B)

As an example, consider $\Theta_1$ from Example 7, a match between $P_1$ and $S_1$:

> **Example 11.** match_graph applied to $\Theta_1$.
> Given:
>    $\Theta_1$=$\langle\varpi_1,\omega_1,\{\langle R,"team1"\rangle,\langle A,"team1"\rangle\}\rangle$ and
>    $\Lambda_1$ from Example 3,
> $R_P(\Theta_1)$= match_graph(G,$\varpi,\omega,\Lambda_1,$B)=
>    $\langle\forall e : e \in \{e_1,e_2\} : C_c$=true where c=match_edge_area(e,$\varpi_1$(e),$\Lambda_1,$B)$\rangle \wedge$
>       $\langle\forall n : n \in \varnothing : C_c$=true where c=match_node($\Lambda_1$(n),$\omega_1$(n),B)$\rangle$=
>    $C_{c1}$=true $\wedge$ $C_{c2}$=true (where:
>       c1=match_edge_area($e_1$,$req_4,\Lambda_1,$B)=$\langle B,true\rangle$ and
>       c2=match_edge_area($e_2$,$appr_{40},\Lambda_1,$B)=$\langle B,false\rangle$) =
>    false

### 4.4.6 Policy application to a system

Violations(P,S) is the set of matches of the policy graph P to the system history S that violate the policy. Violation(P,S) denotes whether a set of policies, P, is violated on a system history S. Upheld(P,S) denotes the opposite case.

> **Definition 31.** Policy semantics for a system.
> For a policy graph P applied to a system history S:
>    violations(P,S)= $\{\Theta \mid \Theta \in$ matches(P,S) $\wedge \neg R_P(\Theta)\}$

> **Definition 32.** Semantics of policy composition.
> For a set of policy graphs P applied to a system S:
>    violation(P,S)= $\langle\exists p : p \in P : $violations(p,S) $\neq \varnothing\rangle$
>    upheld(P,S)= $\neg$violation(P,S)= $\langle\forall p : p \in P : $violations(p,S) $= \varnothing\rangle$

## 4.5 Partial policy to system matches

Section 4.3.3 considered complete matches of a policy graph to a system history, which is what is typically found when considering LaSCO's application from a formal point of view. In some other situations, particularly when implementing LaSCO, we must deal with partial information. When browsing a system history to detect policy matches, a program would naturally notice some parts of the match before other parts. In fact, some events may not have occurred when first checking, e.g., if doing run time monitoring.

Given this, we introduce the concept of a ***partial policy to system match***. This is defined to be a ps map plus variable conditions. These variable conditions represent the requirements on the value of variables that enable a partial match.

> **Definition 33.** Partial policy to system match.
> Let $\Theta_P^* = \langle \varpi, \omega, C \rangle$ denote a partial policy to system match, where:
> $P = \langle G, \gamma, \Lambda, V \rangle$ is a policy graph as defined in Definition 9,
> $\varpi$ is a one-to-one function mapping each edge in $E^* \subset E$ to a system edge ($E{\to}M$),
> $\omega$ is a one-to-one function mapping each isolated node in $N^* \subset N$ to a system node ($N{\to}O$), and
> $C$ is variable conditions as defined in Definition 19.
> $\Theta_P^*$ is shortened to $\Theta^*$ when the policy referred to is clear.

Towards building a complete match, two partial matches may be merged if they agree on which subsets of the system history correspond with which subgraphs of the policy and if their variable conditions are consistent. That is, they don't contradict on what the value of a variable should be. The function unifiable represents whether two partial match may be merged.

> **Definition 34.** unifiable.
> unifiable($\Theta_1^*, \Theta_2^*$) denotes whether the unification of partial matches $\Theta_1^*$ and $\Theta_2^*$ might be valid, where:
> $\Theta_1^* = \langle \varpi_1, \omega_1, C_1 \rangle$
> $\Theta_2^* = \langle \varpi_2, \omega_2, C_2 \rangle$
> $\langle B, c \rangle = \text{merge\_conds}(C_1, C_2)$
> unifiable($\Theta_1^*, \Theta_2^*$) = consistent($\varpi_1, \varpi_2$) $\wedge$ consistent($\omega_1, \omega_2$) $\wedge$ may_be_sat(c)

A more complete partial match is the result of merging partial matches. This has the union of the ps maps. The new variable conditions are the conjunction of the variable conditions of the original partial matches. Definition 35 formally defines this as the function unify.

> **Definition 35.** unify.
> unify($\Theta_1^*, \Theta_2^*$) denotes the unification of the partial matches $\Theta_1^*$ and $\Theta_2^*$:
> $\Theta_1^* = \langle \varpi_1, \omega_1, C_1 \rangle$
> $\Theta_2^* = \langle \varpi_2, \omega_2, C_2 \rangle$
> $\langle \varpi_r, \omega_r, C_r \rangle = \text{unify}(\Theta_1^*, \Theta_2^*)$, where:
> $\varpi_r = \text{map\_combine}(\varpi_1, \varpi_2)$
> $\omega_r = \text{map\_combine}(\omega_1, \omega_2)$
> $C_r = \text{merge\_conds}(C_1, C_2)$

Note that by the point at which a partial match becomes complete, its variable conditions can be extracted into a single set of variable bindings, due to the variable assignment restriction from Section 3.2.4.

# 5   A Framework for Applying LaSCO

In the next four chapters, we discuss the application and enforcement of LaSCO on systems and the implementation of LaSCO. This section focuses on a general architecture for implementing LaSCO policies on any system. Section 5.1 describes the concept and its implementation is described in Section 5.2.

## 5.1 An architecture for implementing LaSCO policies

As LaSCO is independent of any specific system or enforcement mechanism, let us now consider how to apply LaSCO to a particular system. In this chapter, when a *system* is referred to, it is the actual resource that is intended to be protected. An appealing approach is to use a generic policy engine and an interface layer for each particular system, and is in fact the approach we take in our implementation. Using this approach one only need create a new interface layer to apply LaSCO to a new system. One need not rewrite the core policy interpretation elements of an implementation. This architecture is diagramed in Figure 5-1.

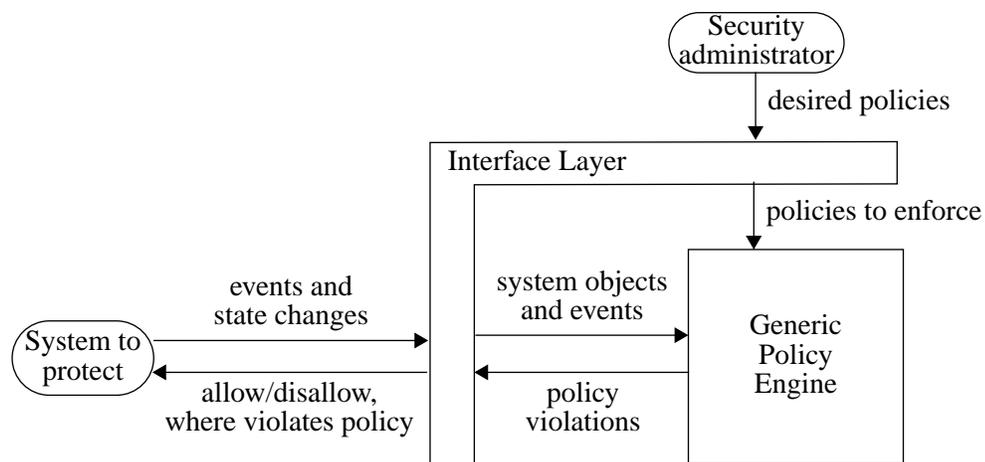

**Figure 5-1.** Diagram of the generic policy engine architecture for applying LaSCO.

### 5.1.1 Generic Policy Engine

In our architecture, a ***generic policy engine*** is a component that interprets the activity on a system (presented to it in terms of the system description for the system) with respect to policy. It is capable of detecting policy violations on any system that meets the LaSCO system model (Section 3.1).

There are two forms of input required to the policy engine, both provided an interface layer (described in the next section). The first is the set of policies, specified in LaSCO, that it must monitor for violations. The second form of input is the system execution upon which it should apply the policy. This execution is the system history for which the policies are being checked. This consists of system events and objects, both with their attributes. How the system history is gathered, whether by report, by query, or both is dependent on the implementation. Both of these input describe the system using the terms from the system description (the system's instance of the system model), which allows the engine to be independent of the monitored system.

The engine would receive sets of new events and new objects (or new attribute values) together over time. In the case of off-line policy checking, all this might be given together at once. In the case of real-time checking, this would the events and object changes for a given moment. In its operation, the engine would build matches of each policy domain to the system history it has seen so far. When a complete policy to system match has been formed, the policy requirement is evaluated. The output of the policy engine is the set of policy to system matches in the provided system history that violate the policies it was given to enforce. One can thus view the policy engine abstractly as a query engine, responding to queries about how a system history violates given policies.

### 5.1.2 Interface Layer

The ***interface layer*** is customized for specific systems. Its task is to be a bridge between the activities and state of a system and the generic policy engine. It interprets the natural form of the system and provides the engine the system description view.

The form of the interface layer depends much on the corresponding system, and there may be different approaches. For real-time monitoring, it may be in-line code, a wrapper around entities of a system, or potentially something in-between. If the monitoring is not done in real time, the layer may run over log files and pass activity reports to the engine. Depending on how the generic policy engine was designed, it may need to push reports to the engine, respond to queries, or both.

In some manner appropriate to the system being protected, the interface layer also needs to determine the proper set of policies to be enforced, and let the engine know this. Presumedly there is an administrator of some sort that decides what policies should be in effect. When the engine reports a match of the policy to the system in which the system violates a policy, the interface layer should respond to it. The response can vary widely from aborting system execution to preventing the offending event to simply making note of the violation or reporting it to the appropriate person or program.

## 5.2 Generic policy engine implementation

We have implemented a generic LaSCO policy interpretation engine in Perl [34], which is described here. The interface to implementation are instances of the class *LaSCO*. Each of these instances serves as an interpreter for a particular LaSCO policy. This class represents a LaSCO policy and is implemented using object orientation in Perl version 5.

We now examine the aspects of the implementation in the following subsections.

### 5.2.1 Initialization and policy representation

An instance of the *LaSCO* class, representing a LaSCO policy, is initialized by reading the policy from in a file or a string. In both cases, the text is in the LaSCO file format (Section A.2). Once an instance is created, policy node and edges can be added or deleted through method calls. Predicates may also be modified in this way.

Predicates are represented within the *LaSCO* class by instances of the class Predicate. Each node and edge in an instance of the *LaSCO* class (a *LaSCO* object) is annotated with two of these, one for the domain, and one for the requirement. To facilitate processing, predicates are represented by a parse tree. A YACC-type source file is converted by *perl-byacc* into a Perl class for translating predicate strings into a predicate parse tree. Each node in the parse tree is an instance of the class *PredTree*.

### 5.2.2 Interacting with an interface layer and representing partial matches

An interface layer would maintain a set of *LaSCO* objects corresponding to the set of policies to enforce. When the interface layer wishes for a policy to be evaluated on a system history, it encodes this history in an instance of the class *SystemGraph*. This *SystemGraph* instance is then passed to the *LaSCO* object for the policy via the method call *check_with_system*. The return value is all the locations in the *SystemGraph* where the policy is violated, encoded as a set of instances of the class *Match*. This method may also be used to find violations involving only the most recent parts of the system history, as is appropriate when the system is being checked incrementally for violations.

The *SystemGraph* class is an implementation of a system graph, a system history viewed as a directed graph annotated with node and edge attribute names and their values. For nodes (representing objects), attributes are noted with their values at various times. This representation is chosen as it allows the search for a policy match to be a search for a graph overlay, as will be seen in Section 5.2.5.

The *Match* class is an implementation of a policy to system match, introduced in Section 4.3.3. As such, it stores a one-to-one map between pieces of a *LaSCO* object and elements of a *SystemGraph* object. In particular, it maps policy edges to system graph edges and isolated policy nodes to system graph nodes. There is also a set of variable bindings that allow the match. Incidental policy node to system graph node bindings, the result of an adjacent edge match, are also stored. All system graph nodes in a *Match* have the time instance of their match recorded.

The *Match* class can represent partial matches of the policy to the system (Section 4.5) in addition to complete matches, so rather than storing a simple set of variable bindings, a set of variable conditions (Section 4.3.4) is stored. These variable conditions are encoded in the class *VarConds*, which consists of a set of separated variable bindings and a condition expression. Recall that the condition expression uses a subset of the predicate language, so a *PredTree* instance is used to represent this expression.

### 5.2.3 Predicate evaluation

When finding domain matches and in evaluating the policy requirement, predicates on nodes and edges are evaluated. In either case, they are evaluated with respect to a certain set of attribute values. For edges these are the parameter values and for nodes these are the object state at a given time. The result of this evaluation is the variable conditions under which the given attribute values cause the predicate to evaluate to true.

The implementation uses the following sequence of four steps to evaluate predicates:

1. ***Attribute substitution.*** The given set of attributes are substituted into the predicate. To accomplish this, the *PredTree* representing the predicate is searched for attribute references. If the attribute name is one that is given (as is typically the case), then the leaf node representing the attribute is replaced by a node representing the corresponding (literal) value. The order in which the substitution takes place is unimportant.

2. ***Undefined attribute expression replacement.*** The *PredTree* that results from attribute substitution is next checked for attributes that were not in the given set. By the language semantics, undefined attributes cause their closest enclosing boolean expression to evaluate to false. To achieve these semantics, we search the *PredTree* for remaining attributes. As can be seen from the language definition, the only operator that can prevent an expression containing a undefined attribute from evaluating to false is "‖", representing logical *or*. So, if the found attribute is contained in an "‖" expression, that expression is replaced by the subexpression on the other side of the "‖". If it is not, then the *PredTree* as a whole becomes false. This step eliminates all attributes from the expression, leaving literals and variables as the only operands.

3. ***Constant folding.*** The compiler optimization technique of constant folding is employed next to simplify the expression. For this, the *PredTree* is searched for subtrees where both operands are literals. In all cases like this, the result of that subexpression can be determined. This result, a literal, is computed and a corresponding *PredTree* node is substituted for the *PredTree*. This search proceeds in a depth first, postfix manner so that only one pass is required over the *PredTree*. The net effect of this step is that all constant expressions are replaced by their value.

4. ***Forming a variable condition.*** The result of the previous step is used as the basis for a conditional expression of the variable condition resulting from this predicate evaluation. For this, established variable bindings are extracted. The constant-folded *PredTree* is searched for subtrees where the operator is "=", one of the operands is a variable, and the other operand is a literal. This represents a case where the required value of a variable is known. The variable operand and the literal operand are added as a variable binding in the variable condition. (If a binding with the same variable name is already present (as a result of a previous extraction), and the value is different, the variable condition is known to be false and is so converted.) In addition, the value of the binding is substituted for the variable name in the rest of the expression. The *PredTree* with "=" is replaced with the literal "true" as a *PredTree*. If any variable bindings were extracted, steps 3 and 4 are repeated until no new variable bindings are extracted.

This process could be achieved in one passes, but are kept sequential to keep the implementation simple.

### 5.2.4 Variable condition satisifiability and unifiablity

A frequent query to the *VarConds* class is whether certain a certain variable conditions are satisfiable by any set of variable bindings. If not, then a considered match can be disregarded. Given this use, the implementation keeps things simple by producing an inexact but conservative answer to the question as to whether the variable conditions are unsatisfiable. For otherwise, it would need to have the logic to determine that expressions such as `$x > $y && $x < 4 && $y > 4` are unsatisfiable. This is sufficient to implement LaSCO correctly due to the requirement mentioned in Section 3.2.4 that all variables in a policy are part of an expression of the form `<variable>=<value>`. Our implementation can extract variable bindings from that expression form. There is no harm from a correctness point of view in retaining partial matches that are could in logically be found to be inconsistent.

In that implementation, satisifiability of a set of variable conditions is determined by looking at the condition expression (it is already known that the variables in the variable bindings are satisfiable). If the condition expression is the literal "false", then result is that it is not satisfiable, otherwise the result is that it might be satisfiable. Recall that the condition expression is kept reduced by constant folding, variable extraction, and substitution.

Another common question for the implementation is whether two variable conditions are mutually satisfiable. We term this question **unifiability**. At a conceptual level, this can be addressed by seeing if the conjunction of the variable conditions is satisfiable. We use this approach in the implementation.

The first step the implementation takes towards merging two variable conditions is it to compare the variable bindings in their corresponding *VarConds* instances. If any variable name is present in both bindings but the value is different, the resulting *VarConds* is "false" as there is no way to unify them. Otherwise the variable bindings are merged. Next the condition expressions are conjuncted by creating a new *PredTree* with "&&" as the operator and the condition expression *VarConds* as the operands. The merged variable bindings are now applied to the conjuncted variable conditions. This is done in a similar manner to attribute substitution in predicate evaluation (Section 5.2.3), by searching the *PredTree* for a variable reference and replacing it with a variable value. Constant folding and variable extraction is then employed. At this point, the combined variable conditions are tested for satisifiability. The result of this is the result of the unifiablity test as well.

### 5.2.5 Finding domain matches

The generic policy engine implementation provides the method *check_with_system* on *LaSCO* instances to find policy violations. The *SystemGraph* instance to check (created by the interface layer) is an argument to this method. The first step it takes is to searches for matches of the policy domain to the system history. Discussion of how the operation is different is when that method is used to find only matches involving previously unseen events is preserved to the last subsection, Section 5.2.5.3.

### 5.2.5.1 Initial partial matches

Every policy edge is considered in the context of every system graph edge as a potential match. Consider one such potential match. The domain predicates on the edge and adjacent nodes are evaluated against the corresponding edge or node in the system graph. The attributes used for predicate evaluation for the edge are its parameter values. For the nodes, their attributes at the time of the event represented by the system edge are used. Each of these predicate evaluations produces variable conditions, and which are then merged. At each step the variable conditions produced are tested for satisifiability. If the implementation finds it to be unsatisfiable, then the potential match is discarded. Otherwise, the variable conditions and the binding between the policy and system are formed into a *Match* instance representing the partial match. The overall result of this step is a set of partial matches for every policy edge.

A similar mechanism is employed to find partial matches between isolated policy nodes and system graph nodes at various times. For each potentially satisfiable variable condition formed by a binding a policy node to a system graph node at a time, a partial *Match* instance is created representing the match. The result of this is a set of partial matches to each of the isolated policy nodes.

Sometimes the interface layer may know that only certain system graph nodes or edges might match certain policy nodes or edges. This knowledge may be passed as an argument to *check_with_system* in the form of a hint. The hint takes the form of a list of system edges for certain policy edges and a list of system nodes for certain isolated policy nodes. The policy engine takes these hints as truth, and uses only the list members as the source of potential matches.

### 5.2.5.2 Growing full matches

From these partial matches, the implementation produces a set of full matches. The general approach is to merge together the initial partial matches for the different semantic pieces of the policy graph (as defined in Section 4.4.2, i.e., the edges and isolated nodes of the policy), until the entire policy graph is covered. At this point we have a complete match.

To form a full match, one initial match from each semantic piece must be used. For the match to be legal their graph bindings must agree and their variable conditions must be unifiable. This is handled pairwise by the unification operation on partial matches (Section 4.5). An addition to a partial match is disregarded if the partial matches are not unifiable.

A depth first recursive algorithm is employed to investigate all possible full matches. At each recursive invocation, all initial matches to a new semantic piece are tried with the new accumulated

```
// edgematches[e] the initial matches for edge e
// nodematches[n] the initial matches for isolated node n
// Order is a queue with policy semantic pieces, sorted to the proscribed search order

fullmatches= grow_match(<empty match>,Order)

list-of-matches grow_match(match accmatch,queue Order) {
    if Order is empty {
        return accmatch
    }
    pc= dequeue_first(Order)
    if pc is an edge {
        totry= edgematches[pc]
    } else {
        totry= nodematches[pc]
    }
    newmatches={}
    for each initmatch on totry {
        form a new match newmatch by combining accmatch and initmatch
        if newmatch is legal {
            newmatches= newmatches + grow_match(newmatch,Order)
        }
    }
    return newmatches
}
```

**Figure 5-2.** Algorithm employed to find all matches from initial matches.

match. This branched is followed, recursing with it if the match succeeds. This is displayed algorithmically in Figure 5-2.

The order of the semantic pieces for which the initial matches are tried can make a large difference in the number of partial matches considered. We wish to constrain the search as early as possible. For this reason, the order is carefully considered before the search begins. We describe this by the order in which initial matches for the semantic pieces are added. For reasons described below, the search order is sorted primarily by reverse order of the size of the connected graph it is a part of (keeping the pieces together for a connected graph) and secondarily by the number of initial matches they have (smaller is earlier).

The first heuristic used in ordering the search come from the insight that graph topology creates significant constraints on which partial matches can be added later. Once a match is found for an edge, the bound system node for each adjacent node is fixed. This in turn restricts when system edges can match adjacent policy edges -- only those with the given system node might match. Thus the considered initial matches for policy edges adjacent to these nodes must have a certain system node present. This is likely to be a subset of the matches, and thus we save on following up further in that direction. If we add a semantic entity that is disconnected with our accumulated match, then we achieve no topological constraint.

For this reason, we ensure that the set of semantic pieces of connected piece of the policy graph are kept adjacent in the search order. These sets are placed on the search order in reverse order of set size. Thus larger connected pieces are placed first, taking the most advantage of topological con-straints early. We ensure that isolated policy nodes are placed at the end of the search order, as they have the weakest constraints.

The question remains as how to order the search among the semantic pieces in a set. The heuristic we use for this is to sort them by the number of initial matches they have. Placing the pieces with the least number of initial matches earlier decreases the potential number of branches, sometimes to a high degree[1]. This technique is also used to order the isolated nodes amongst themselves at the end of the search order.

A feature of this generic policy engine used in specialized situations is that an interface layer can request that certain sets of system edges be considered to be the same in terms of finding a match. The meaning of this is that distinct matches should not be made for the edges in a given set. If one of the edges, *e*, in such a set *{d,e},* has already been used in a match, then another match with the same pieces except for *d* instead of *e* should not be formed. However, *d* and *e* may be used in the same match, and this is the only purpose of having them both in the system graph. The edges that are in the same set are identified by having the same value for a certain given attribute.

### 5.2.5.3 Finding new matches

This section considers how finding a match differs when *check_with_system* is asked to only find new matches of a policy to a *SystemGraph*. The expected use of this feature is when when additions are made to a *SystemGraph* object repeatedly, with the call being made at certain points along the way. A new match in this context is one involving system edges and isolated node attributes that have never before been present in the *SystemGraph* when the call was previously made.

A cache is maintained of initial matches of a policy edges and isolated nodes to system edges and nodes, for a particular *SystemGraph*. Since something that was in a *SystemGraph* previously is assumed not to change, these initial matches remain valid indefinitely. Thus, maintaining the cache removes the need to find this again, once they have been found once.

---

1. Consider a simple case where the are three members of the set, where two have 1 initial match and the other has 100. If it is placed in the order suggested here, then there potentially 101 new partial matches formed, whereas if the 100 is placed first, there are 200. One might also expect that the restrictions that led to fewer matches to start with may be stronger and possibly have stricter variable conditions resulting, which increases the chances of aborting a branch earlier.

The first step the implementation takes is to identify the new additions to the given *System-Graph*. *SystemGraph* contains a feature to make this efficient. The initial matches for these additions are determined as described in Section 5.2.5.1. These are added to the cache.

The next step is that full matches are grown that contain these new initial matches. The procedure described in Section 5.2.5.2 is varied only in how it is used. The new initial matches are segregated by which policy semantic piece (edge or isolated node) they are a match for. The procedure is used repeatedly with almost the complete set of initial matches. Each iteration though, the full set of initial matches for a particular semantic piece is replaced with the new initial matches for that piece. This is done once for each semantic piece. The union of the new matches found for each iteration is the set of new matches sought. Through this methodology, only matches that contain a new initial match (and therefore a new system node or edge) are built and discovered.

### 5.2.6 Checking the requirement

Once a domain match is found, then the policy requirement is evaluated on it. The match has a corresponding *Match* instance and set of required variable bindings. For each policy edge, the edge that it is bound to in the *Match* is located. Its attributes and the variable bindings are used to evaluate the policy edge's requirement predicate. (This evaluation is similar to predicate evaluation described in Section 5.2.3, with variable values substituted for their names, attribute values for their names, and constant folding to produce the boolean result.) The adjacent nodes are also evaluated with respect to the requirement predicate and the variable bindings. (Recall that attribute names are not allowed in node requirement predicates.) This evaluation is also done for isolated nodes. If any of the requirement predicates evaluate to false, then the given *Match* violates the policy and this is returned to the interface layer.

# 6 LaSCO Implementation for Java

We have implemented LaSCO for Java, which is described in this section. The toolkit consists of a program to extract a schema from a Java program, a user interface to create and edit policies for the program facilitated by a schema, and a compiler to add policy checks to the program. The components of this system and how they operate are depicted in Figure 6-1. This shows the steps supporting a user as he or she writes a security policy for a Java program and sees it through to linking to a program. The implementation is approximately 15,000 lines of code (mostly Perl [34]), including the generic policy engine. We summarize the functional aspects of the implementation completed for this dissertation in Figure 6-2. Note that some programs make use of more than one of these, as depicted in Figure 6-3.

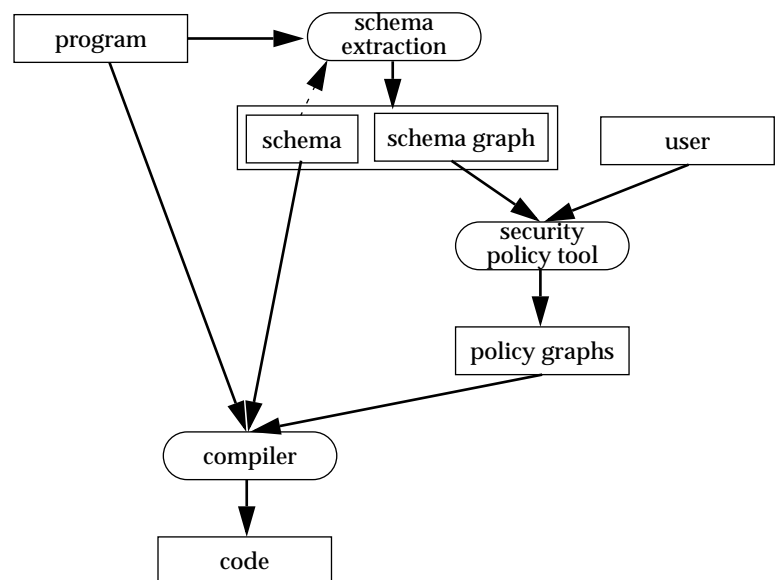

**Figure 6-1.** Specifying and enforcing LaSCO policies on Java programs.

| Function | Summary |
|---|---|
| Annotated graph | The AnnGraph, AnnEdge, AnnNode, and AnnElement classes implements a annotated graph and is the base of the LaSCO and SchemaGraph classes. Built on the Graph::Node and Graph::Edge Perl modules. |
| Generic Policy Engine | The LaSCO, Match, VarConds, Predicate, PredTree, PredParse, and SystemGraph classes implement a generic policy engine for LaSCO. See Section 5.2. |
| Schema representation | The SchemaGraph module represents a Java schema and schema graph. See Section A.4.1. |
| Schema Extraction | The extract_schema.pl script accepts Java source and produces schema and schema graph files. See Section 6.3. |
| Java Parsing | Parse Java source code into the Java representation. See Section A.4.2. |
| Java representation | About 75 classes in the ParseNode module to represent Java source code in a parse tree and to help navigate and analyze it. See Section A.4.3. |
| Policy editing | Graphical user interface to edit LaSCO policies in the context of a Java program schema. See Section 6.4. |
| Policy insertion | Compiler to add code to enforce policies on Java source. See Section 6.5.1. |
| Policy check run time system | Run time system to enforce policies in a Java program. See Section 6.5.2. |

**Figure 6-2.** Summary of the implementation for this dissertation, by function.

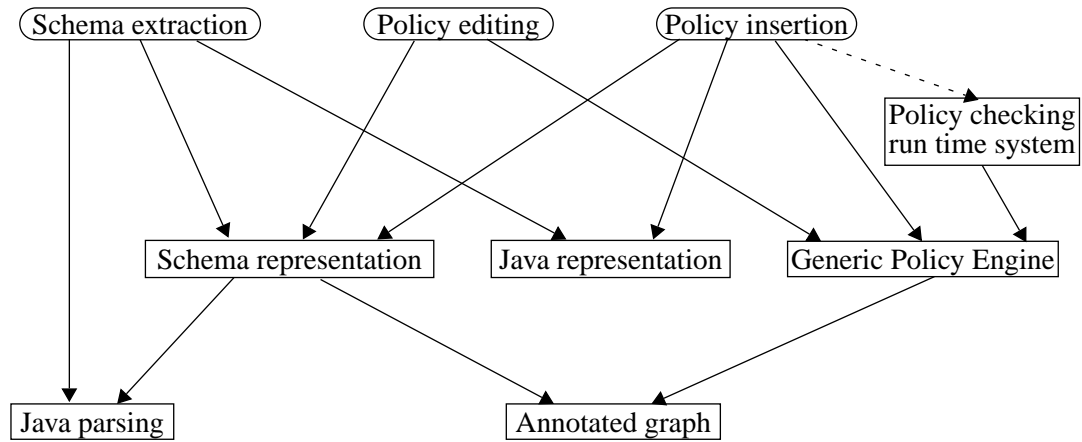

**Figure 6-3.** Relationship among functional aspects of the implementation. Arrows indicate the source making use of the destination. Broken lines indicate an indirect relationship.

The Java instance of the LaSCO system model is presented in Section 6.1. Java schemas are described briefly in Section 6.2 and the various parts of the toolkit in Sections 6.3 to 6.5. Applying LaSCO to Java and the effect of LaSCO policy checks on a Java program is discussed in Section 7.

## 6.1 Java system description

This section describes how Java programs can fit within the LaSCO system model for this implementation.

**Attributes for class objects**
- type: always "class"
- class: the name of the class the object represents
- an attribute for each static member of the class, with the declared name and the value at any given time

**Attributes for instance objects**
- type: always "instance"
- class: the name of the class of the instance that the object represents
- an attribute for each member of the instance's class, with the declared name and the value on the instance at any given time

**Attributes common to all objects:**
- classes: a set of the names of all the superclasses of the class named in the *class* attribute plus *class* itself
- id: unique to an object, but opaque in value

**Attributes common to all events:**
- name: the name of the method called
- time: the time of the call
- arg1, arg2, …: the argument in position 1, 2, … in the call, with the value of the actual parameter

**Attributes for calls to static code:**
- an attribute for each argument in the call, with the names the same as their name as formal parameters for the method being invoked (might be in a base class) and with the value of the actual parameter

**Attributes for calls to instance code:**
- an attribute for each argument in the call, with the names the same as their name as formal parameters for the method in the class statically determined as a base class of the target object (this will be the method called unless the target object is of more specific type and overrides the method) and with the value of the actual parameter

**Figure 6-4.** Attributes defined on objects and events in the Java system description for LaSCO.

The events in a Java program execution that are covered by the system description are method invocations and constructor invocations (either through a class instance creation expression (i.e., "new")). For convenience these will collectively be called *method invocations*. These are the events in the system description, and the objects are the class instances in a program and the classes themselves.

Consider the section of code that contains the method invocations to be the source of the call. Similarly, consider the destination of the method invocation to be the actual section of code that a method call causes to execute. Either end of the call might be instance code or static code. For this reason, there are two type of system objects: ones representing instances and ones representing the static aspects of the class (i.e., the class itself).

Now, if the code that contains the call is static (either it is part of a static method, a constructor body, a static member initialization, or a static initializer), the *LaSCO* object that is the source of the event is the one representing the class itself. In the alternate case, where the code that contains the call is associated with an class instance, the LaSCO object representing that instance is the source of the event. If the target of a method invocation is a static method (as is sometimes the case for regular method invocation and as is always the case for constructor invocations), then the system object representing the class that contains that method is the target of the event. Otherwise, the system object representing the class instance the method is invoked upon is that target of the event.

Figure 6-4 describes the attributes in the model for different object and events. How variables (including class members and formal parameters) are presented as LaSCO attribute values depends

on its type. All numeric primitive types and their corresponding classes in *java.lang* are viewed as numbers. The *boolean* type and the java.lang.Boolean are interpreted as booleans. Strings are viewed as strings. All other objects are viewed as strings, with their value being defined as being identical to the defined result of the method "toString" being called on the object. (At present there are no Java types that are converted into sets.)

## 6.2 Java Program Schemas

The concept of a Java schema has been developed to provide an abstract view of a Java program, particularly with respect to the methods invoked, as these are the events of interest to a policy writer. A visual depiction of a Java schema is provided by a Java schema graph, which includes an edge for method invocation, between the classes involved. More details on Java program schemas can be found in Section A.3.

## 6.3 Schema extraction tool

This tool takes files representing a Java program and produces representations of its schema. The input is a set of files containing Java compilation units, schemas, and schema graphs. The output is file representations of the corresponding schema as a schema, a schema graph, or both. Schemas, schema graphs, and the file formats are described in Section 6.2 and Section A.3. Details of this implementation are described in Section A.4.

## 6.4 Graphical LaSCO policy editor for Java

The policy editor for Java is a graphical user interface that facilitates writing LaSCO policies for a Java program. In addition to an editing pane for creating and manipulating LaSCO policies, a Java schema graph may be displayed and used for a user's reference and in an automated manner to facilitate policy construction.

A brief tour of the user interface is presented in Section 6.4.1. Generic policies are introduced in Section 6.4.2. The major functions of the editor are presented in Section 6.4.3 and Section A.5 provides some implementation notes.

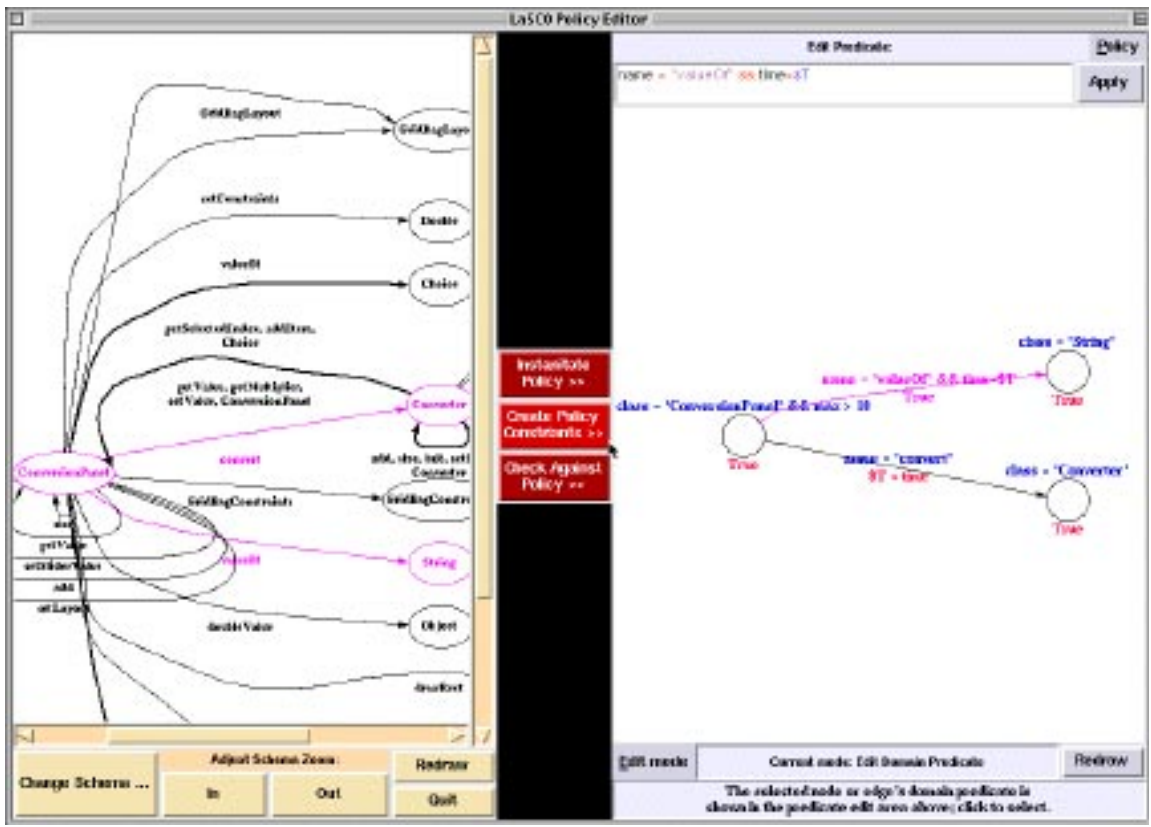

**Figure 6-5.** LaSCO policy editor interface main window screen snapshot.

### 6.4.1 User interface overview

A screen snapshot of the policy editor is shown in Figure 6-5. The right side of the main window

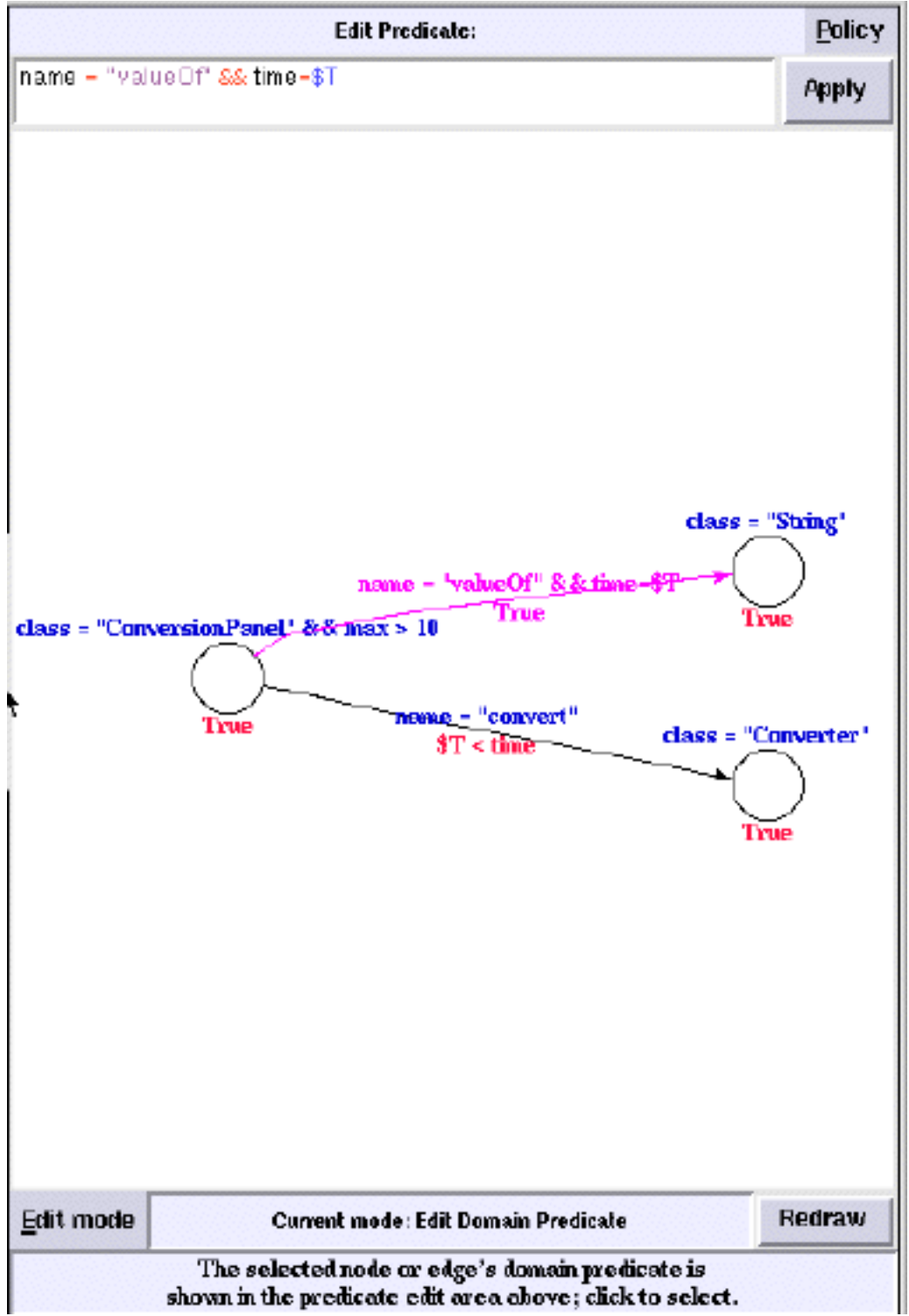

**Figure 6-6.** Right side of LaSCO policy editor interface main window screen snapshot.

(Figure 6-6) is devoted to policy editing. The largest area is where policy graphs are displayed and manipulated. Above that is an entry box for editing predicates. Syntax coloring is employed here to ease viewing. In the upper right corner there is a menu for interface features for the policy such as saving and loading  policies from files, and causing the policy to be syntax checked. A pull-down menu towards the bottom on the left selects amongst the editing (interaction) modes available: add node, add edge, edit domain predicate, edit requirement predicate, delete, and selection. Mode-sensitive text in the middle bottom on the left explains how to accomplish different tasks for each of the modes. In all modes, point-and-click style interaction is used for selecting nodes, edges, and their predicates.

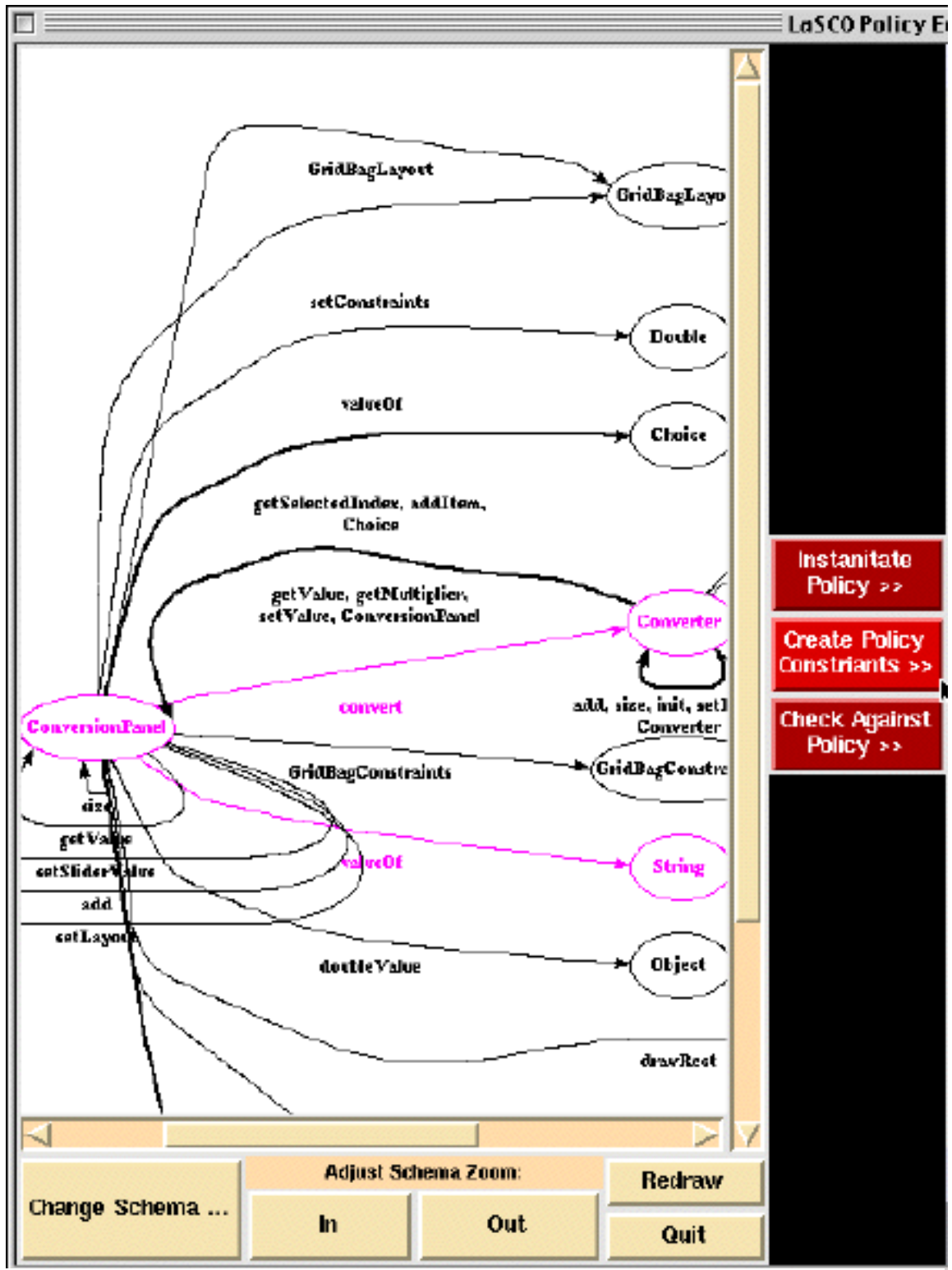

**Figure 6-7.** Left side of LaSCO policy editor interface main window screen snapshot.

The left side of the main window (Figure 6-7) is devoted to schema graph viewing. The largest area is where the schema graph is displayed. Scroll bars adjust the portion of the graph that is displayed, and the zoom in and out button below the viewing area adjusts the scale of the graph. A file to load a schema from for display can be selected after using the "Change Schema ..." button in the lower left. The center of the main window has buttons for accessing the policy/schema integration features.

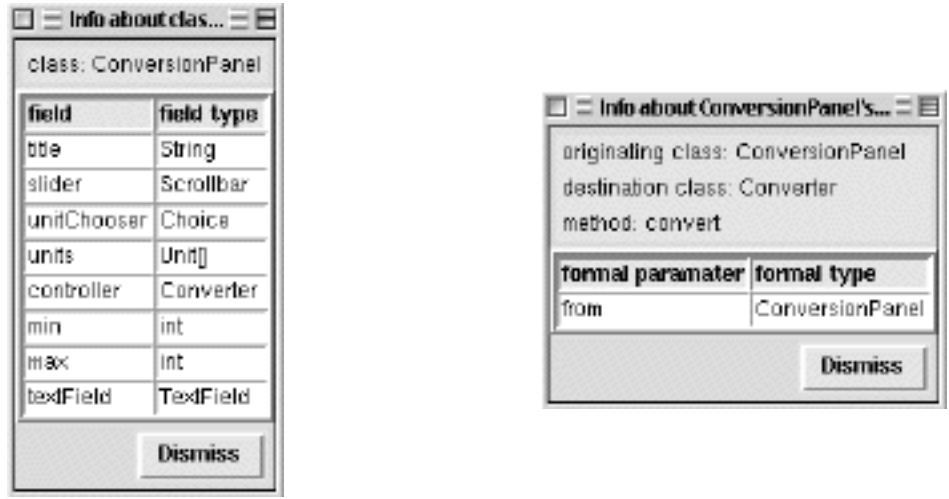

**Figure 6-8.** LaSCO policy editor interface class and method information window.

Details of the schema graph nodes and edges are available as separate information windows (e.g., those in Figure 6-8). Matches are displayed in a specialized match browsing window, which

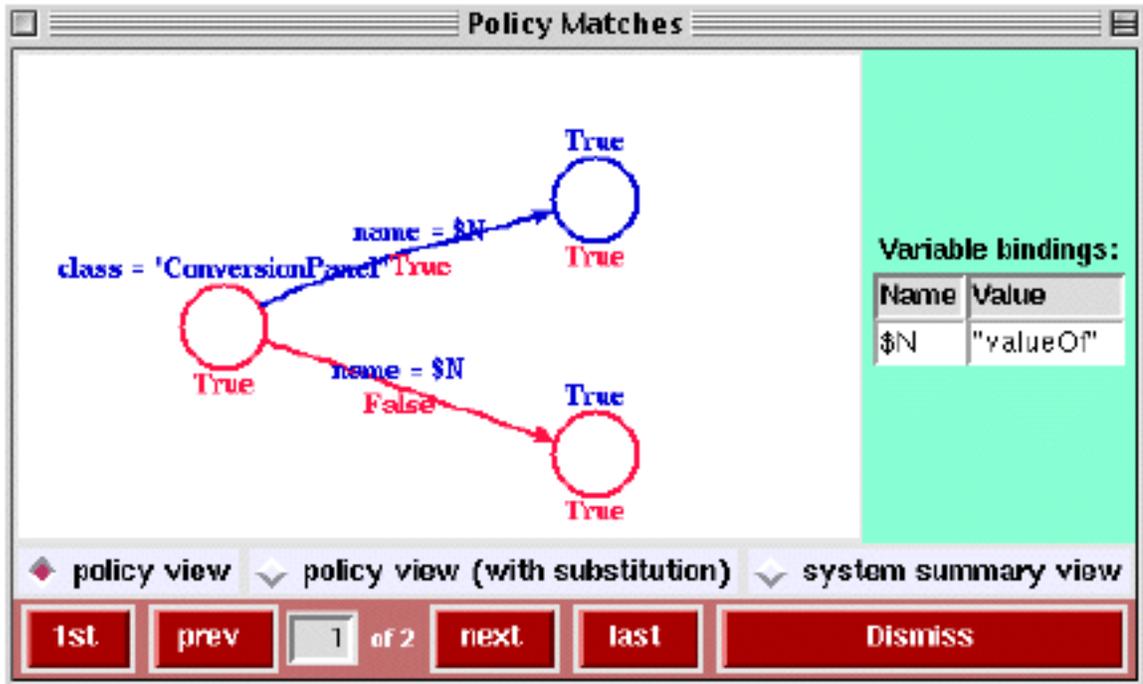

**Figure 6-9.** Policy display in the LaSCO policy editor interface match browser, screen snapshot.

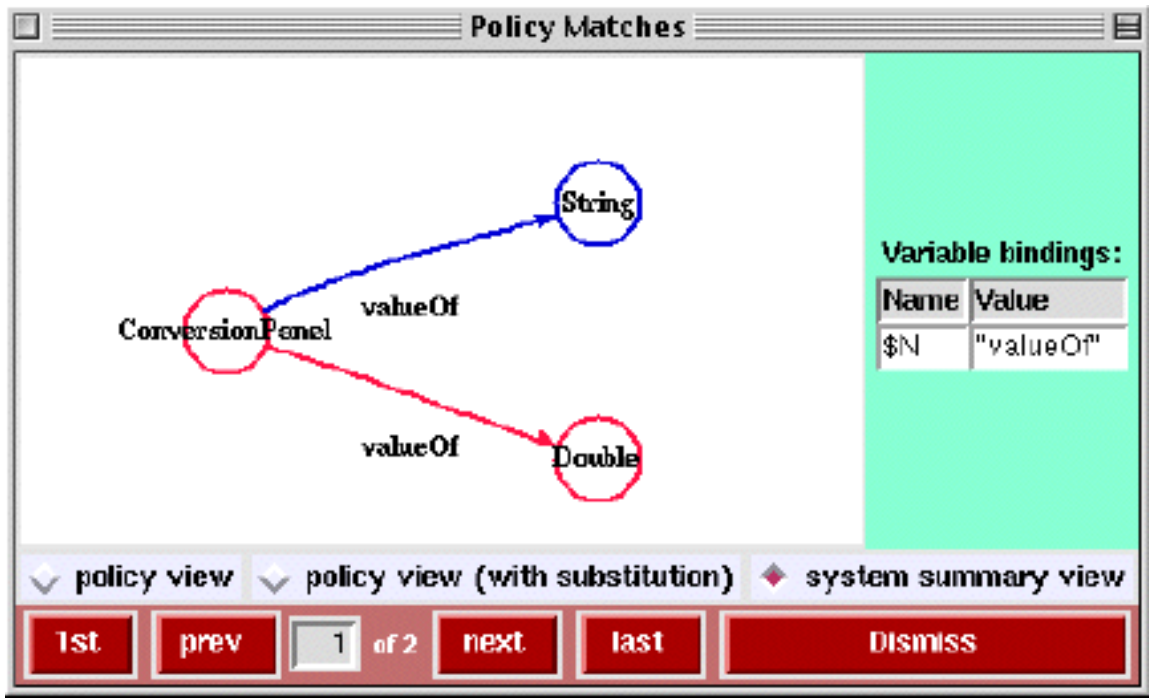

**Figure 6-10.** System display in the LaSCO policy editor interface match browser, screen snapshot.

depicts the policy (Figure 6-9), the system that matched it (Figure 6-10), match variable bindings,

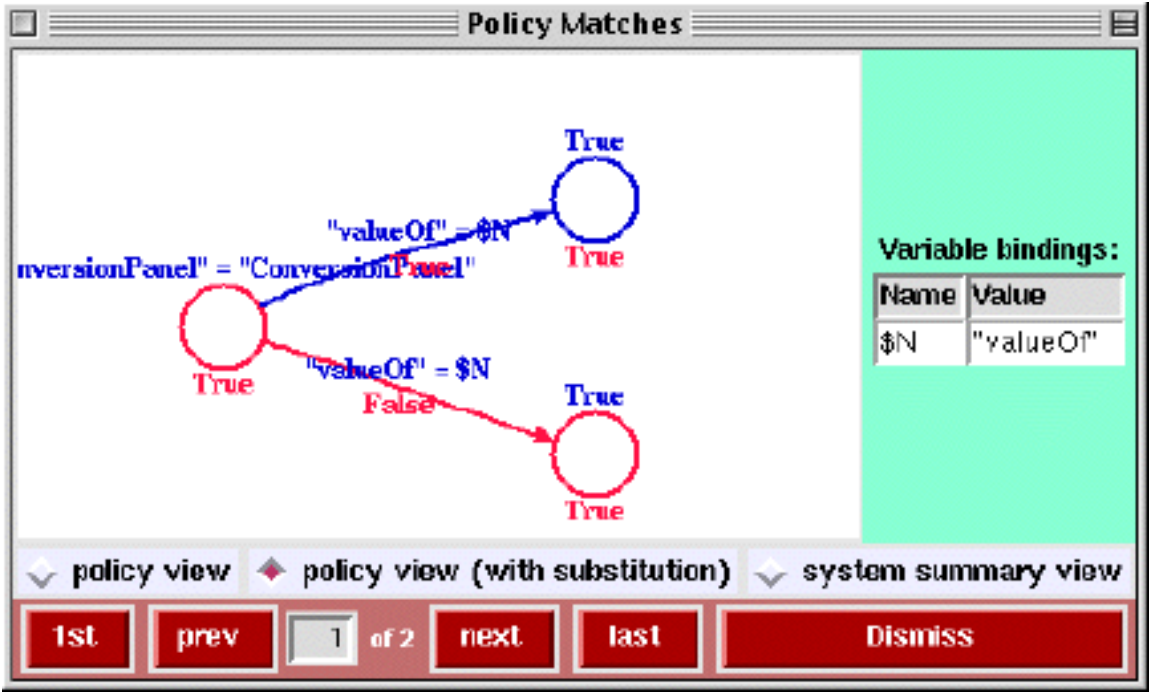

**Figure 6-11.** Policy substitution display in the LaSCO policy editor interface match browser, screen snapshot.

and how the policy predicates evaluated (Figure 6-11).

Keyboard accelerators are available for most of the menu and button items.

### 6.4.2 Generic policies

Generic policies may be loaded and created using the interface. These are LaSCO policies with place-holders in parts of their predicates. This interface allows these place-holders to be filled in when applied to a part of a schema graph. This facilitates policy construction as generic policies can be created, then filled in repeatedly in different contexts. A library of generic policies might be provided for an end user to make use of for programs that they are enforcing security constraints on, without that end user needing to go though the steps of formulating the policy structure.

Generic predicates contain template labels, which is an attribute name surrounded by angle brackets ('<' and '>'). When instantiated in a context where the named attribute has a defined value, a template label is replaced by that value. In this implementation, the substitution is textual, though more sophisticated methods may be desirable.

### 6.4.3 Some editor functions

This section describes the major functions available through the user interface. This is divided into schema viewing functions, policy editing function, and schema and policy integrated functions.

### 6.4.3.1 Schema viewing functions

***Load schema graph.*** Schema graphs may be extracted from a Java source file, a schema file, or a schema graph file. This replaces any displayed schema graph in the viewing area.

***View class information.*** After clicking on a displayed schema graph node, an information window is displayed for the corresponding class, which shows the class name, its data fields, and their types.

***View method invocation information.*** After clicking on a displayed schema graph edge, an information window is displayed for the corresponding method invocation, which shows the method name, the class the invocation was located in, the class the method invoked is located in, its method formal parameters, and their types.

***Adjust schema view.*** By using scrollbars, and the zoom in and out buttons, the perspective on the schema may be altered for the user's needs.

***Node compression.*** To save space or for more conceptual reasons, several nodes may be selected and compressed into a representative supernode. This supernode represents the original nodes for selection, edge placement, and integrated functions. If a subset of the nodes that make up the supernode are selected, just their names are displayed. Otherwise all the class names are displayed.

***Edge compression.*** To reduce visual complexity or for more conceptual reasons, the edges between a pair of nodes may be compressed into a representative superedge. This superedge represents the full set of edges for selection and integrated function purposes. If a subset of the edges that make up the supernode are selected, just those method names are displayed. Otherwise all the method names are displayed.

### 6.4.3.2 Policy editing functions

***Empty policy creation.*** A new, empty, policy may be created, replacing the current one displayed.

***Load a policy.*** A policy or a generic policy may be loaded from a file to replace the current one.

***Insert a policy.*** A policy or a generic policy may be inserted from a file adding to the current one. If the nodes have the same names when saved to a file, then they are considered the same and merged.

***Save a policy.*** A policy or a generic policy may be saved to a file in the LaSCO file format.

***Add a node.*** A new node may be added to the policy.

***Add an edge.*** A new edge may be added to the policy between a selected pair of nodes.

***Delete a node.*** An indicated node may be deleted from the policy. This causes the edges incident to it to also be deleted.

***Delete an edge.*** An indicated edge may be deleted from the policy.

***Edit a predicate.*** The domain or requirement predicate for a node or edge may be edited. When a node or edge is selected, one of its predicates may be displayed in the predicate editing area, depending on the editing mode. There it may be edited and applied to the policy.

***Check syntax and type use in predicates.*** A static analysis may be performed on the predicates of selected nodes and edges. Syntax checking ensures that a predicate is legally formed. Type use check on a predicate determine whether the type of the operands may be valid given the operator throughout the predicate. This is done to the extent possible, without type knowledge of the attributes and given the polymorphic nature of some operands. For example, $a + "b"$ would be reported as invalid. Any problems found by these checks are reported to the user.

### 6.4.3.3 Schema and policy integrated functions

***Instantiate a generic policy.*** Given the selected schema graph node or edge, instantiate the generic aspects of the predicates in the policy selection. If a policy edge is selected, the predicates on the nodes are also instantiated with the corresponding nodes in the schema graph selection.

***Create domain constraints.*** Nodes and edges may be inserted into the policy graph that would lead to domain matches of the currently selected schema graph nodes and edges. The domain predicates for the new policy edges are of the form *name=‹methodname›*, and *class=‹classname›* for new policy nodes. The requirement predicates in both cases are *true*.

***Check policy against the schema graph.*** The current policy may be analyzed against the selected part of the schema or the schema as a whole. This is in simulation of the policy being applied at

```
                                           …
                                           double tempVar1= x*y;
                                           A tempVar2= a;
                                           int tempVar3= i;
…                                          double tempVar4= x+y;
// want to promote f                       // check of call to f could go here
z=(x*y+a.f(0,i,x+y))/y;      ⟹           double tempVar5= tempVar2.f(0, tempVar3, tempVar4);
…                                          z=(tempVar1+tempVar5)/y;
                                           …
```

**Figure 6-12.** Example of method invocation promotion. The call to *f* is promoted to allow a check.

run time. The extend to which this may be done is limited to what is known statically. That is, the value of class names, method names, and knowledge of which attributes are defined.

## 6.5 Policy compiler

The remaining part of the LaSCO implementation for Java, the policy compiler, enforces LaSCO

policies on a Java program. To do this, it consists of two distinct parts: a static program modifier

(Section 6.5.1) and a run time policy interpretation system called *JavaLaSCO* (Section 6.5.2). The program modifier takes a Java program and adds calls to the *JavaLaSCO*. At run time, these calls to *JavaLaSCO* monitor the program for violations. Section 6.5.3 describes some limitations in the implementation. More details about this implementation can be found in Section A.6.

### 6.5.1 Program modification

To instrument Java source code for policy checks, a script is run. In addition to the policies to enforce and the source files to enforce them on, the script also accepts as input schemas as a refer-ence about classes. In addition to modifying the source file for method invocations as described below, certain initialization code is added to each source file with a "main" method.

This script processes one source file at a time and one method invocation and constructor invo-cation at a time. The basic information about the invocation is determined, such as method name, source class, and destination class. The implementation then makes an effort to find which policy edges might match the invocation based on static knowledge. A special method in the generic pol-icy engine is used for this.

Assuming that any policy edges might match the call, the code surrounding the call is prepared

for a policy check. To allow a check to be added, the method invocation must be a statement and

all actual parameters must be computed before that statement. Consider the example in Figure 6-

12. The Java Language Specification [15] describes the proper order of execution of Java source in great detail, and this constraint is adhered to in manipulating the code.

Next, a call a method of *JavaLaSCO* that checks new events is added just prior to the method invocation. The particular policy edges that might match the event is given as an argument as an optimization. Also passed is the argument values for invocation and the class member values for the source and destination. These are the attributes for the system event and system nodes corresponding to the call. Only the attributes that are of interest to any of the policy edges that might match are passed on the call.

### 6.5.2 Run time policy checking

Policies are checked as a program executes by the *JavaLaSCO* run time policy checking system. *JavaLaSCO* acts as the interface layer between the Java program being monitored and the generic policy engine (Section 5.2). The generic policy engine does all the work of interpreting policies; *JavaLaSCO* stores the necessary data, calculates certain attribute values, and routes the invocation being checked to the appropriate policy(ies). All data associated with the interface is stored on in Perl. A system history is maintained in a *SystemGraph* for each policy. Policies are represented by the *LaSCO* class, which is used when checking the policy.

### 6.5.3 Implementation notes

The prototype implementation has a few limitations at present, in order to facilitate the implementation. These include:

- Policies involving isolated nodes can't be enforced. This is since, at present, changes in instance and static member data are not reported to *JavaLaSCO*. In terms of monitoring, what is needed is to look for changes in member data values. Since Java doesn't have pointers, passes non-array parameters by reference, and since the attribute values needed can be inferred easily from the policy, finding locations where relevant attributes change is not a challenge -- it is an explicit assignment to the member.

- Calls from class member initialization expressions are not checked. This is since the modifications to the program cannot be handled in exactly the same manner. There is no place inside the variable declaration to make the program changes needed to promote the method invocation to a statement and to compute its actual parameters beforehand. The best solution to deal with this seems to be to create a new method that is called instead of the original. This new method would call *JavaLaSCO* to check for violations before calling the original method and returning the result.

- Threads and interfaces are ignored. This was done to reduce the amount of development work for the various operations performed on Java source code. It seems that the only change necessary to properly deal with threads is to force monitor semantics by using the "synchronized" statements around policy checks and other parts of *JavaLaSCO* that make changes to state. The modifications made to a Java program never cross a "synchronized" boundary, so that is not an issue. Since interfaces have no run time implementation by Java, the harm in ignoring them is limited to static analysis of a Java program. The interface might contain a method declaration or class data members that is sought after when analyzing the type of expressions, possibly leading to a compile-time error for the modified code or to a method invocation being thought of as possibly matching a policy edge when it can be statically reasoned that it cannot.

- When an overloaded method (one that is declared multiple times with different formal parameters) is called, the particular one that would actually be called is not resolved at present; instead a guess is made. This might result in edge attributes being given the wrong labels (the parameter names differ) or to an incorrect type being computed (if the return types differ). To implement this correctly, we would need to add the logic described in the Java Language Specification [15] to find which method declaration is (in the specification's terms) applicable, accessible, and most specific.

- Method invocation involving chained expressions (i.e., of the form ‹expression› ("." ‹expression›)+ "." ‹method-name› and (‹package-name› ".")* ‹type› "." ‹expression› ("." ‹expression›)+ "." ‹method-name›) are not resolved as to the class invoked. To solve this, additional logic would need to be added to follow the chain of expressions through to the eventual class that results.

- When wrapping method calls, not all statically knowable information, e.g., literals as method arguments and superclasses of a class, is employed in deciding which policy edges may match. The only effect of this is that unnecessary checks (checks in a situation where it will always fail) could sometimes be made of a method invocation against a policy edge at run time.

The implementation is generally verbose when attempting to handle situations beyond its abilities.

We analyze the application of LaSCO to Java in the next chapter.

# 7 Analysis of Applying LaSCO to Java

In this section, the application of LaSCO to Java is analyzed. Section 7.1 presents the setup of various experiments run with LaSCO policies added to a Java program. Sections 7.2 presents experiments with one-edged policies, Section 7.3 those with two-edged policies, and Section 7.4 policies with two or more edges. The analysis of the results are presented in these sections and the main results summarized. Section 7.5 takes a step back from the current implementation, considering the application of LaSCO to Java in general.

## 7.1 Experimental setup

These experiments were run on a Sparc Ultra 10 with a 333Mh UltraSparc2 processor and 256MB of real memory, running Solaris 7. Java version 1.1, Perl version 5.005_61, and the August 20th, 1999 release of JPL were used.

Specially designed Java programs are used to explore different aspects of policy enforcement. These programs are then modified using the policy compiler tool described in Section 6.5 to enforce one or more specially designed policies. The modified programs are executed and their run times monitored. Specifically, the amount of time Solaris credits to the user for the process execution (as opposed to system time) is used as the run time for analysis purposes. Due to the manner in which Unix measures user time (sampling the executing process every 10 milliseconds), the amount of user time reported for a process can vary from one execution to another, even for deterministic processes such as were designed for these experiments. Therefore, a particular modified program is run multiple times and the mean used as an estimate of the actual time the program uses. Since there is a relatively greater range of variation in shorter running processes, these are generally run more times than longer running processes.

As a general note, the run times experienced in the experiments in this dissertation are particular to the computer used, the Java compiler used, the LaSCO implementation, etc. Thus in our analysis, we do not focus on the precise time obtained as much as on relative times and other more portable results. Sometimes we make use of detailed knowledge of the implementation in the analysis.

## 7.2 One edge policies

The results in this section are from experiments applying LaSCO policies with exactly one edge to a Java program. The goal of these experiments was to determine the effect on run time of different policies being applied. The policies were chosen to explore this effect. We summarize our main results then present our experiments.

### 7.2.1 One edge policy main results

These are our main results of applying one edge LaSCO policies to Java programs:

**Result 1.** When any policy edge is determined by the policy compiler to possibly match a particular method call, the time taken to make the method call increases greatly.

**Result 2.** The start-up time of a program is somewhat slower if it has been modified with policy checks.

**Result 3.** The content of the predicates found in a policy domain and requirement affect the execution time of program with them compiled in. Even predicates that are logically equivalent in application can have significantly different run times. There is cost associated with each object or event attribute mentioned and with each variable mentioned. Different operators have different impact on run time.

**Result 4.** For a single edge policy, the cost of checking a call remains unchanged regardless of how many times it has been done before.

**Result 5.** The average time taken to check a policy decreases when fewer invocations match it.

**Result 6.** The overhead of enforcing multiple policies is a certain amount less than the cost of enforcing the policies separately, i.e., there is a certain reduction in overhead for additional policies enforced.

All these results except Result 4 are equally valid for policies regardless of the number of edges present.

```
class Ana {

static boolean static_run= true;

public static void main(String argv[]) {
    Ana2 ana2= new Ana2();
    for (int i=0; i < 750000; i++) {
        if (static_run) {
            Ana2.aux(i);
        } else {
            ana2.aux2(i);
        }
    }
}

}
```

```
class Ana2 {

Ana2() {
}
static void aux(int i) {
    if (i % 10000 == 0) {
    }
}
void aux2(int i) {
    if (i % 10000 == 0) {
    }
}

}
```

**Figure 7-1.** Java program for experiment round 1. The number of iterations was varied as was the static_run variable.

| Number of calls (millions) | no policy applied | | | |
|---|---|---|---|---|
| | instance | | static | |
| | user time (s) | cost/call (μs) | user time | cost/call |
| 0 | 0.24 | | 0.25 | |
| 500 | 57.62 | 0.115 | 26.895 | 0.053 |
| 750 | 86.57 | 0.115 | | |
| 1000 | 115.36 | 0.115 | 53.222 | 0.053 |
| 1500 | 172.81 | 0.115 | 79.387 | 0.053 |

| Number of calls (thousands) | *attrs000* applied | | | |
|---|---|---|---|---|
| | instance | | static | |
| | user time (s) | cost/call (ms) | user time | cost/call |
| 0 | 0.98 | | 0.93 | |
| 25 | 114.56 | 5.74 | 133.32 | 5.38 |
| 50 | 276.17 | 5.50 | 259.84 | 5.18 |
| 75 | 417.49 | 5.55 | 392.65 | 5.22 |

**Figure 7-2.** Round 1 measurements. Cost per call calculations discount start-up time (the cost with 0 calls).

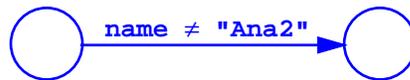

**Figure 7-3.** *attrs000*

## 7.2.2 Round 1 of experiments: the cost of checking policies

The Java program shown in Figure 7-1 was used for the first round of experiments. Note the heavy use of method calls. Experimental results are shown in Figure 7-2.

To get a basis for comparison, the Java program was run unmodified for different number of iterations of the main loop and as either making static or instance calls. For both static and instance results, the average cost per call (which includes the per-iteration time in the main loop but not execution outside of the main loop) was invariant with the number of calls made. This cost did vary between the static and instance calls, with instance calls taking over twice as long.

The policy *attrs000* (Figure 7-3) was compiled into the Java program using the policy compiler and the result compiled with the Java compiler. The policy is specially designed to match the calls in the main loop. As before, the program was run with different numbers of iterations and different static/instance settings. The compiler adds initialization code to the *main* method as well as policy check code to calls in the loop. Due to this and having to load in different libraries, there was a small additional cost in starting the program modified (Result 2).

The most profound result (Result 1) is that, with the policy checks in place, it takes much longer to make the method calls (up to a few milliseconds from a fraction of a microsecond). This should not be too surprising when comparing the amount of work performed. In the unmodified case, the cost is just making the call and a couple comparisons. In the wrapped case, the call to the run time system must be set up and made (including the transition from Java to Perl), the run time system must record the new event in a *SystemGraph*, and the run time system must call the generic policy engine to look for new matches to the policy and check the requirement.

The cost per check (TPC) for both instance and static calls varied amongst themselves to a small degree, but the static calls consistently took a little shorter on average. This is since the hashcode and class name needs to be calculated at run time in the instance case. This is not entirely offset by the need to get member modifiers in the static case. Note that the Java-inherent difference in the call is not significant here. However, it is important to pay attention to whether static or instance calls are being executed in the program.

```
class Ana {

static boolean static_run= false;
int ii1=1;
int ii2=2;
int ii3=3;
static int si1=1;
static int si2=2;
static int si3=3;
double id1=1;
double id2=2;
static double sd1=1;
static double sd2=2;
String is1="1";
String is2="2";
static String ss1="1";
static String ss2="2";

public static void main(String argv[]) {
    Ana2 ana2= new Ana2();
    for (int i=0; i < 50000; i++) {
        if (static_run) {
            Ana2.aux(i,0,"0",0);
        } else {
            ana2.aux2(i,0,"0",0);
        }
    }
}

}
```

```
class Ana2 {

int ii1=1;
int ii2=2;
int ii3=3;
static int si1=1;
static int si2=2;
static int si3=3;
double id1=1;
double id2=2;
static double sd1=1;
static double sd2=2;
String is1="1";
String is2="2";
static String ss1="1";
static String ss2="2";

Ana2() {
}

static void aux(int i,double d,String s,long l) {
    if (i % 10000 == 0) {
    }
}
void aux2(int i,double d,String s,long l) {
    if (i % 10000 == 0) {
    }
}

}
```

**Figure 7-4.** Java program for experiment round 2. The number of iterations was varied.

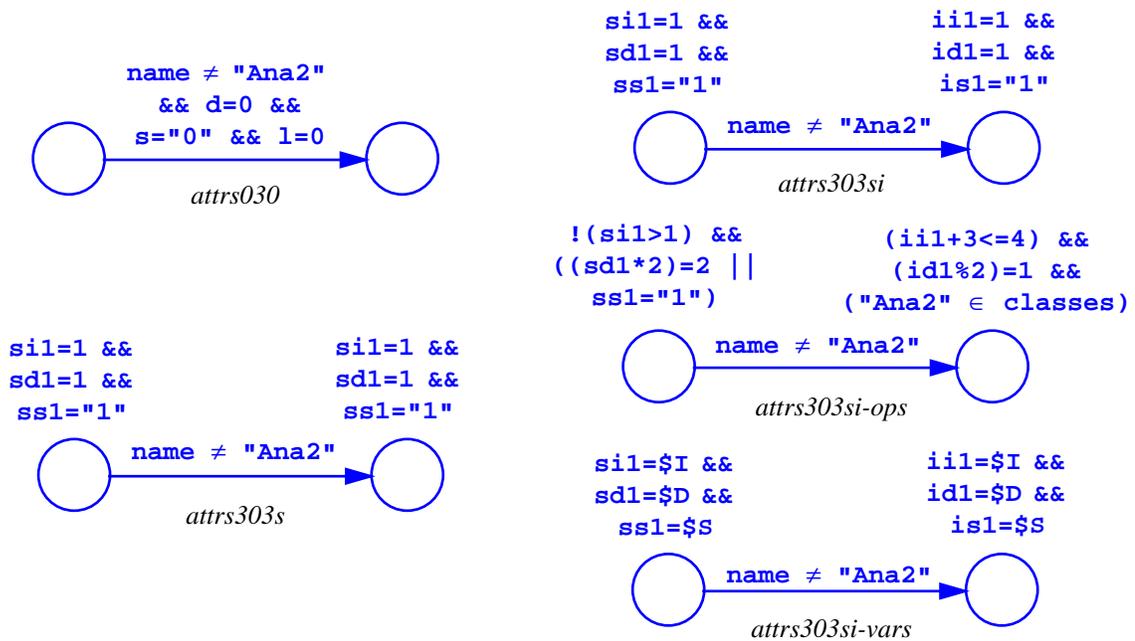

**Figure 7-5.** *attrs303si* and *attrs303si-ops*

## 7.2.3 Round 2 of experiments: different predicates

Figure 7-4 presents the Java program for round 2. The policies used are shown in Figure 7-5.

| policy name | user time (s) | | TPC (ms) |
|---|---|---|---|
| | 0 its | 50,000 its | |
| attrs000 | 0.93 | 639.16 | 12.76 |
| attrs030 | 0.91 | 951.46 | 19.01 |
| attrs303si | 0.95 | 1347.76 | 26.94 |
| attrs303s | 0.92 | 1207.59 | 24.13 |
| attrs303si-ops | 0.91 | 1531.38 | 30.61 |
| attrs303si-vars | 0.92 | 2090.13 | 41.78 |

**Figure 7-6.** Round 2 policy descriptions and measurements.

The results are shown in Figure 7-6. This round of experiments varies predicates of a single edge policy in order to see the effect on the run time. All the policies here have the same logical meaning when applied the program; it always matches the call to *aux2* inside the loop and is never violated. The policy *attrs000* is used as a base case for this, being as simple as can be while only matching the intended method invocations. Each policy modified program is run with 0 and 50,000 iterations.

Consider the run times of *attrs030*, *attrs303si*, and *attrs303s* with respect to that of *attrs000*. Each of these three cases takes longer than the base. The increase in time taken is roughly proportional to the number of attributes mentioned in any of the predicates. (Result 3) (It is not clear why having three static attributes rather than instance ones mentioned on the target node's domain predicates should cause it to run a little faster. The difference must be in Java or the reflection mechanism as the checking code makes no explicit differentiation.)

*Attrs303si-ops* is a variation on *attrs303si* than incorporates operands other than "=" and "!=". Additionally, the classes attribute on a object is mentioned, necessitating looking up that there are no base classes for class Ana2. This takes a little longer per call. *Attrs303si-vars* is another variation on *attrs303si*, but making use of 3 variables used twice each to compare source and destination attributes. This use has a significant time penalty as shown in the results and each variable use has greater impact than the attribute uses earlier.

### 7.2.4 Round 3 of experiments: varying selectiveness of policy

The Java program for round 3 is shown in Figure 7-7. For each iteration of the main loop, there are 5 calls to method *a*, with second argument equal to 1 through 5. Figure 7-8 describes the policies and shows the measured run times. In this experiment, the program modified with the policy is run with 0, 50, 100, 200, 300, 400, and 500 iterations. The program compiled with any of these policies checks all calls to a.

```
class Ana {

static boolean static_run= true;
static int num_its = 1000;
static int si1=1;
static int si2=2;
static int si3=3;
static double sd1=1;
static double sd2=2;
static String ss1="1";
static String ss2="2";

public static void main(String argv[]) {
    Ana2 ana2= new Ana2();
    for (int i=0; i < num_its; i++) {
        if (static_run) {
            Ana2.a(i,1);
            Ana2.a(i,2);
            Ana2.a(i,3);
            Ana2.a(i,4);
            Ana2.a(i,5);
        } else {
            ana2.b(i,1);
            ana2.b(i,2);
            ana2.b(i,3);
            ana2.b(i,4);
            ana2.b(i,5);
        }
    }
}

}
```

```
class Ana2 {

int ii1=1;
int ii2=2;
int ii3=3;
double id1=1;
double id2=2;
String is1="1";
String is2="2";

Ana2() {
}

static void a(int i,int d) {
    if (i % 10000 == 0) {
    }
}

void b(int i,int d) {
    if (i % 10000 == 0) {
    }
}

}
```

**Figure 7-7.** Java program for experiment round 3, 4 and 5. The num_its variable was varied.

| policy name | description of policy | user time (s) with various number of iterations | | | | | | |
|---|---|---|---|---|---|---|---|---|
| | | 0 | 50 | 100 | 200 | 300 | 400 | 500 |
| *a* | matches all calls | 0.91 | 4.46 | 8.04 | 15.22 | 22.28 | 29.23 | 36.52 |
| *a0* | matches no calls | 0.94 | 3.37 | 5.78 | 10.64 | 15.48 | 20.39 | 25.21 |
| *a1* | matches all `arg2=1` calls | 0.91 | 3.66 | 6.39 | 11.82 | 17.26 | 22.64 | 28.15 |

**Figure 7-8.** Round 3 measurements and policy descriptions.

| policy name | TPC (ms) for various number of iterations | | | | | |
|---|---|---|---|---|---|---|
| | 50 | 100 | 200 | 300 | 400 | 500 |
| *a* | 14.20 | 14.26 | 14.31 | 14.24 | 14.16 | 14.24 |
| *a0* | 9.72 | 9.69 | 9.70 | 9.70 | 9.73 | 9.71 |
| *a1* | 10.97 | 10.95 | 10.91 | 10.90 | 10.86 | 10.90 |

**Figure 7-9.** Round 3 time per policy check. There are 5 policy checks per iteration.

Figure 7-9 shows the run times per policy check. Note the consistency in the time taken per check, which supports Result 4.

The policy *a* forms a match on every policy check, i.e., every call inside the loop, since the domain always matches. The difference with *a0* and *a1* is that they have an additional subexpression on the edge that *arg2* equals 0 or 1, respectively. The former occurs only once per loop iteration (20% of the checks) and the later never matches. As would be expected, the more often a match is formed, the longer it took to run on average, leading to Result 5. It is apparent that there is a certain time attributable to checking and creating initial matches and a certain amount to forming matches.

| names of policies enforced | user time (s) with various number of iterations | | | names of policies enforced | TPC (ms) with various number of iterations | | average TPC (ms) |
|---|---|---|---|---|---|---|---|
| | 0 | 500 | 1000 | | 500 | 1000 | |
| *a1* | 0.94 | 29.03 | 57.94 | *a1* | 11.23 | 11.40 | 11.32 |
| *a1*, *a2* | 0.94 | 55.67 | 109.90 | *a1*, *a2* | 21.89 | 21.79 | 21.84 |
| *a1* to *a3* | 0.95 | 81.62 | 161.88 | *a1* to *a3* | 32.27 | 32.19 | 32.22 |
| *a1* to *a4* | 0.97 | 108.32 | 215.77 | *a1* to *a4* | 42.94 | 42.96 | 42.95 |
| *a1* to *a5* | 0.96 | 134.41 | 268.33 | *a1* to *a5* | 53.38 | 53.48 | 53.43 |

**Figure 7-10.** Round 4 measurements and time per check.

### 7.2.5 Round 4 of experiments: different numbers of policies enforced

The Java program for round 4 is shown in Figure 7-7. Policies *a1*, *a2*, *a3*, *a4*, and *a5* are used in this round. The number in the policy name is the value of the second argument of *a* that the policy requires. Different size sets of policies are compiled into the program, which is run with 0, 500, and 1000 iterations. *A1* through *a5* are identical except for a single number. Matches to these policies are equally occurring and make no difference in the execution time. Thus these policies can be assumed to take the same amount of time. Consider the results shown in Figure 7-10. The average TPC is considered.

The increase in TPC between enforcing a certain number of policies and one additional policy remains nearly the same for all the sets. However the increment is less than the initial, so enforcing the additional policies had a cost savings, which is Result 6. This is due to the fact that, even when multiple policies are in force, only one call from the Java program to the policy checking mechanism is needed. Note also that the start up time increases only slightly with additional policies loaded.

## 7.3 Two edge policies

The goal of the experiments in this section is to explore the cost considerations in applying LaSCO policies with exactly two edges to a Java program. The particular two edge policies we use represent a range of possibilities. We summarize our main results then present our experiments.

### 7.3.1 Two edge policy main results

These are our main results of applying two edge LaSCO policies to Java programs:

| policy name | description of policy | user time (s) with various number of iterations | | | | | | |
|---|---|---|---|---|---|---|---|---|
| | | 0 | 50 | 100 | 200 | 300 | 400 | 500 |
| *aa* | matches all pairs of calls | 0.91 | 321.41 | 1285.03 | 5138.73 | 11646.59 | 20552.40 | 32431.18 |
| *aa1* | matches all pairs of calls that includes one with `arg2=1` | 0.91 | 87.53 | 343.25 | 1370.40 | 3092.00 | 5475.48 | 8558.51 |
| *aav* | matches all pairs of matches with same *arg2* | 0.92 | 182.89 | 732.47 | 2956.58 | 6704.83 | 11932.28 | 18624.91 |
| *a1a1* | matches all pairs of `arg2=1` calls | 0.94 | 25.09 | 90.58 | 353.49 | 805.91 | 1435.14 | 2264.58 |

**Figure 7-11.** Round 5 run times and policy descriptions.

**Result 7.** For a policy applied to a program execution in which there is a large number of matches formed in comparison to the number of policy edges being checked against a method invocation, the number of matches dominates the time per check.

**Result 8.** The selectiveness of edges in a two edge policy can make a large difference in run time, even if the selectiveness is not realized until run time.

**Result 9.** The number of initial matches formed makes a significant difference in run time; if the same number of complete matches are formed but with more initial matches, then the run time will be higher.

**Result 10.** The shape of a multi-edge policy makes only a small difference in its associated run time per match.

**Result 11.** The cost of checking multiple policy edges in a single call to *JavaLaSCO* is noticeably lower than checking them in the different calls.

These can all be generalized to the case of more than two edges.

### 7.3.2 Round 5 of experiments: two edge policy edge selectiveness

The Java program shown in Figure 7-7 was used for round 5. Figure 7-11 describes the policies used in this round and shows the run times measured for the various number of iterations of the main loop. All policies check both policy edges for all calls to *a*. Figure 7-12 shows run times per policy check and per complete match.

| policy name | TPC (ms) for various number of iterations | | | | | |
|---|---|---|---|---|---|---|
| | 50 | 100 | 200 | 300 | 400 | 500 |
| *aa* | 1282.01 | 2568.24 | 5137.82 | 7763.79 | 10275.75 | 12972.11 |
| *aa1* | 346.49 | 684.68 | 1369.49 | 2060.72 | 2737.29 | 3423.04 |
| *aav* | 735.89 | 1436.10 | 2955.66 | 4469.27 | 4469.27 | 5965.68 |
| *a1a1* | 96.62 | 179.29 | 352.55 | 536.65 | 717.10 | 925.46 |

| policy name | # of matches per N iterations ($M_N$) | TPM (ms) for various number of iterations | | | | | |
|---|---|---|---|---|---|---|---|
| | | 50 | 100 | 200 | 300 | 400 | 500 |
| *aa* | 5N*(5N-1) | 5.15 | 5.15 | 5.14 | 5.18 | 5.14 | 5.19 |
| *aa1* | N*(5N-1) | 6.96 | 6.86 | 6.85 | 6.87 | 6.85 | 6.85 |
| *aav* | N*(5N-1) | 14.78 | 14.66 | 14.79 | 14.91 | 14.92 | 14.91 |
| *a1a1* | N*(N-1) | 9.86 | 9.06 | 8.86 | 8.97 | 8.99 | 9.07 |

**Figure 7-12.** Round 5 time per policy check and per complete match. There are 5 policy checks per iteration.

As can be seen in the TPC table in Figure 7-12, the run times for the policies in this round these are superlinear with respect to the number of checks. This is since there is an increasing number of matches formed with later checks. For m events, the number of matches to a two edge policy can be as high as m*(m-1). This is indeed the case of the policy *aa*, which forms 5N*(5N-1) matches for N iterations of the main loop. This factor comes to dominate the cost of checking more than the cost of checking an edge for initial matches and forming initial matches, which is linear with the number of checks. Consider the table in Figure 7-12 of cost per complete match (TPM) for the different policies and number of iterations. This shows an approximately constant cost per complete match, particularly when there is a large number of matches. This is Result 7.

*Aa1* is like *aa*, which matches for every pair of calls to *a*, except that a certain one of these calls must be with second argument equal to 1. This means there are about one fifth fewer matches for the same number of checks. This accounts for the lower TPC and higher TPM. The situation is similar for *a1a1*, which matches with every pair of calls to *a* with second argument 1. This matches in about 1/25 of the cases *aa* matches in. Generally note how much less time it takes to enforce *a1a1* and even *aa1*. This is since the policy come into effect less often due to the policy edges being more restrictive. This is as opposed to *aa*, which very frequently comes into effect. Thus reducing the number of matches through more selective matching of events can make a large difference in execution time, which brings us Result 8. A suggestion for a policy developer then is to have the policy apply less often unless they are willing to pay the price.

In the policy *aav*, in which a policy matches wherever the are two calls with the same second argument, every method invocation in the main loop causes an initial policy match to be formed for both policy edges. All pairs of these must be checked to see if a complete match is formed, which happens one fifth of the time. This explains why, while *aa1* has the same number of matches as *aav*, *aav* takes longer per match. We generalize this to Result 9, that the number of initial matches present is important in the run time.

```
class Ana {                              Ana2() {
                                         }
static boolean chain= false;             static void call() {
static boolean parallel= false;              if (Ana.chain) { Ana3.c(0,"Ana2"); }
static boolean in= false;                }
static boolean out= false;               static void a(int i,String s) {
static boolean wide= true;               }
static int num_its= 250;                 void b(int i,String s) {
static int sil=1;                        }
static int si2=2;
static int si3=3;                        }
static double sd1=1;
static double sd2=2;
static String ss1="1";                   class Ana3 {
static String ss2="2";
                                         int ii1=1;
public static void main(String argv[]) { int ii2=2;
    for (int i=0; i < num_its; i++) {    int ii3=3;
        Ana2.a(i,"Ana");                 double id1=1;
        if (Ana.wide) { Ana2.a(i,"Ana"); } double id2=2;
        if (Ana.out) { Ana3.c(i,"Ana"); } String is1="1";
        Ana2.call();                     String is2="2";
        Ana4.call();
    }                                    Ana3() {
}                                        }
                                         static void c(int i,String s) {
}                                        }
                                         void d(int i,String s) {
class Ana2 {                             }

int ii1=1;                               }
int ii2=2;
int ii3=3;                               class Ana4 {
double id1=1;
double id2=2;                            static void call() {
String is1="1";                              if (Ana.parallel) { Ana3.c(0,"Ana4"); }
String is2="2";                              if (Ana.in) { Ana2.a(0,"Ana4"); }
                                         }

                                         }
```

**Figure 7-13.** Java program for experiment round 6. Which of chain, parallel, in, out or wide was set to true and the num_its variable was varied.

| policy name | user time (s) with various number of iterations | | | | | |
|---|---|---|---|---|---|---|
| | 0 | 50 | 125 | 250 | 375 | 500 |
| *chain* | 0.92 | 17.35 | 99.99 | 394.43 | 887.07 | 1573.98 |
| *in* | 0.93 | 17.22 | 101.44 | 397.06 | 888.10 | 1593.58 |
| *out* | 0.93 | 17.10 | 100.24 | 395.84 | 887.29 | 1585.69 |
| *wide* | 0.93 | 51.85 | 323.18 | 1291.77 | 2881.55 | 5125.18 |
| *parallel* | 0.92 | 18.09 | 105.57 | 417.46 | 934.58 | 1660.24 |

| policy name | TPM (ms) for various number of iterations | | | | |
|---|---|---|---|---|---|
| | 50 | 125 | 250 | 375 | 500 |
| *chain* | 6.57 | 6.34 | 6.30 | 6.30 | 6.29 |
| *in* | 6.52 | 6.43 | 6.34 | 6.31 | 6.37 |
| *out* | 6.47 | 6.36 | 6.32 | 6.30 | 6.34 |
| *wide* | 5.14 | 5.18 | 5.17 | 5.13 | 5.13 |
| *parallel* | 6.87 | 6.70 | 6.66 | 6.64 | 6.64 |

**Figure 7-14.** Round 6 measurements.

### 7.3.3 Round 6 of experiments: policies with different shapes

The Java program used in round 6 is shown in Figure 7-13 and Figure 7-14 shows the run time measurements. Figure 7-15 depicts the various policies applied to the program. The program variable with the same name as the policy applied was set to true with the others false. The number of iterations of the main loop, N, was varied between 0, 50, 125, 250, 375, and 500. Regardless of the flag set, there are 2N calls checked in all cases. These calls are either to *a* or to *c*. These methods may be regarded as identical as they have the same parameters, the same (empty) method body, and are both static. Except for *wide*, the policy compiler was able to statically resolve which one policy edge should be checked for each method call. In the *wide* case, both edges match on the same occasions, so both edges are checked in all cases.

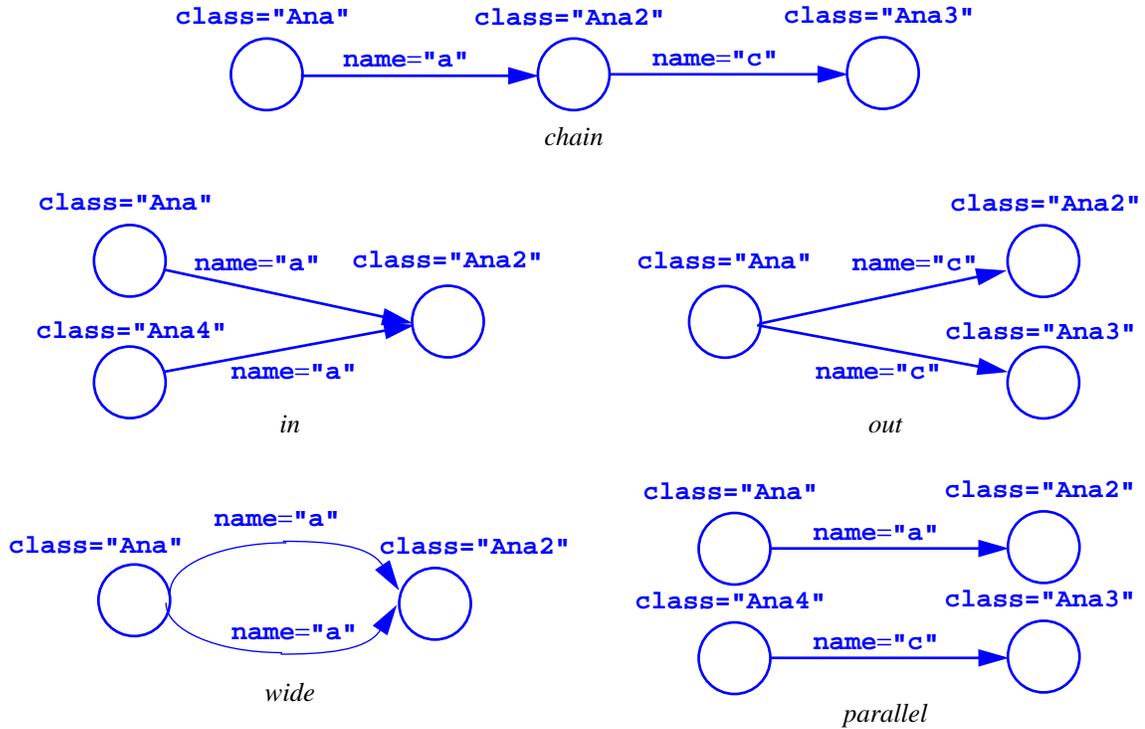

**Figure 7-15.** Round 6 policies

Supporting Result 7 from above, the TPM is fairly consistent across the number of iterations for each policy, particularly with the larger number of iterations where the cost overhead per check is less significant. The number of matches is $N^2$ except for *wide*, which is 2N*(2N-1).

The results from applying chain, in, out, and parallel are close despite the difference in policy shapes. There is a slight tendency for policies with more nodes present to run slightly slower, presumedly due to the increased number of predicates to evaluate. (Result 10)

*Wide* runs decidedly more efficiently per match than does the other policies. A part of the reason is that there are fewer nodes in the policy. However, based on scale of the difference found between the different number of policy nodes above, this is only a fraction of the difference. What seems more responsible is that two different policy edges are being checked with the same call to the enforcement mechanism. Thus we have Result 11.

## 7.4 Several edged policies

The goal of the experiments in this section is to explore the cost considerations in applying LaSCO policies with two or more edges to a Java program. We summarize our main results then present our experiments.

### 7.4.1 Several edged policy main results

These are our main results of applying LaSCO policies with two or more edges to Java programs:

**Result 12.** When more system nodes (derived from more instances of classes) are present in an executing system, the result can be fewer matches and much faster policy checking.

**Result 13.** The costs associated with checking policies can increase dramatically for more iterations for policies with several edges, but this is not a general inevitability. It depends on the characteristics of the system (Result 12) and the selectiveness of the policy (Result 5 and Result 8).

### 7.4.2 Round 7 of experiments: several edged policies

In this round, three slightly different programs were run, one with two run options. Each of the programs produces a chain of calls between different classes for some number of iterations. The depth of the chain is specified by the depth variable. The same set of policies were applied in all cases, some of which are shown in Figure 7-16. All the policies follow the same pattern as the shown policies, with one edge matching a call to a method *r* on a certain class for each "r" in the name. When running a program with a certain policy applied, the depth variable is set to the number of edges in the policy; this results in the precise number of calls in a chain to match the policy. For all the policies, each method called to any method named "r", a policy check is in place, looking for a match to a particular edge in the policy. All  checks of a method invocation against a policy edge result in an initial match.

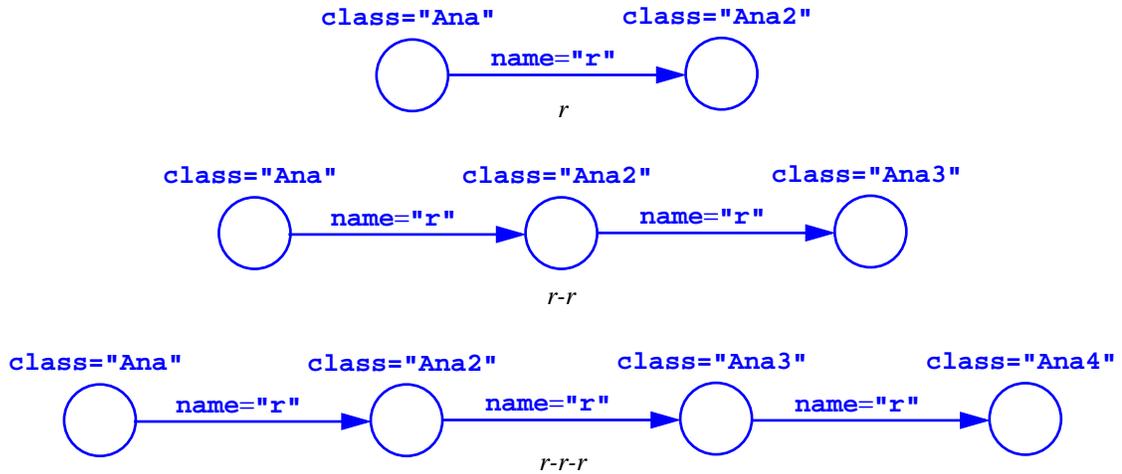

**Figure 7-16.** Some round 7 policies

```
class Ana {

static int num_its= 20;
static int depth= 8;
static int si1=1;
static int si2=2;
static int si3=3;
static double sd1=1;
static double sd2=2;
static String ss1="1";
static String ss2="2";

public static void main(String argv[]) {
    for (int i=0; i < num_its; i++) {
        Ana2.r(depth);
    }
}

}

class Ana2 {
```

```
static void r(int d) {
    if (d < 1) return;
    Ana3.r(d);
}

}

class Ana3 { static void r(int d) {
    if (d <= 2) return;  Ana4.r(d);}}
class Ana4 { static void r(int d) {
    if (d <= 3) return;  Ana5.r(d);}}
class Ana5 { static void r(int d) {
    if (d <= 4) return;  Ana6.r(d);}}
class Ana6 { static void r(int d) {
    if (d <= 5) return;  Ana7.r(d);}}
class Ana7 { static void r(int d) {
    if (d <= 6) return;  Ana8.r(d);}}
class Ana8 { static void r(int d) {
    if (d <= 7) return;  Ana9.r(d);}}
class Ana9 { static void r(int d) {
    if (d <= 8) return;}}
```

**Figure 7-17.** Java program for experiment round 7a. The num_its and depth variables were varied.

| policy name | user time (s) with various number of iterations | | | | | | | | | |
|---|---|---|---|---|---|---|---|---|---|---|
| | 0 | 4 | 8 | 10 | 12 | 16 | 20 | 30 | 40 | 50 |
| *r-r* | 0.93 | | | 1.86 | | | 3.85 | 7.04 | 11.51 | 17.19 |
| *r-r-r* | 0.96 | | | 7.65 | | | 50.61 | 166.05 | 391.78 | 761.57 |
| *r-r-r-r* | 0.93 | 3.39 | 31.87 | 75.282 | 153.43 | 485.04 | 1175.63 | | | |
| *r-r-r-r-r* | 0.95 | 10.79 | 285.63 | 858.272 | 2138.75 | 9235.89 | | | | |
| *r-r-r-r-r-r* | 0.95 | 43.38 | 2605.33 | | | | | | | |

| policy name | number of matches for various number of iterations | | | | | | | | |
|---|---|---|---|---|---|---|---|---|---|
| | 4 | 8 | 10 | 12 | 16 | 20 | 30 | 40 | 50 |
| *r-r* | | | 100 | | | 400 | 900 | 1600 | 2500 |
| *r-r-r* | | | 1000 | | | 8000 | 27000 | 64000 | 125000 |
| *r-r-r-r* | 256 | 4096 | 10000 | 20736 | 65536 | 160000 | | | |
| *r-r-r-r-r* | 1024 | 32768 | 100000 | 248832 | 1048576 | | | | |
| *r-r-r-r-r-r* | 4096 | 262144 | | | | | | | |

| policy name | TPM (ms) with various number of iterations, N | | | | | | | | |
|---|---|---|---|---|---|---|---|---|---|
| | 4 | 8 | 10 | 12 | 16 | 20 | 30 | 40 | 50 |
| *r-r* | | | 9.33 | | | 7.29 | 6.79 | 6.61 | 6.51 |
| *r-r-r* | | | 6.68 | | | 6.21 | 6.11 | 6.11 | 6.08 |
| *r-r-r-r* | 9.62 | 7.56 | 7.44 | 7.35 | 7.39 | 7.34 | | | |
| *r-r-r-r-r* | 9.60 | 8.69 | 8.57 | 8.59 | 8.81 | | | | |
| *r-r-r-r-r-r* | 10.36 | 9.93 | | | | | | | |

**Figure 7-18.** Round 7a measurements, number of matches, and time per match.

The program used in round 7a is shown in Figure 7-17 and the experimental results from applying policies to that program are shown in Figure 7-18. The names of the different policies applied and the different number of iterations they were run with are displayed there. Note that due to the construction of program, every call in every iteration after the first one causes an edge to be added

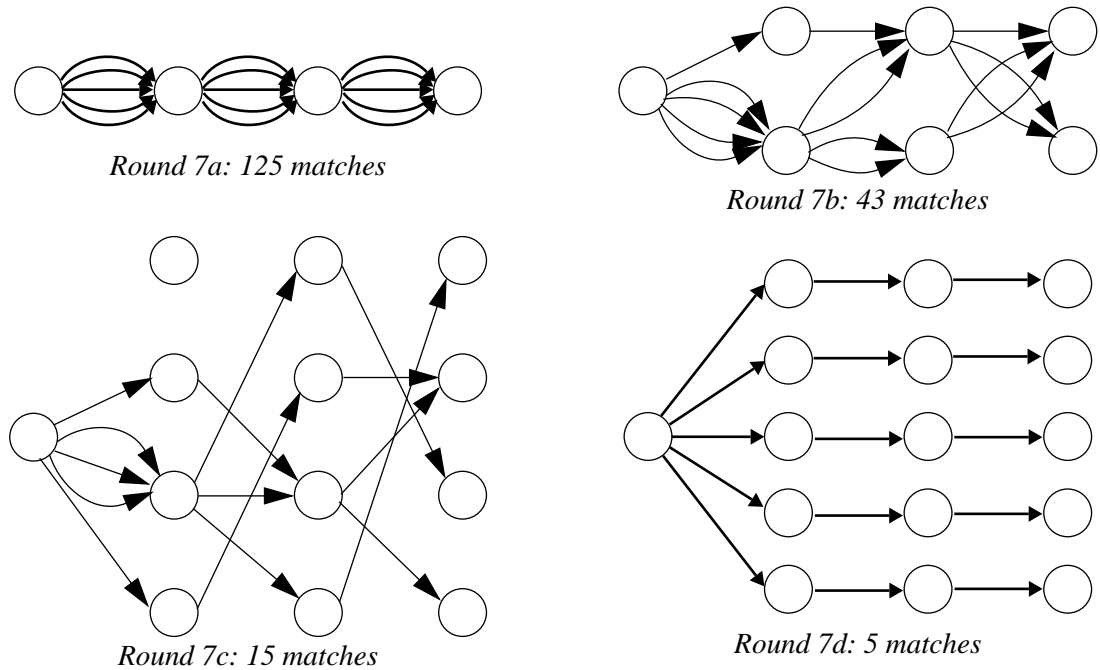

*Round 7a: 125 matches*

*Round 7b: 43 matches*

*Round 7c: 15 matches*

*Round 7d: 5 matches*

**Figure 7-19.** Example system graphs for round 7a, 7b, 7c, and 7d with *r-r-r* after the 5th iteration. The edges are invocations of a method *r*. The columns of nodes represent the class *Ana* and different instances of the classes *Ana2*, *Ana3*, and *Ana4* (or for round 7a, the classes themselves), respectively. Complete matches can be located by finding paths from the *Ana* node to *Ana4* nodes.

to the system graph alongside the previous one, i.e., it is placed between the same two system edges as the same call in a previous iterations. This is depicted in Figure 7-19. This forms a worst case in terms of the number of matches that result from chained calls with these policies.

```
import java.util.*;                                        }
                                                      }
class Ana {                                            static int getRand() {
                                                          return (int)Math.floor(Ana.randgen.next-
static int num_its= 40;                               Float()*Ana.buckets);
static int depth= 4;                                  }
static int buckets= <buckets>;                         }
static Ana2 a2[]= new Ana2[<buckets>];
static Ana3 a3[]= new Ana3[<buckets>];
static Ana4 a4[]= new Ana4[<buckets>];                class Ana2 {
static Ana5 a5[]= new Ana5[<buckets>];                Ana2() {}
static Ana6 a6[]= new Ana6[<buckets>];                void r(int d) {
static Ana7 a7[]= new Ana7[<buckets>];                    if (d < 1) return;
static Ana8 a8[]= new Ana8[<buckets>];                    Ana.a3[Ana.getRand()].r(d);
static Ana9 a9[]= new Ana9[<buckets>];                }
static Random randgen= new Random(0L);                }

public static void main(String argv[]) {              class Ana3 { Ana3() {} void r(int d) {
    for (int k=0; k < Ana.buckets; k++) {             if (d <= 2) return; Ana.a4[Ana.getRand()].r(d);}}
        Ana.a2[k]= new Ana2();                        class Ana4 { Ana4() {} void r(int d) {
        Ana.a3[k]= new Ana3();                        if (d <= 3) return; Ana.a5[Ana.getRand()].r(d);}}
        Ana.a4[k]= new Ana4();                        class Ana5 { Ana5() {} void r(int d) {
        Ana.a5[k]= new Ana5();                        if (d <= 4) return; Ana.a6[Ana.getRand()].r(d);}}
        Ana.a6[k]= new Ana6();                        class Ana6 { Ana6() {} void r(int d) {
        Ana.a7[k]= new Ana7();                        if (d <= 5) return; Ana.a7[Ana.getRand()].r(d);}}
        Ana.a8[k]= new Ana8();                        class Ana7 { Ana7() {} void r(int d) {
        Ana.a9[k]= new Ana9();                        if (d <= 6) return; Ana.a8[Ana.getRand()].r(d);}}
                                                      class Ana8 { Ana8() {} void r(int d) {
    }                                                 if (d <= 7) return; Ana.a9[Ana.getRand()].r(d);}}
    for (int i=0; i < num_its; i++) {                 class Ana9 { Ana9() {} void r(int d) {
        Ana.a2[Ana.getRand()].r(depth);              if (d <= 8) return;}}
```

**Figure 7-20.** Java program for experiment round 7b and 7c. The num_its, depth and buckets variables were varied.

| policy name | user time (s) with various number of iterations | | | | | |
|---|---|---|---|---|---|---|
| | 0 | 10 | 20 | 30 | 40 | 50 |
| r-r | 0.95 | 1.75 | 3.20 | 5.60 | 8.69 | 12.80 |
| r-r-r | 0.95 | 3.65 | 18.31 | 54.44 | 123.90 | 238.64 |
| r-r-r-r | 0.95 | 15.11 | 191.01 | 974.18 | 3128.39 | 8848.40 |
| r-r-r-r-r | 0.96 | 151.15 | 3064.66 | 19530.88 | | |
| r-r-r-r-r-r | 1.15 | 566.89 | | | | |

| policy name | number of matches for various number of iterations, N | | | | | TPM (ms) with various N | | | | |
|---|---|---|---|---|---|---|---|---|---|---|
| | 10 | 20 | 30 | 40 | 50 | 10 | 20 | 30 | 40 | 50 |
| r-r | 58 | 202 | 468 | 808 | 1282 | 13.81 | 11.14 | 9.94 | 9.58 | 9.24 |
| r-r-r | 268 | 2248 | 7140 | 16656 | 32500 | 10.07 | 7.72 | 7.49 | 7.38 | 7.31 |
| r-r-r-r | 1380 | 20408 | 107148 | 348064 | 824488 | 10.26 | 9.31 | 9.08 | 8.99 | 10.73 |
| r-r-r-r-r | 14792 | 259360 | 1652664 | 6758144 | 21151000 | 10.15 | 11.81 | 11.82 | | |
| r-r-r-r-r-r | 40040 | 2256876 | 25526000 | 154780480 | 546383312 | 14.13 | | | | |

**Figure 7-21.** Round 7b measurements, number of matches, and time per match.

The program used in round 7b and 7c is shown in Figure 7-20. For round 7b, it the buckets variable is set to 2 and for round 7c it is set to 4. The results from round 7b are shown in Figure 7-21.

| policy name | user time (s) with various number of iterations | | | | | |
|---|---|---|---|---|---|---|
| | 0 | 10 | 20 | 30 | 40 | 50 |
| *r-r* | 1.13 | 1.89 | 3.33 | 5.59 | 8.52 | 12.33 |
| *r-r-r* | 1.15 | 3.11 | 11.67 | 29.32 | 66.18 | 127.01 |
| *r-r-r-r* | 1.15 | 7.10 | 58.61 | 281.58 | 862.86 | 1959.09 |
| *r-r-r-r-r* | 1.15 | 35.78 | 544.46 | 3151.41 | | |
| *r-r-r-r-r-r* | 1.15 | 56.28 | 2130.13 | | | |

| policy name | number of matches for various number of iterations | | | | | TPM (ms) with various number of iterations | | | | |
|---|---|---|---|---|---|---|---|---|---|---|
| | 10 | 20 | 30 | 40 | 50 | 10 | 20 | 30 | 40 | 50 |
| *r-r* | 30 | 102 | 236 | 406 | 654 | 25.29 | 21.55 | 18.88 | 18.19 | 17.11 |
| *r-r-r* | 75 | 730 | 2005 | 4897 | 9648 | 26.07 | 14.40 | 14.05 | 13.28 | 13.05 |
| *r-r-r-r* | 226 | 3078 | 15987 | 50688 | 114522 | 26.33 | 18.67 | 17.54 | 17.00 | 17.10 |
| *r-r-r-r-r* | 1968 | 29272 | 161052 | 594420 | 1782768 | 17.60 | 18.56 | 19.56 | | |
| *r-r-r-r-r-r* | 2738 | 124513 | 1248415 | 5775560 | 18271876 | 20.13 | 17.10 | | | |

**Figure 7-22.** Round 7c measurements, number of matches, and time per match.

```
class Ana {                                    void r(int d) {
                                                   if (d < 1) return;
static int num_its= 50;                            Ana3 a=new Ana3();
static int depth= 5;                               a.r(d);
static int si1=1;                              }
static int si2=2;                              }
static int si3=3;
static double sd1=1;
static double sd2=2;                           class Ana3 { Ana3() {} void r(int d)
static String ss1="1";                         {if (d <= 2) return; Ana4 a=new Ana4(); a.r(d);}}
static String ss2="2";                         class Ana4 { Ana4() {} void r(int d)
                                               {if (d <= 3) return; Ana5 a=new Ana5(); a.r(d);}}
public static void main(String argv[]) {       class Ana5 { Ana5() {} void r(int d)
    for (int i=0; i < num_its; i++) {          {if (d <= 4) return; Ana6 a=new Ana6(); a.r(d);}}
        Ana2 a=new Ana2();                     class Ana6 { Ana6() {} void r(int d)
        a.r(depth);                            {if (d <= 5) return; Ana7 a=new Ana7(); a.r(d);}}
    }                                          class Ana7 { Ana7() {} void r(int d)
}                                              {if (d <= 6) return; Ana8 a=new Ana8(); a.r(d);}}
                                               class Ana8 { Ana8() {} void r(int d)
}                                              {if (d <= 7) return; Ana9 a=new Ana9(); a.r(d);}}
                                               class Ana9 { Ana9() {} void r(int d)
class Ana2 {                                   {if (d <= 8) return;}}
Ana2() {}
```

**Figure 7-23.** Java program for experiment round 7d. The num_its and depth variables were varied.

The results from round 7c are shown in Figure 7-22. For these, the bucket size corresponds to the number of instances created of all each class (except *Ana*). See Figure 7-19. Note that while which instance to jump to for the next invocation is chosen randomly as a means to spread out the edges in the graph, the execution is deterministic and repeatable due to the constant seed.

| policy name | user time (s) with various number of iterations | | | | | |
|---|---|---|---|---|---|---|
| | 0 | 10 | 20 | 30 | 40 | 50 |
| *r-r* | 0.94 | 1.50 | 2.47 | 3.86 | 5.70 | 7.94 |
| *r-r-r* | 0.94 | 2.16 | 5.02 | 10.92 | 21.30 | 37.54 |
| *r-r-r-r* | 0.95 | 3.11 | 8.51 | 20.02 | 40.41 | 72.76 |
| *r-r-r-r-r* | 0.95 | 4.36 | 12.67 | 30.06 | 61.09 | 110.13 |
| *r-r-r-r-r-r* | 0.95 | 6.94 | 26.40 | 72.19 | 158.22 | 297.35 |

**Figure 7-24.** Round 7d measurements.

Round 7d uses the program in Figure 7-23 and Figure 7-24 shows its run times. For this program, there is a new class instance created for every call to *r*. Only one match is formed per iteration.

Compare the run times for the different policies from the different rounds with different number of iterations. There can be a large difference in the amount of time taken. For example, consider *r-r-r-r* with 20 iterations. For round 7a, it took 1175.63 seconds, 974.18 seconds in round 7b, 58.61 seconds in round 7c, and only 8.51 seconds in round 7d. The difference is the number of matches: 160,000, 20,408, 3078, and 20, respectively. As shown, the number of system nodes present can have a large impact on the number of matches. (Result 12) We can further conclude that large amount time that can be taken in checking policies with several edges is not inevitable, but depends on the execution. From prior results, we also know the selectiveness of policy edges can also help. (Result 13)

## 7.5 Qualitative analysis

In this section, applying LaSCO to Java is considered from a point of view beyond the implementation described in this chapter. In the following subsections, performance issues are discussed, whether LaSCO can state useful policies for Java, the ease of writing policies, and LaSCO is compared to the Java Security Architecture.

### 7.5.1 Performance

Whether LaSCO can be applied to Java in a practical setting might often depend on how much of a performance impact checking policies has on an executing program. As the quantitative analysis has shown, this is could be an issue. However, the current implementation is hardly optimized. It is written in Java, Perl, and JPL The focus was on the convenience of development over the efficiency of execution. Some ideas that might make LaSCO policy checking more efficient are mentioned here:

- A large amount of savings can be had if a method is known at compile time to have no need to be checked. This is the case if the statically available information implies that no policy edge will match regardless of state at run time. So, an implementation can take advantage of this by doing as much static analysis as possible.

- Building complete matches can be improved for policies with more than two semantic entities by remembering all known partial matches, rather than just initial matches. Thus to develop new complete matches, new initial matches need only consider the stored ones, rather than trying to build all possible complete matches. (There is a trade-off here for increased memory requirements.)

- Custom-generating code to check a particular policy or set of policies should improve run time efficiency.

- Implement it in a language with more efficient execution and with more efficient data structures, e.g., C or C++ and import it into Java using the Java Native Interface.

Regardless, Java has a reputation for executing slowly as is, so it is not clear that performance-critical application would be written in it.

A more complete implementation of LaSCO for Java would be able to accept policies with isolated nodes in them. Section 6.5.3 described how this might be done. A change in data members might even be infrequent, reducing the overhead of checking. When an portion of code executes that contains a relevant member change, a snapshot of the relevant data members for the class or instance must be added to the *SystemGraph* and this new instance checked against isolated policy nodes that might match. This is clearly less work than checking a method, so should run faster.

In general, when the overhead for checking policies becomes overwhelming, it is when there are a large number of matches to follow up on. Without changing the policy there is little that can be done about that -- the policy says each of those cases must be checked. If the policy can be rewritten such that the domain matches less often, that would help.

The worst case number of matches of a certain policy to a certain system history is $\theta(|M|^{|E|}|O|^{|L|})$, where $|M|$ is the number of events, $|E|$ is the number of edges in the policy graph, $|O|$ is the number of instances of objects, $|L|$ is the number of isolated nodes in the policy graph. This occurs if every policy edge matches every event and every isolated policy node matches every object instance.

### 7.5.2 Stating useful policies

For the setting in which one is wishing to add assertions about method invocations or object state to their code or to check invariants, this is a straightforward matter in LaSCO. The nodes represent objects and the edges events. Even those involving context (history or state) can be easily handled. The situation is similar where a program is being run by persons not necessarily trusted, e.g, at a kiosk. Also consider an environment in which an implementation is being used by a group of persons with different roles. Here one might make assertions about role usage. Provided the role is denoted in object state, this can be handled.

In an environment where one is wishing to guard oneself against malicious code, it would seem that one wants to guard access to a resource or a set of resources. One can state policies in which the methods or sources that may access a certain method or a certain resource can be delineated. If one has in mind a broader idea of access that is irrespective of the sequence of method calls, this can handled as well. The most convenient manner is to define the system model differently than is done in Section 6.1. One can have any event be the transitive closure of method invocations. That is, when A calls a method on B and B calls a method on C, then A is also accessing C by the later invocation. One might use such a system model if one was worried about information flow.

### 7.5.3 Ease of writing policies

Nodes and edges are a natural and common means of expressing classes and their relationship through method calls. LaSCO policies are a sort of template based on this, a natural extension of this. Through the user interface one can write policies using the the program to which they will be applied as a guide and reference, creating the template from the instance. Generic policy may also be instantiated from a program and generic policies created and saved. If the user prefers they can also create policies using direct manipulation without reference to a program.

### 7.5.4 Comparison to the Java Security Architecture

Here, some of the key differences between LaSCO and policy based protection in the Java Security Architecture for JDK version 1.2 [14] are discussed. This architecture provides some primitives such as signed code to the application developer. It also provides access control services to classes that employ it, such as the standard file and network I/O classes. These access decisions are based on policies -- sets of allowed permissions. Permissions generally consist of a target and an action. An access is allowed if the caller is allowed the action on the target. Each of these permissions are defined by the class employing the access control.

One advantage that the Java Security Architecture has is that it can monitor and control activities that LaSCO cannot. For example there is no way to describe a restriction on what is allowed to create a thread, at least in the system model described in Section 6.1 and the current implementation.

However, as [14] indicates, the Java Security Architecture (at least at present) does not provide the correct mechanisms to provide access control within a system or application, since it does not understand the internal semantics. For example, its controls are not fine grained enough to prohibit invocation of a method where such denial is based on parameters' values or the history of events and state.

There is a bit of work involved in applying the Java Security Architecture to a new system. Actions and targets must be defined and access checks must be strategically placed. If this is done manually, it is subject to errors (with respect to an intended a higher level policy) and it does not express the intended policy. This could presumedly be conducted by a higher level mechanism similar in function to the LaSCO complier for Java. Permissions are similar to access control entries and do not convey higher level intent.

# 8   LaSCO Application to GrIDS

This chapter presents our application of LaSCO to networks as observed by the GrIDS intrusion detection system. This includes both the actual activity on the network and reports by data sources and intrusion detection systems. A GrIDS network is a good contrast with a Java program as an application environment for LaSCO since it is distributed, has different types of objects and events and a difference notion of state, and policies are enforced in a different manner. A result of this study is that policies and IDSs are mutually beneficial, but there are some design considerations for both to keep in mind to enable policies to be enforced using IDSs.

We begin this chapter with a description of GrIDS (Section 8.1), present a LaSCO system description for a GrIDS network (Section 8.2), present a design for monitoring networks for policy violations using GrIDS (Section 8.4), give some example policies (Section 8.3), and conclude with few observations about applying policies to IDSs (Section 8.5).

## 8.1 GrIDS

The Graph-based Intrusion Detection System (GrIDS) [8,32] is an intrusion detection system for large networks. GrIDS was designed at UC Davis to explore the issues involved in performing large scale aggregation of traffic patterns. It features a hierarchical decomposition of a protected organization and its networks. GrIDS puts together reports of incidents and network traffic from various sources into graphs, and is able to aggregate those graphs into simpler forms at the higher levels of the hierarchy. A prototype version of GrIDS has been implemented and is deployed at several organizations.

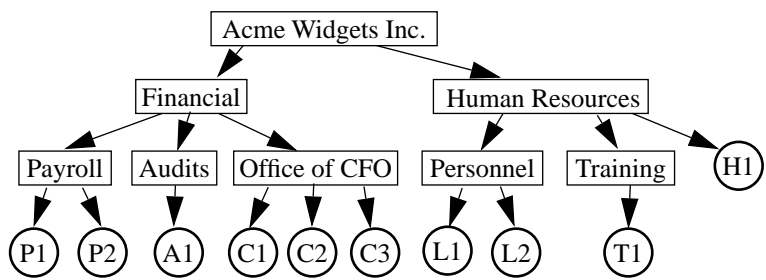

**Figure 8-1.** A GrIDS hierarchy of departments.

### 8.1.1 Hierarchy of departments

GrIDS models an organization as a hierarchy of departments, such as the one depicted in Figure 8-1. Each department consists of a set of hosts and departments. Viewed as a graph, this forms a tree with hosts as the leaf nodes. This hierarchy may be reconfigured while GrIDS operates. Each department in the hierarchy has an aggregation engine, which builds and evaluates graphs of activity within that department. The hierarchy of departments is used to help GrIDS scale and to manage how different parts of the network should be monitored.

### 8.1.2 Detecting intrusions

Intrusion detection systems that observe the activity on a single host or on a single network segment are limited in the scope of activity they observe. As a result, they may not be able to observe all the activity needed to determine that an attack, for example from a worm, is underway. IDSs that are not for a single host or network segment but that are centralized have scalability issues that cause it to be limited in the scope of observed activity as a practical matter. GrIDS addresses this by decentralizing initial detection to low level departments, and summarizing this for higher departments.

So as to allow a department to decide what types of attacks to look for and to establish thresholds, each department possess a set of rulesets. A ruleset describes what type of activity to look for, how to combine them, and when to send alerts. Thus, a ruleset might be created for a department to meet its needs.

### 8.1.3 GrIDS system overview

The software infrastructure for a department consists of two components: a graph engine and a software manager. The engines uses the departmental hierarchy as the means for aggregation of graphs, as described in Section 8.1.4. The software managers control software operation in their local department, and maintain that department's sense of the hierarchy around it (parent departments, child departments, and hosts directly contained in the department).

The whole GrIDS hierarchy requires an Organizational Hierarchy Server (OHS), which is the only place guaranteed to have a complete view of the hierarchy. It is used as a central place to guarantee the consistency of the hierarchy, and to implement such functions as locking a portion of a hierarchy when a reconfiguration is underway. In contrast, the software managers for individual departments have only a local view of the parts of the hierarchy immediately adjacent to their department.

Additionally, every host that runs GrIDS software (aside from user interfaces and the OHS) has a module controller, which is responsible for starting and stopping the GrIDS modules on that host and for configuring them. Any piece of software (most commonly other intrusion detection systems and state reporting programs) can serve as a GrIDS data source. The reports from a data source take the form of a graph that gets fed into the engine for the department to which it belongs. A passive network monitor is a standard GrIDS data source.

Finally, user interfaces acquire views of the hierarchy and use them to help users visualize it and request changes to its structure and to manage what software is running in a department. All changes are performed as transactions on the hierarchy. They are initiated by a user interface, but require the involvement of instances of the GrIDS components. In addition, the user interface may be used to manage the rulesets for a department. The user interface receives and displays alerts from the engine in the department it monitors and may be used to query the engine regarding its current graphs.

### 8.1.4 GrIDS engine

This section describes aspects of the GrIDS graph engine. The engine takes basic activity reports and converts them into graphs. The overall purpose of the graph engine is to build graphs of network activity and analyze them to see if they are suspicious. The engine implements a set of rulesets. When a ruleset detects what it considers to be suspicious activity, it sends an alert message to the user interfaces that are monitoring the engine.

### 8.1.4.1 Report processing and resulting communication

Reports come into the engine from either a child engine or from a GrIDS data source. In either case, the reports take the form of a graph consisting either of a single node or a single edge with adjacent nodes. At the point the reports come in, it is passed through three buffering mechanisms with different goals:

- one sorts reports (which could arrive out of order) by time,

- one removes redundant reports, and

- one renames edges for a particular engine's point of view.

All incoming reports are assumed to have a *time* attribute which represents the true time at which the event to which they refer occurred. After passing though the last buffer, the nodes in the report represent either children of the engine's department (either hosts or departments) or hosts external to the department.

The reports are then given to each of the rulesets for processing (unless the report is targeted to a particular ruleset). A report might be accepted by a ruleset for incorporation or not. An accepted report might be merged with the graphs already maintained by that ruleset. If one of the nodes in an edge report was external to the department, at this point, that edge (in its new graph context) is passed to the same ruleset in the parent engine (assuming there is one), with nodes representing children renamed to the name of the department. If the attributes of a graph have changed, the graph is reported to the same ruleset in the parent engine in summary form, as a single node with the global attributes of the graph as the node's attributes. (See [8] for more details of this.) If a graph is idle long enough, it times out. Should a graph formed by a ruleset be determined to be suspicious by the ruleset, it is sent to the user interface for the department as an alert, along with an alert string and alert level.

Note that by using the approach of passing reduced graph to the parent engine (rather than the whole graph), GrIDS is able to infer the suspicious nature of large graphs, while still reducing drastically the amount of information which must be considered at the top of the hierarchy. It is this that makes GrIDS a scalable design.

### 8.1.4.2 GrIDS graphs

Graphs built in an engine, reports output from the engine, and reports to the engine consist of nodes, edges, attributes of nodes, attributes of edges, and global attributes of the graph as a whole. All attributes consists of a pair *(name,value)*. Names are identifiers, while attribute values may either be scalars (interchangeably strings, numbers, or logical values), sets of scalars, or ordered lists

**Global Attributes:**
- **gids**: a set of graph ids associated with this graph, any of which can be used as a unique identifier.
- **ruleset**: the name of the ruleset this graph is in.
- **nnodes**: the total number of nodes in a graph.
- **nedges**: the total number of edges in a graph.

**Node Attributes:**
- **name**: the host or department the node represents.

**Edge Attributes:**
- **source**: the departments and host associated with the source of this edge that are within this engine's scope, in a list starting with the name of the source node, and ending with the particular host.
- **dest**: same as *source* except pertaining to the destination side of the edge.
- **id**: a string unique identifier for this edge.

**Figure 8-2.** Graph auto-computed attributes

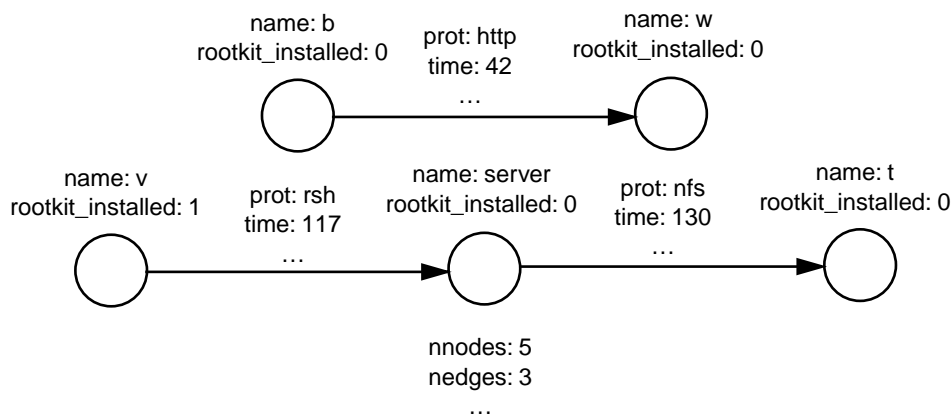

**Figure 8-3.** Example GrIDS graph.

of scalars. Certain attributes on a graph in an engine are maintained by the engine implicitly. These are described in Figure 8-2.

A ruleset might produce a graph such as depicted in Figure 8-3.

### 8.1.4.3 Rulesets

The rulesets consist of several sections that are used for different tasks. These are used in a particular order to process incoming reports and to aggregate graphs. There are rules to filter incoming reports to type wanted and rules to transform the attributes of a report into the attributes maintained by the ruleset. Certain rules decide whether a graph should combine with other graphs with which it overlaps and how the overlapping attributes should be updated upon combining. Finally, there are rules to evaluate whether a graph warrants sending an alert Each of the rules in the different sections contain expressions on attributes and many update a particular attribute.

## 8.2 System description for GrIDS

In this section, we describe what a LaSCO system description for a network might look like, when it is being used by GrIDS. There is a natural correspondence, but there are a few challenges.

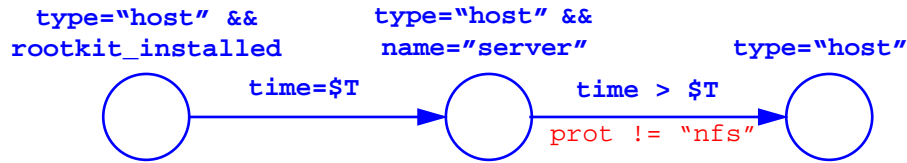

**Figure 8-4.** Policy graph for server compromised connection policy.

An active attempt is made to keep the description close to what is present in GrIDS in order to keep the implementation simple. Keeping it close would also make it simpler for the policy writer familiar with GrIDS or just trying to learn GrIDS.

The objects in the system are hosts on a network. Events correspond to the events described in the network monitor chapter of the GrIDS design document [8]. In short, several events are reported for each TCP connection or sequence of UDP packets. These always contain a *begin* and an *end* event. The events that correspond to the middle of of a connection are particular to the protocol. For telnet, they correspond to option negotiation, authentication, data, and reset. If data sources other than the network monitor generate reports containing events, then each of the types of edges that generates is also an event. In any case, the event attributes are identical to those defined for that type event (i.e., those that are found in the report). The time attribute is standard on edges and represent seconds as Unix reckons time. This is used for the required time attribute on edges. The *id* attribute is standard and unique. (It would be nice to be able to have a certain attribute, call it *data_source*, that is a string indicating the entity that generated the report. However, at present, this information is not available as part of the report, at least in any standard way.)

The attributes on nodes correspond to those reported by the data sources active for the corresponding host. For example, *alert* from the execution monitor. The *id* and *name* attributes are both always the fully qualified name of the host. Unlike with objects and classes in the Java case, the state of a host at a particular time is not directly observable in GrIDS. That is, there is nothing that can be done to actively check the state. So, for each attribute, the last reported value before the end of a second is the value at that time. We could also introduce an attribute *type* which always has value "host" to make the objects more self-descriptive.

## 8.3 Example Policies

Two example policies for networks were presented in Figure 3-14 on page 26 and Figure 3-15 on page 26. A policy that makes use of the special information found in GrIDS is presented in Figure 8-4. The policy says that if a host has been flagged as having RootKit installed on it and it connects to the host named "server" followed by another connection to another host, then the second connect cannot use the NFS protocol. The network activity described in Figure 8-3 would match this policy and violate it.

## 8.4 Enforcing LaSCO on GrIDS

Enforcing LaSCO on a GrIDS hierarchy is a collaborative effort among the different aggregation engines. The cleanest manner in which to integrate this capability is to expand the capability of the engine to enforce LaSCO policies natively, with the assistance of the generic policy engine (Section 5.2). Thus LaSCO policies would be peer to rulesets, receiving the same type of reports and producing alerts with the policy is violated. This section describes key aspects of how this extension would operate.

An alternate approach, in which LaSCO policies are translated into rulesets for enforcement has also been considered. That approach proved to be difficult, primarily due to the peculiar programmatic approach that rulesets require. Even with extensions to the ruleset, constructing a translation for the general case was quite a challenge.

Contingent matches (defined below) maintain the state of how policies match what has been seen thus far. Each engine maintains a pool of these for each policy. These are passed among engines in a certain manner to detect all policy matches and violations. Within a engine, processing for each new report or new contingent match consists of two parts: producing new contingent matches and evaluating these new contingent matches. Those steps are discussed in the following subsections. Before that though, contingent matches are described and the idea of locality of policy nodes and edges is introduced.

### 8.4.1 Contingent matches

The idea of a partial policy to system match (Section 4.5) is generalized to what is termed a ***contingent match***. Contingent matches are partial matches that may contain ***contingent nodes*** whose domain predicate has not yet been evaluated at certain time(s). These nodes have an adjacent edge that matched at each of those times. However, not enough information was available about the node to evaluate it. Thus, the partial match's validity is *contingent* upon the successful evaluation of the contingent nodes at the given times. A contingent match is never complete nor considered a partial match until all the contingent nodes have been successfully evaluated.

In terms of representation, a contingent match may be a partial match augmented with a set. The set is a group of ***contingent conditions***. A contingent condition is a tuple consisting of a policy edge and which node of the edge (either the source or the destination) needs to be evaluated (with respect to the system node that is bound to that location).

Merging contingent matches is similar to merging partial matches. However, a contingent condition may (and should) be removed from either contingent match prior to merging the partial matches if the policy edge has an event bound in the other match and does not list the same contingent condition in its set. (The requirement that policy edge needs to be bound to the same event and that the variable conditions must be unifiable are taken care of when merging the partial matches.)

### 8.4.2 Locality of policy nodes and edges

Due to the way in which the GrIDS hierarchy is organized, reports about a certain host are only sent to one engine. Specifically, it is sent to the engine that contains the host. Therefore, that is the only place that has the full state attributes for the corresponding system object. Here it is considered a ***local host***; elsewhere it is ***non-local***. The only thing known about the object state at the engines in which it is non-local is its *name* (and therefore its *id*). (As a side note, non-locality is why contingent matches are needed.)

It is presumedly not uncommon for a policy to be referring to a particular host by name in one or more of its node domain predicates. In such cases, a policy node may be considered to inherit the locality of the host it represents. A policy node is termed ***local*** if it is known to match only local hosts. A policy node is termed ***non-local*** if it is known that it will never match local hosts. If a policy node is neither local nor non-local, it is termed a ***general node***.

Policy edges may then also be classified in terms of their locality, based on the locality of the nodes they are adjacent with. A policy edge is ***local*** if both its adjacent nodes are local, and is ***non-local*** if both its adjacent nodes are non-local. If only one node adjacent a policy edge is local, then the policy edge is termed to be ***half-local***. Otherwise, the policy edge is ***general***.

### 8.4.3 Producing new contingent matches

New input into the engine is processed according to whether it is a report from a general data source or a report of a contingent match from a lower level engine. These are discussed separately in the following subsections.

### 8.4.3.1 New activity reports

When a new activity report arrives from a data source, it is first sent through the buffers described in Section 8.1.4.1. For rulesets it is sent through all the buffers, as described there. For processing by policy enforcement, however, it is just passed through the time reordering buffer and the redundant report elimination buffer.

After buffering, all the reports for a given second are considered together. The new reports are converted into system nodes and system edges of the form described in Section 8.2. For edges, this is a straightforward process. For certain nodes, the values of the attributes for the time needs to be derived. These nodes are the ones that are on the new reports, either by itself or as an incident node in an edge report. To start the determination of attribute values, the ones from the last time instance available are used. The node is then updated with attributes mentioned in node reports for that node. As an optimization, only those node and edge attribute mentioned in any policy need be recorded.

These system nodes and edges are added to a *SystemGraph* maintained for all policies collectively, to represent local activity. (Unlike in the Java case in which separate *SystemGraph*s are maintained, there is little advantage to doing so in this case. The difference here is that there is no static foreknowledge of what edges a given event will match. So there are no hints to pass to the generic policy engine. Since cached matches are used, there is little extra cost in having edges around that didn't match anything, when evaluating new activity.)

The generic policy engine would be extended to a degree to allow a call that reports back all the contingent matches found between a policy and a *SystemGraph*. These contingent matches are not just the maximally sized ones, but include subsets of others, to include even initial matches. However, contingent nodes are only established for non-local and general policy nodes; the local policy nodes can always be evaluated. Note that at most one end of a policy edge is contingent since if both ends of a edge report were not local, it would not be found in the engine. The particular option used with this call would be to only report back the contingent matches including a new system edge or an updated system node in the *SystemGraph*. These new contingent matches are each associated with one of the policies in effect.

### 8.4.3.2 New contingent match reports

When contingent matches are received by an engine, they are handed directly to the part of the engine responsible for processing them. Each is associated with a particular policy. Let us consider a particular contingent match that has arrived. An attempt is made to combine (as described in Section 8.4.1) the contingent match with all the partial matches in the pool. Those matches that succeed are the contingent matches for the next step.

Actually, there are ways to organize the pool of contingent matches so that they need not all be compared with the incoming match -- some of the ones that will fail can be skipped. A simple way is, for each of the semantic entities of the policy, to keep the contingent matches indexed by the system node or edge that matches, if any do. Then to find the candidate contingent matches a set is created for each of the established bindings to semantic entities in the new contingent match. This set corresponds to the contingent matches that have the same system node or edge plus those that had no binding for that policy piece. The intersection of these sets is the set of candidate contingent matches sought after. This process eliminates the contingent matches that disagree with the bindings for the semantic entities. There still can be disagreement for the incident nodes, with system nodes being multiply used, and with the variable bindings. There are likely ways to further index the matches by these criteria as well. It also seems like that there are results from database theory on efficient ways to organize the data to avoid the need to take the intersection. It is not clear how worthwhile further optimizations are in actual use.

### 8.4.4 Evaluating new contingent matches

For each new contingent match with all local isolated policy nodes matched and all local and half-local policy edges matched, further processing is done. If the match is complete, then the requirement is evaluated. (Actually, to do this, any attributes found in the requirement need to be passed along with the contingent match, as this information may not be local. Alternately, the requirement predicates may be evaluated early with just the attributes, and the result passed along.) If the requirement is satisfied, then the match may be discarded since the policy was not violated. If any of the requirement predicates evaluates to false, then the policy is reported to have been violated by sending an alert. A graph describing the violation may be produced from the match.

However, if the contingent match is not complete, the contingent match is sent up to the engine in the parent department, if any. Furthermore, if all general semantic pieces have been matched in the contingent match (i.e., only non-local semantic pieces remain), the match may be discarded since there is no further growth than can be made in the local engine.

In any case, all of the new contingent matches except those that were discarded and incoming match are added to the pool for later growth.

## 8.5 Implementing policies using an IDS

The study of applying LaSCO to GrIDS has inspired several observations about both policies and intrusion detection systems.

**It is advantageous for an IDS to be able to enforce policies.** (Enforcement in this case typically means monitoring for policy violations and perhaps responding to them.) Although a particular policy language might not be able to express all that is expressible in a particular IDS's native methods, policies represent restrictions in a more structured manner. Besides being able to convey the intention of the restriction more clearly, this can mean that the policy can be easier to write. If the policy is easier to write or if the policy (possibly used elsewhere) is already written in a policy language, then the native means of writing policies might not need to be learned or used. The desire to state policies without having to write rulesets for GrIDS was what initially motivated LaSCO's development.

**There are benefits from enforcing policies with an IDS.** Despite the fact that IDSs typically cannot directly prevent policy violations, certain benefits can be found in enforcing policies in this manner. Chief of these is that the IDS can provide information not typically available at the level of the system, e.g., inferences that the IDS has made regarding suspicious activity on the system and even knowledge that policies have been violated. This information could be used in making a policy decision. For example, a policy might come into effect only "/bin/sh" has been found to be modified. As described in Section 8.2, this information can be incorporated in LaSCO by attributed defined on the system objects and events. The advantage provided by GrIDS in particular to policies is multiplied since GrIDS incorporates the output of other IDSs. Another advantage to enforcing policies with an IDS is that the IDS might already have the infrastructure in place to do the monitoring and communication of system events.

**An IDS needs to provide a practical means of expressing the policy in native terms.** The IDS should either be able to accept expressions of policy directly, it should provide a target language that is convenient to generate from policy specification, or it should allow modules to be provided for detection analysis. This may be in addition to another native method of expression.

As discussed earlier, GrIDS rulesets are not particularly suitable for automated generation of security constraints (at least for those coming from an expressive a source as LaSCO policies). (To be fair, it was not designed to be; the ruleset language was intended to have its writer think in a certain manner and to keep detection linear (e.g., no graph searches).) Although GrIDS as originally con-ceived does not accept analysis modules, it was not a difficult extension to add LaSCO enforcement. This was due to the modular design.

**A policy language must be able to state policies for the target system.** In order for an IDS to enforce a policy, the policy language used must be able to state policies for the system the IDS is monitoring. Moreover, it must be able to state it at a sufficient level of detail that it can be enforced. For LaSCO and GrIDS, LaSCO was able to model the network that GrIDS observes, including the information provided by arbitrary data sources.

**A policy language should be amenable to decomposition for IDS monitoring, especially if the IDS is distributed.** In a distributed setting, all events that compose a certain application of policy may not be observed at the same location. Thus, there must be a degree of coordinated analysis in the general case. In order for this to not involve sending all events to a central location, one must be able to decompose the policy into individual parts to look for. It is observed instances of these parts that can be passed during the coordinated analysis. Thus is should be possible to readily infer the individual parts to monitor for from the the policy statement. This is decomposition of policy. If a policy language is amenable to decomposition, then it would be useful even in the non-distributed case if the IDS can use it for incremental analysis.

We find LaSCO to be quite amenable to decomposition. From its formal semantics, we know that the semantic pieces defined in Section 4.4.2 (i.e., policy edges and isolated nodes) are what needs to be monitored for atomically (i.e., must be observed at the same instance and is the basic element of composition). A minor challenge in applying this to GrIDS was that the state attributes for both ends of an event might not be centrally located, but introducing contingent matches addressed that.

# 9 Comparisons with other work

When considered with respect to the desirable characteristics of a security policy language we introduced in Section 1.3, LaSCO stands out among other policy languages and security statement mechanisms. It is the only known work to be strong on clarity, executability, and the precision with which an intended policy may be expressed. Furthermore, only a few other languages share user friendliness as an attribute. Those efforts that are not strong on clarity of course cannot be strong in amenability for formal reasoning. The attribute of being able to express policies for many different systems and the attribute of being able to express policies at multiple levels of detail are strongly related.

The traditional decision model of access control in computers has access decisions based on the subject, object, and privilege requested. In an access matrix [23], what is allowed or denied is recorded for each of these triples individually. This clearly does not convey intent and does not allow for much expressiveness in policy decisions. It is easier to understand a single policy than a set of access rights. For example, this does not allow for decisions based on the context of the system (states of objects and other events that are occurring or have occurred) or policies restricting the state of objects. In access control lists and capability-based systems (see [11]), each triple need not be specified, but it maps down to the same model and has the same expressiveness. On the positive side of this model, as it amounts to a table lookup, it can be readily enforced. Additionally, the model is general and can be applied to many systems and perhaps to different levels of detail. Some extensions to this model have been proposed, such as BEE [25], which can refer to global attributes in evaluating an access. This allows more expressiveness, but falls short of LaSCO in terms of precision. TAM [28] extends access matrices to model certain safety properties. This increases its ability in terms of formal reasoning, with respect to the general model. However the model is built on system primitives without distinct semantics, which, as Woo and Lam [33] argue, is less clear.

Deeds, developed by Edjlali, Acharya, and Chaudhary [12], is a history-based access control mechanism for Java whose goal is to mediate accesses to critical resources by mobile code. This resembles the implementation of LaSCO for Java described in Chapter 6 in that they both insert code into Java programs. However, whereas their basis for access control decisions is the result of dynamically executing Java code provided by the user, LaSCO's basis is formal policies. We argue that our approach better captures the intention of the policy writer, is more clear, is more friendly to inexperienced users, and is more amenable to formal reasoning. Clearly, their mechanism is tied to Java programs and can only speak at the level of detail of a program. The benefit to their approach is immediate executability since the specifications are code and that it can express any intended policy (though not necessarily concisely).

Some policy languages are based on logic. A clear benefit of an approach based on logic is the unambiguity of meaning and potential amenity toward formal reasoning. A general issue though is how user friendly such languages are, especially to those without a strong logic background. Woo and Lam present an appealing approach in [32]. They encode authorization requirements in a policy base, a structured set of statements in a particular logical language. They use three strategies to keep the overall specification succinct: closure (sets of authorizations that are on or off together), default (implicit rules), and inheritance (rules for translating authorization from a group to a user). Implementability one of their goals and they can state policies for different systems. Their approach aims for management of sets of policies, whereas this has not been a focus of LaSCO thus far. LaSCO has been more focussed on being able to express more types of policies (Woo and Lam do not attempt to express policies based on history such as those in Section 3.3.2) and to facilitating use of language by humans.

Another policy language based on logic is the Authorization Specification Language (ASL) [16]. ASL introduces groups and roles in the access control model and, like Woo and Lam, provides rules to express policy beyond a single access control policy. These consist of different types of rule predicates to express explicit authorizations, derived authorization, default subject-object authorization, general access control rules, access performed, active roles, subject membership, type membership, and integrity checks. Though authorizations expressed resemble Prolog [8], there is clearly a learning curve. There does not appear to be any way to express constraints on objects as we do in LaSCO, as an object's state at a given time is not modeled. It seems the language can be applied to a variety of systems.

Cholvy and Cuppens [6] take a novel approach to expressing all the policies on a site. They employ deontic logic, a logic that expresses obligations. Through this, liveness properties can be stated, which is beyond the present capability of LaSCO. An enlightening result from Schneider [29] proves that without strong knowledge of the possible executions of a system, non-safety policies cannot be enforced. This leaves the question as to whether policies expressed in using [6] can in general be enforced, especially in a system-independent manner. To those familiar with logic expressing policies with this approach would seem to be straightforward (provided the right primitives are present); it is less user friendly for other cases. This approach rates highly on clarity, amenable to reasoning (they demonstrate contradiction detection), the ability to express policies for multiple systems, and the ability to express policies at different levels of detail.

Ponder [10], developed at the Imperial College, is a promising recent policy language for specifying security and management policies for distributed systems. This declarative language is for specifying different types of policies, grouping them into roles and relationships, and defining configurations of roles and relationships as management structures. Its scope is RBAC and general purpose management policies. Ponder is intended to be extensible and is platform-independent. The types of policies it includes are: positive and negative authorization policies; refrain policies (logically equivalent to negative authorization but enforced on the source); obligation policies; positive, negative, and cascade delegation policies; conditional authorization policies; and filtered authorization policies. Its syntax is largely adopted from OCL [26] and efforts are made to keep it implementable. It can be applied to different systems. The language does not seem to be targeted to casual users. Formal semantics and formal reasoning mechanisms are not currently available though it is a main goal and is planned for future work.

The Adage architecture [31], developed at the Open Group Research Institute, focuses on creating and deploying security policies stating access control on roles in a distributed environment. The developers argue that security products that a user cannot understand will not be used and focus on usability through enabling the user to build policy from pieces that the user understands. This emphasis towards user friendliness is shared with LaSCO. As discussed in Section 3.3.4, LaSCO can express any Adage policy and more. Formal semantics have not been defined for Adage and sometimes the implication of a policy is not clear, which implies that it is does not fare well on the amenability to formal reasoning desired characteristic.

The policy language presented in this dissertation took some inspiration from the Miró work of Heydon, Tygar, Wing, *et. al.* at Carnegie Mellon University [20]. Using their constraint language, one can state allowable access control states and restricting file and group nesting for file systems. They employ visual means, an annotated graph based on Harel's hierarchical graphs [18], which promotes user friendliness. However, Miró can only express allowable states (a snapshot of a system), whereas LaSCO states constraints based on events and object state. A policy checker for the language has been implemented [21], demonstrating its executability. The approach seems general but has only been applied to file systems. Semantics have been defined for their language for representing system snapshots [24] but not for their constraint language, hindering formal reasoning.

Object Oriented Domain and Type Enforcement (OO-DTE) [33] is a technology whose focus is organizing, specifying, and enforcing access control in a distributed object environment such as CORBA. In the DTEL++ policy language's support for distributed object systems, types are declared and assigned to interface methods (possibly through inheritance) and domains (roles for processes) are declared and assigned permissions (e.g., invoke and implement) over types. This is enforced using a run time mechanism. Implementations of it exist for at least three different ORBs (platforms), demonstrating its executability. However, as indicated in [33], unlike LaSCO, the semantics of the language are ambiguous and difficult to implement and OO-DTE can be difficult to use by a human. There does not appear to be a way change object domains at run time, which means that, unlike LaSCO, OO-DTE cannot state policies that depend on the dynamic state of an object nor apparently is there a way to make policy decisions based on historical context.

As contrasted with a policy, a system specification describes the bounds under which a system should execute. This captures more than what is important for security to be maintained; it can capture other system properties as well. However, it does not capture the intention of the policy nor does it express constraints in a system-independent manner. A security policy may be used to help express the desired security and so help derive a specification for a system.

# 10 Conclusion

In this dissertation, we presented a formal policy language based on graphs. This language may be used to express policies for any system that can be modeled as objects that interact through events. The details of objects and events are described by their named attributes. LaSCO policies separate into two components: the domain and and requirement. The domain of a policy matches when objects and events in a system's execution correspond to nodes and edges in the policy graph, given their domain predicate. Predicates are expressions on the attributes of a object or event. A policy is violated if any predicate in the requirement section of the policy is not satisfied. Variables may be used to impose relationships between the objects and events matching a policy. LaSCO may express standard notions of systems safety such as Bell-LaPadula [3] and Chinese Wall [4]. In addition, custom policies maybe be created to fit the particular needs of a system. In particular, policies that are dependent on system historical context and that restrict the state of objects may be expressed. We emphasize that customizability promotes security since the policy writer can express the policy they want.

The formal operational semantics of the language were presented. These map a set of formally represented policies and a formally represented system execution into locations where a policy matches, locations where a policy is violated, and whether an execution violates the set of policies. We presented an architecture for implementing LaSCO policies on any system that LaSCO can write policies for, consisting of a policy engine that is system-independent and an interface layer specific to a system. An implementation in Perl of the generic policy engine was described, allowing policy violations to be checked incrementally against an executing system or against an entire system history at once.

Two detailed studies of applying LaSCO on particular systems were described in this dissertation. We modeled a Java program for use in LaSCO and implemented a policy enforcement mech-anism. The enforcement mechanism introduces checks of policy that at run time use the generic policy engine to prevent method invocations that would violate policy. The policy writing aid in this implementation is a GUI that facilitates writing policies, especially for a program for which an abstract representation is presented in the interface. Experiments were conducted and the use of LaSCO on Java analyzed quantitatively and qualitatively.

The second study presented was in writing policies for networks as viewed from the GrIDS intrusion detection system. In addition to normal network traffic, GrIDS also receives reports from data sources such as host-based IDSs. We model this using the LaSCO system model and present a design for modifying GrIDS to enforce LaSCO policies. This involves a modification to the GrIDS aggregation engine, a key component in the distributed IDS. Through this modification, LaSCO policies can be enforced natively through the generic policy engine, making use of the GrIDS hierarchy to pass messages and find violations of policy that might be widespread. Based on the experience with this design (as well as with LaSCO and GrIDS), we described our conclusion that IDSs and policies are mutually beneficial. We also observe that a few design decisions for IDSs and policy languages are important towards this: an IDS needs to provide a practical means of expressing the policy in native terms, a policy language must be able to state policies for the target system, and a policy language should be amenable to decomposition for IDS monitoring, especially if the IDS is distributed.

LaSCO has characteristics that establish its strength in a list of desirable properties of a security policy language we described. Through its formal semantics, policy statements in LaSCO are clear and unambiguous. As demonstrated by the implementation, LaSCO policies can be executed. The generality of the system model allows it to be used in different settings, avoiding someone having to learn a new language or invent one (and its policy interpreter) for different systems. That it can characterize policy entities by the attributes of object or event described, that it quantifies its application over an entire system, and that it is able to make use of historical context promotes LaSCO's abilities to represent the precise policy the writer wanted. LaSCO can state policies at the level of detail appropriate for an application. Its graphical basis and several other characteristics promote its user friendliness in creating, modifying, and understanding policies represented in it, for both casual and expert users. Though not a deeply investigated area or research yet, LaSCO's formal semantics should support efforts at formally reasoning about LaSCO policies.

Future research directions for LaSCO include:

- Formal reasoning with different goals may be conducted about policies specified in LaSCO. The correctness and completeness of LaSCO policies may be analyzed. There are potential applications which compositional semantics other than the conjunction semantics presented in this dissertation (e.g., disjunction and prioritization) are useful, so this is an area that may be beneficial to pursue. In an setting in which a policy is being translated from one level of abstraction to another, it would be advantageous to be able to verify the correctness of the mapping.

- A formal characterization of the policies that LaSCO can express is an interesting area of research. It would be interesting to see what policies LaSCO can express either in the current language or through extensions. Extensions might allow, for example, policies of obligation to be expressed directly. The types of policies it should be able to express is an open issue.

- LaSCO can be implemented for GrIDS as described in Chapter 8. This involves modifying the GrIDS engine to interface it with the generic policy engine of Chapter 5.

- The policy editor might be generalized beyond its current implementation for Java. This would allow abstract views of other systems to be included. An extension with the potential for great benefit is one to allow the execution of a system be monitored and the policy applications that take place in the execution to be visualized through animation.

# A Supplemental Implementation Details

This appendix provides details that are in addition to the description of the implementation of LaSCO for Java presented in Chapter 6.

## A.1 Predicate text format

This text format for LaSCO predicates is needed to represent predicates as an ASCII string, for storage in a file (see the next section) and for other computer processing. Figure A-1 depicts the EBNF for the text format. The unit for a predicate is ‹predicate›.

1. ‹predicate› : ‹pred-expr›
2. ‹pred-expr› : ‹literal› | ‹attr-ref› | ‹var-ref› | '(' ‹pred-expr› ')' | '!' ‹pred-expr› | ‹pred-expr› ‹binary-op› ‹pred-expr›
3. ‹binary-op› : '&&' | '||' | '=' | '!=' | '<' | '>' | '<=' | '>=' | 'in' | 'pcont' | 'cont' | 'union' | 'intersect' | '+' | '-' | '*' | '/' | '%'
4. ‹attr-ref› : ‹name›
5. ‹var-ref› : '$' ‹name›
6. ‹name› : [A-Za-z0-9_]+
7. ‹literal› : ‹boolean-lit› | ‹string-lit› | ‹numeric-lit›
8. ‹boolean-lit› : [Tt][Rr][Uu][Ee] | [Ff][Aa][Ll][Ss][Ee]
9. ‹string-lit› : '"' [^\n]* '"'
10. ‹numeric-lit› : [0-9]+ | [0-9]+ '.' [0-9]+ | '-' [0-9]+ | '-' [0-9]+ '.' [0-9]+

**Figure A-1.** EBNF of the predicate text format.

1. ‹file› : (‹policy› "\n')* ‹policy›
2. ‹policy› : (‹node› | ‹edge›) *
3. ‹node› : ‹node-name› ‹predicates› '\n'
4. ‹edge› : ‹node-name› '->' ‹node-name› ‹predicates› '\n'
5. ‹predicates› : | '\t' ‹predicate› | '\t' ‹predicate› '\t' ‹predicate›
6. ‹node-name› : [A-Za-z0-9_]+

**Figure A-2.** EBNF of the LaSCO file format

## A.2 LaSCO file format

The LaSCO file format is defined for the purposes of storing LaSCO policies in a file. Figure A-2 depicts the EBNF for the file format, where ‹predicate› is as defined in Figure A-1. The unit for a file is ‹file›.

The first predicate is the domain predicate for the node or edge and the second predicate (if it appears) if the requirement predicate. If a predicate does not appear, its implicit value is 'True'. Node names have no semantic meaning to the policy and are used solely for cross referencing purposes within a policy. ‹edge› must appear for each edge in the policy and ‹node› for each node that has at least one predicate that is not 'True'.

## A.3 Java program schemas details

This section provides detailed information about Java program schemas and schema graphs. These were briefly introduced in Section 6.2.

### A.3.1 Java schema

The information elements of a Java schema are listed here:

- the classes defined in a program, with:
  - class data fields, types, and modifiers
  - class methods and constructors, with:
    - formal parameters and types
    - return value type
    - modifiers
  - superclass name

1. ‹file› : (‹node› | ‹edge›) *
2. ‹node› : ‹class-name› ‹attribute-info› '\n'
3. ‹edge› : ‹class-name› '->' ‹class-name› '\t' ‹method-name› ‹attribute-info› '\n'
4. ‹attribute-info› : ('\t' ‹attribute-name› ':' ‹attribute-type›)*
5. ‹class-name› : [A-Za-z0-9_]+
6. ‹attribute-name› : [A-Za-z0-9_]+
7. ‹attribute-type› : [A-Za-z0-9_]+ ( '[' ']' )*

**Figure A-3.** EBNF of the schema graph file format

- the (possible) method invocations and object instantiations in the program, with the following data items:

  - the method name (same name as the class for constructors)

  - the class whose method is being invoked

  - the class and method that contain this invocation

- type import declarations

A file format is defined for a Java schema. It shares its syntax with a Java compilation unit, but with only the above data elements present and with different semantics. The method invocations are placed within the method that contained the invocation, but at the top level. For that, the method name is qualified with the class it is invoking and there are no arguments listed.

### A.3.2 Java schema graph

A Java schema graph is a view of a subset of a Java schema as a graph. The schema is sometimes viewed in this form so as to provide a visual depiction of the schema, which facilitates human use. The nodes of a schema graph denote classes the program. Edges represent method invocations or object instantiations identified in the program source, and originate on the calling class and terminate on the called class. Annotations on a node contain the fields declared and their types, and annotations on edges depict the formal parameters and types for the method invoked.

The EBNF of the schema graph file format is shown in Figure A-3. Lines beginning with a "#" are treated a comments and ignored, as are empty lines. The first class name on an edge line is the source of the method invocation and the second is the destination class. The optional sequence square braces in an attribute type indicates that number of dimensions in an array of the given type.

## A.4 Schema extraction tool implementation details

The schema extraction too was introduced in Section 6.3. This section describes more detail of our implementation. The main script "extract_schema.pl" is simple and leaves the core functionality to the *SchemaGraph* class, described in the first subsection. This script accepts file names and output options on the command line. For each given Java source file, it creates a new *SchemaGraph* from the file and invokes methods on the *SchemaGraph* to create files in the specified formats. Schema files given on the command line are uses as information about other classes. "merge_schema_graphs.pl" takes a number of schema graphs in files and merges them together into a single schema graph.

### A.4.1 SchemaGraph class

*SchemaGraph* is a Perl module whose instances represent an abstract view of a Java program. It possesses enough knowledge of a Java program to represent a schema and a schema graph.

Input for its Java program representation comes from source files, schema files, and schema graph files. (If the input is from a schema graph file, it naturally only has enough knowledge to represent a schema graph, and not a full schema.) Input from Java source files and schema files are parsed by the *JavaParser* class, described in the next section. This produces a parse tree representation, from which the desired data is extracted and stored internally. For the most part, it is assumed by the implementation that a given source file is a legal compilation unit.

It should be noted that a only a limited method call analysis is performed on the method invocations in the source file. There is no analysis as to whether a method invocation is ever reachable in execution. A reasonable effort is made to determine the class upon whose method is invoked. Per the Java Language Specification [15], this cannot always be determined prior to execution. Other Java source code input provided with the one that possessed the method invocation being analyzed is used to help resolve the class executed. If the right side of a method invocation is a qualified name, the type of the variable or the class upon which the method is invoked is the class. If the right side contains an expression (i.e., the form is *Primary.<Name>(...)*, using the Java specification terminology), an effort is made to determine the result type, which is the class that is invoked. In case the class could not be determined precisely, Object (the base class of all Java classes) is reported.

### A.4.2 JavaParser module

The *JavaParser* Perl module takes a string representing a Java compilation unit (e.g., from a source file), and produces a parse tree representing all (semantically significant) aspects of the compilation unit. This module is generated by *perl-byacc* (a version of Berkeley YACC modified by Rick Ohnemus) from a YACC-like specification, but for Perl rather than C. The parse tree in an instance of a tree produced by the ParseNode module, described in the next section.

### A.4.3 ParseNode module

The *ParseNode* module is a collection of about 75 Perl classes that collectively can represent Java source code in a parse tree. Each of the tree nodes has the ParseNode class as base class, so may each be referred to as a ParseNode. The schema extraction mechanism uses a fraction of the functionality available in these classes. It uses it to build the tree, locate instances of certain elements in the tree, find the types of variables, and determine the result type of expressions. This module is are also used in the policy compiler (Section 6.5).

## A.5 Graphical LaSCO policy editor for Java implementation notes

The policy being viewed is maintained using the *LaSCO* class, and the generic policy engine (Section 5.2) is used for the integrated function mentioned last above. The schema displayed is a representation of an instance of the *SchemaGraph* class. This class processes a Java source file if it is given as input.

The graph layout tool DOT [18] is used for constructing the node and edge graph layout for both the schema graph and the policy graph. The layouts that DOT produces are not necessarily stable across even minor changes to the graph and there is a noticeable delay while DOT lays out the graph, so it is not used after every change to the graph. A simple and unoptimized modification to the layout is made in most cases instead. DOT is called whenever a new graph is formed and on demand from the user.

More details of how a *LaSCO* policy is statically checked against a schema graph (Section 6.4.3.3) is now presented. A system graph is created from the schema graph by repeatedly inserting edges corresponding to method invocations from the schema graph. It is repeatedly inserted since a method may be invoked any number of times. However, we only need to insert it an limited number of times given the size of the policy. An upper bound on the number of inserts required is the maximum number of edges in a given direction between any two nodes in the policy graph. The edges inserted take advantage of the *LaSCO* class match option (Section 5.2.5.2) to only report a match with to a particular method invocation once, although it may appear multiple times as edges in the system graph. The *id* attribute on objects is given a value, and the *time* attribute on events is artificially set to zero. *check_with_system* is called on the LaSCO graph. The matches of the system to the policy domain and the matches that violate the policy are reported.

## A.6 Policy compiler implementation

Details about the policy compiler implementation (Section 6.5) are presented here.

### A.6.1 Program modification

To instrument source code for policy checks, the user invokes the program "wrap_invs.pl". This accepts three type of inputs on the command line:

- java source files to be modified,

- LaSCO policies in the LaSCO file format (Section A.2), such as those produced by the LaSCO policy editor (Section 6.4), to be enforced, and

- schema files, for use as reference about classes not in the source files.

LaSCO policies are represented using the *LaSCO* class. Java source and schema files are represented using the various classes in the *ParseNode* module (Section A.4.3). This module is used extensively to query and manipulate the source code.

Wrap_invs.pl processes one source file at a time. For each method invocation and constructor invocation found in a given compilation unit, the following steps are taken:

1. The method name and class invoked are determined. Depending on the form of method invocation expression, this can involve calculating the result type of an expression. In cases such as this, the actual class of the object invoked might be different than the one computed statically, but it is always has the computed class as a base class. This does not cause any particular difficulties.

2. Every edge on every provided policy is considered with respect to the method invocation. Those that might match the given invocation are noted. To determine this, a one edge *SystemGraph* is constructed representing such an invocation (using the information known statically) and is passed to a special function of the generic policy engine. If no policy edges might match the invocation, then processing of this call is terminated and the invocation left unchanged.

3. The code surrounding the method invocation expression is modified (if needed) to permit a policy check to be inserted prior to the invocation but after argument computation. Specifically, the method invocation (which might be deeply buried in an expression) is pro-moted to be a statement. Any actual parameters that need computation or whose values might change as a result of a computation (anything except literal) are promoted to variable assignment statements before the invocation. Recall the example presented in Figure 6-12.

4. A call to the run time system is added. The method invocation is replaced with a block containing the method invocation preceded by a call to *JavaLaSCO.checkNewEvent*. This call is necessarily preceded by statements that set up the arguments, including determining the value of certain attributes of the source and destination and the event. As this check is limited to certain policy edges (determined in step 2), the attributes whose values are passed are those mentioned in the corresponding place in any of those policy edges. For edges, a statically determined expression finds the value of their attributes and is inserted. Nodes however, employ calls to *JavaLaSCO.getObjInfoList* and *JavaLaSCO.getClassInfoList* to determine their attribute values at run time. Which one is used depends on the situation. For source, *JavaLaSCO.getClassInfoList* is used from static code and *JavaLaSCO.getClassInfoList* from instance code. For the destination, the determining factor is if the method called is static or instance.

For each class in every source file that contains a "main" method, some initialization statements are added to that at the start. This is done so that the policy checking will be set up properly regardless of what class is ran. The initialization loads policies into the run time system from a string and labels policy edges (see next section).

### A.6.2 Run time policy checking

*JavaLaSCO* is called at run time to check policies against new events. As the source of calls to the run time system is written in Java and the policy checker is written in Perl, language-level interfacing is needed as well. The Java-Perl Lingo[1] (JPL) is used to bridge this gap. JavaLaSCO.jpl contains some methods written in Java and some in Perl. Additional Perl routines are in JavaLaS-

---

1. JPL is a new package by the developers of Perl. In its present form, JPL source files are like Java source files where method bodies can optionally be written in Perl. These methods are made available to Java through the Java Native Interface (JNI).

- **init():** initializes the class data structures and loads needed Perl modules
- **add_policy(name,val):** introduces a policy with the given name to the class and initializes the system history. The second argument is either the text of the policy or a file name.
- **delete_policy(name):** causes the policy with the given name to be deleted
- **clear_history(name):** resets the history for the policy with the given name
- **label_policy_edge(label,name,source,dest,pred):** give certain edges (those whose domain predicates equal the text given by the last argument in the policy with the given name and with the given source and destination) the given label
- **getObjInfoList(object,wanted-attrs):** for each of the given wanted attributes that are defined for the object, add it to an info list which is returned. This includes static and instance class data members and *class*, *classes*, and *id*.

- **getClassInfoList(classname,wanted-attrs):** for each of the given wanted attributes that are defined for the given class, add it to an info list which is returned. This includes static classes members and other defined attributes.
- **add_new_event(name,time,edgeinfo,srcinfo,destinfo):** add a new event with given time and the attributes in the given info lists to the history of the policy with the given name
- **checkAllHist(name):** check the entire history for the given policy against the policy and throw an exception if it is violated
- **checkNewEvent(labels,time,edgeinfo,srcinfo,destinfo):** add a new event to history of the policies whose edges are labeled by the given labels and check the new event against the policies and throw an exception if any policy is violated. The first argument can also be a policy name.

**Figure A-4.** Some methods *JavaLaSCO*.

CO.pm. Some of the interesting methods of the *JavaLaSCO* class that are available to Java programs are described in Figure A-4.

To identify the policy edges that could match a particular invocation from Java, a label must be assigned to the various policy edges (edges have no labels in the LaSCO file format and Java has no pointers). The Java side calls *label_policy_edge* to assign a label. The edge is identified by the policy name, source and destination label, and domain text. All edges that are identified by those same four items are identical as far as the locations in which they match, so are given the same label.

Certain *JavaLaSCO* methods require a list of attributes with values associated with them. This is represented in a certain format that was chosen for easy passing between Java and Perl. This is called an info list. In Java it is an array of Strings and in Perl it is a list of scalars. Information about a particular attribute uses three adjacent elements of the list. The first is the name of the attribute. The second is either "str", "num", "bool", or "set", indicating whether the attribute value is a string, number, boolean, or set, respectively. The third element is the value of the attribute, encoded as a string. To obtain this value for nodes using the *getObjInfoList* and *getClassInfoList* methods, the Java reflection mechanism is used. Reflection is used to determine whether a certain class member is defined and to access its value It is also used to find the base classes of a given class.

As described in Figure Figure A-4., *checkNewEvent* adds a new event and checks it against policies. The checking is conducted by the *check_with_system* method of the *LaSCO* class (see Section 5.2.5). As the policy edges that might match the new system edge are known, *checkNew-*

*Event* takes advantage of this when calling *check_with_system* by passing that association as an edge hint.